\pdfoutput=1

\documentclass[numberedappendix,twocolappendix,apj]{emulateapj}
\usepackage{apjfonts}
\usepackage[colorlinks=true,urlcolor=blue,linkcolor=blue,citecolor=blue]{hyperref}
\usepackage{amsmath}
\usepackage{natbib}
\usepackage{graphicx}

\usepackage{etoolbox}
\makeatletter
\patchcmd{\NAT@citex}
  {\@citea\NAT@hyper@{\NAT@nmfmt{\NAT@nm}\NAT@date}}
  {\@citea\NAT@nmfmt{\NAT@nm}\NAT@hyper@{\NAT@date}}
  {}
  {}
\patchcmd{\NAT@citex}
  {\@citea\NAT@hyper@{%
     \NAT@nmfmt{\NAT@nm}%
     \hyper@natlinkbreak{\NAT@aysep\NAT@spacechar}{\@citeb\@extra@b@citeb}%
     \NAT@date}}
  {\@citea\NAT@nmfmt{\NAT@nm}%
   \NAT@aysep\NAT@spacechar%
   \NAT@hyper@{\NAT@date}}
  {}
  {}
\patchcmd{\NAT@citex}
  {\@citea\NAT@hyper@{%
     \NAT@nmfmt{\NAT@nm}%
     \hyper@natlinkbreak{\NAT@spacechar\NAT@@open\if*#1*\else#1\NAT@spacechar\fi}%
       {\@citeb\@extra@b@citeb}%
     \NAT@date}}
  {\@citea\NAT@nmfmt{\NAT@nm}%
   \NAT@spacechar\NAT@@open\if*#1*\else#1\NAT@spacechar\fi%
   \NAT@hyper@{\NAT@date}}
  {}
  {}
\makeatother

\graphicspath{{./}}

\def\Msf{{M}_{\rm sf}}
\def\Mg{{M}_{\rm g}}

\def\MH2{M_{\rm H_2}}

\def\SFR{\dot{M}_\star}
\def\dotMsf{\dot{M}_{\rm sf}}

\def\Fp{F_{\rm +}}
\def\Fm{F_{\rm -}}
\def\Fmdyn{F_{\rm -,d}}
\def\Fmfb{F_{\rm -,fb}}

\def\rhoSFR{{\dot{\rho}}_\star}
\def\rhoH2{\rho_{\rm H_2}}
\def\epsff{\epsilon_{\rm ff}}

\def\avir{\alpha_{\rm vir}}
\def\avirsf{\alpha_{\rm vir,sf}}
\def\cs{c_{\rm s}}
\def\st{\sigma_{\rm t}}
\def\stot{\sigma_{\rm tot}}

\def\nsf{n_{\rm sf}}

\def\SH2{\Sigma_{\rm H_2}}

\def\SSFR{\dot\Sigma_\star}

\def\fsf{f_{\rm sf}}
\def\fH2{f_{\rm H_2}}

\def\b{b}

\def\Nc{N_{\rm c}}
\def\tsf{t_{\rm sf}}

\def\tglob{\tau_{\rm dep}}
\def\tH2{\tau_{\rm dep,H_2}}
\def\tL{\tH2(L)}
\def\tdep{t_\star}
\def\taust{\tau_\star}
\def\tp{\tau_{+}}
\def\tm{\tau_{-}}
\def\tmd{\tau_{\rm -,d}}
\def\tmfb{\tau_{\rm -,fb}}
\def\tff{t_{\rm ff}}
\def\tauff{\tau_{\rm ff}}
\def\taufflow{\tau_{\rm ff}^{\rm dr}}
\def\tauffhi{\tau_{\rm ff}^{\rm sr}}

\def\sigtot{\sigma_{\rm tot}}

\def\pc{{\rm\;pc}}
\def\kpc{{\rm\;kpc}}
\def\Myr{{\rm\;Myr}}
\def\Gyr{{\rm\;Gyr}}
\def\Msun{\;M_\odot}
\def\Msunyr{\;M_\odot{\rm\;yr^{-1}}}
\def\Msunpc2{\;M_\odot{\rm\;pc^{-2}}}
\def\kms{{\rm\;km\;s^{-1}}}
\def\cc{{\rm\;cm^{-3}}}
\def\K{{\rm\;K}}

\def\ns{$n$--$\sigtot$ }
\def\const{{\rm const}}
\def\Lstar{$L_\star$}

\def\Fmolp{\Fp^{\rm H_2}}
\def\Fmolm{\Fm^{\rm H_2}}
\def\tmolp{\tp^{\rm H_2}}
\def\tmolm{\tm^{\rm H_2}}

\defcitealias{Semenov.etal.2017}{I}
\def\P1{Paper~\citetalias{Semenov.etal.2017}}

\begin{document}


\shorttitle{Connection between local and global star formation}
\shortauthors{Semenov, Kravtsov, Gnedin}
\slugcomment{Accepted for publication in the Astrophysical Journal}

\title{How galaxies form stars:\\ the connection between local and global star formation in galaxy simulations}

\author{Vadim A. Semenov\altaffilmark{1,2,$\star$}, Andrey V. Kravtsov\altaffilmark{1,2,3} and Nickolay Y. Gnedin\altaffilmark{1,2,4}}

\keywords{galaxies: evolution -- ISM: kinematics and dynamics -- stars: formation -- methods: numerical}

\altaffiltext{1}{Department of Astronomy \& Astrophysics, The University of Chicago, Chicago, IL 60637 USA}
\altaffiltext{2}{Kavli Institute for Cosmological Physics, The University of Chicago, Chicago, IL 60637 USA}
\altaffiltext{3}{Enrico Fermi Institute, The University of Chicago, Chicago, IL 60637 USA}
\altaffiltext{4}{Fermilab Center for Particle Astrophysics, Fermi National Accelerator Laboratory, Batavia, IL 60510-0500 USA}
\altaffiltext{$\star$}{semenov@uchicago.edu}


\begin{abstract}
Using a suite of isolated \Lstar~galaxy simulations, we show that global depletion times and star-forming gas mass fractions in simulated galaxies exhibit systematic and well-defined trends as a function of the local star formation efficiency per freefall time, $\epsff$, strength of stellar feedback, and star formation threshold. We demonstrate that these trends can be reproduced and explained by a simple physical model of global star formation in galaxies. Our model is based on mass conservation and the idea of gas cycling between star-forming and non-star-forming states on certain characteristic time scales under the influence of dynamical and feedback processes. Both the simulation results and our model predictions exhibit two limiting regimes with rather different dependencies of global galactic properties on the local parameters. When $\epsff$ is small and feedback is inefficient, the total star-forming mass fraction, $\fsf$, is independent of $\epsff$ and the global depletion time, $\tglob$, scales inversely with $\epsff$. When $\epsff$ is large or feedback is very efficient, these trends are reversed: $\fsf \propto \epsff^{-1}$ and $\tglob$ is independent of $\epsff$ but scales linearly with the feedback strength. We also compare our results with the observed depletion times and mass fractions of star-forming and molecular gas and show that they provide complementary constraints on $\epsff$ and the feedback strength. We show that useful constraints on $\epsff$ can also be obtained using measurements of the depletion time and its scatter on different spatial scales. 
\end{abstract} 

\section{Introduction}
\label{sec:introduction}
\setcounter{footnote}{0}

Understanding how galaxies build up their stellar component is a key to understanding galaxy evolution. Formation of stars in galaxies is a complex multiscale process, as stars are formed from gravitationally bound gaseous cores on subparsec scales, while the formation of such cores is aided by bulk gas motions of the interstellar medium (ISM) on hundreds of parsec scales. Despite this complexity, the star formation rate (SFR) per unit gas mass on kiloparsec and larger scales appears to be surprisingly universal: the gas depletion time, $\tglob = \Mg/\SFR$, has a characteristic value and exhibits a relatively small scatter \citep[see, e.g.,][for a review]{Kennicutt.Evans.2012}. This universality is manifested in the tight Kennicutt--Schmidt relation \citep{Schmidt.1959,Kennicutt.1989,Kennicutt.1998} between the surface densities of gas and the star formation rate. 

Existence of such a tight relation implies that the small-scale star formation, averaged over all individual star-forming regions, is closely related to the total gas mass in galaxies. Numerical simulations and semianalytic models of galaxy evolution show that this gas mass is controlled by (1) the net galactic gas supply rate, determined by the rates of inflow and outflow, and (2) the star formation rate or, alternatively, the global depletion time. Over the past decade, our understanding of inflows and feedback-driven outflows has improved dramatically, although qualitative and quantitative details of the relevant physical processes are still the subject of an active and lively debate \citep[see, e.g.,][for recent reviews]{Somerville.Dave.2015,Naab.Ostriker.2017}. Likewise, understanding global star formation rates and depletion times requires insight into the interplay between ISM gas flows and local star formation and feedback processes. Understanding of this interplay can be greatly aided with numerical simulations of galaxies, as we will illustrate in this paper.

Modeling of local star formation and feedback processes in galaxy simulations is admittedly rather crude. With some variations and few exceptions, star formation prescriptions usually follow ideas introduced for the first generation of simulations \citep{Cen.Ostriker.1992,Katz.1992}: star formation occurs only in star-forming gas, defined using some conditions, e.g., that gas density (temperature) is larger (smaller) than some threshold, that gas within some region is gravitationally bound, that gas is in molecular phase, etc. \citep[see, e.g.,][]{Hopkins.etal.2013}. Star-forming gas is then converted into stellar particles using a stochastic Poisson process with the rate 
\begin{equation}
\rhoSFR = \frac{\rho}{\tdep},
\label{eq:rhoSFR1}
\end{equation}
where $\rho$ is the density of the gas that is deemed to be star-forming according to the adopted criteria, and $\tdep$ is its {\it local} depletion time. In most recent studies, this time is parameterized as $\tdep = \tff/\epsff$, where $\epsff$ is the star formation efficiency per freefall time, $\tff \equiv \sqrt{3 \pi/32 G \rho}$. Likewise, the stellar feedback is modeled by simply injecting thermal and kinetic energy and momentum into gas resolution elements adjacent to a young star particle \citep[e.g.,][]{Hopkins.etal.2011,Hopkins.etal.2017b,Agertz.etal.2013,Simpson.etal.2015} or using a subgrid prescription with a specific model of ISM on scales below resolution \citep[e.g.,][]{Yepes.etal.1997,Springel.Hernquist.2003,Braun.Schmidt.2012}. 

Despite a rather simplistic modeling of star formation and feedback on scales close to the spatial resolution, modern galaxy formation simulations  generally predict $\tglob$ and the Kennicutt--Schmidt relation on kiloparsec and larger scales in a reasonable agreement with observations \citep[e.g.,][]{Governato.etal.2010,Stinson.etal.2013,Hopkins.etal.2014,Hopkins.etal.2017,Agertz.Kravtsov.2015,Agertz.Kravtsov.2016,Grand.etal.2017,Orr.etal.2017}. Although in certain regimes the normalization and slope of the Kennicutt--Schmidt relation on galactic scales simply reflect the adopted value of $\tdep$ on small scales (Equation~\ref{eq:rhoSFR1}) and its assumed density dependence \citep{Schaye.DallaVecchia.2008,Gnedin.etal.2014}, in other regimes there is no direct connection between $\tdep$ and the global Kennicutt--Schmidt relation \citep{Hopkins.etal.2017,Orr.etal.2017,Semenov.etal.2017}. The fact that simulations in the latter regime still result in the global depletion time scale close to the observed values is nontrivial. This agreement indicates that such simulations can be used to shed light on the physical processes connecting local parameters of star formation and feedback to the global star formation in galaxies.  

This connection and associated processes are the focus of this paper, and our goal is to extend and make sense of the results of other recent studies of this issue \citep[see, e.g.,][]{Hopkins.etal.2011,Hopkins.etal.2017,Agertz.etal.2013,Agertz.Kravtsov.2015,Benincasa.etal.2016,Li.etal.2017b,Li.etal.2017}. We use a suite  of isolated \Lstar-sized galaxy simulations with systematically varied $\epsff$ value, star formation threshold, and feedback strength to show that the global depletion time and the star-forming gas mass fraction in simulated galaxies exhibit systematic and well-defined trends as a function of these parameters. We also demonstrate that these trends can be reproduced both qualitatively and quantitatively with a physical model presented in \citet[][hereafter \P1]{Semenov.etal.2017} that explains the origin of long gas depletion times in galaxies. 

Our model is based on the mass conservation equations relating the star-forming and non-star-forming components of the ISM and the idea of gas cycling between these components on certain characteristic time scales under the influence of dynamical and feedback processes \citep[such gas cycling was also envisioned by][]{Madore.2010,Kruijssen.Longmore.2014,Elmegreen.2015,Elmegreen.2018}. Our model explicitly relates the global depletion time to the parameters of local star formation and feedback. 

The success of this relatively simple framework in explaining the long depletion time scale of observed galaxies (\P1) and in reproducing the trends exhibited in the simulations presented in this paper and other recent studies (see Section~\ref{sec:disc:prev}) implies that conversion of gas into stars in real galaxies is a result of dynamic gas cycling between the star-forming and non-star-forming states on short time scales. The long time scale of gas depletion is partly due to the low efficiency of star formation in star-forming regions and partly due to rather short lifetime of these regions limited by stellar feedback. The combination of these factors results in gas going through many cycles before complete conversion into stars.

Some of the previous studies \citep[e.g.,][]{Hopkins.etal.2017} argued that in galaxy simulations low local efficiency of star formation is not required for global inefficiency because stellar feedback disperses star-forming gas before it is converted into stars. Global star formation rate in this regime becomes independent of the local efficiency but scales with the feedback strength. This phenomenon is usually referred to as ``self-regulation,'' and our model explains its origin in simulations with high local efficiency and strong feedback. At the same time, observations indicate that star formation efficiency is low in star-forming regions. Thus, we show that our model also explains the global depletion time in the case of low local efficiency.

We will also discuss how the trends identified in simulations and our analytic model can be used to guide the choice of star formation and feedback parameters in high-resolution galaxy simulations. In particular, we will show that both the global depletion times and the star-forming gas mass fractions of observed galaxies should be used on kiloparsec and larger scales, while the measurements of the depletion time and its scatter on smaller spatial scales provide additional constraints on the local efficiency of star formation.

The paper is organized as follows. In Section~\ref{sec:sims}, we describe our simulation suite and the adopted star formation and feedback prescriptions. In Section~\ref{sec:results}, we present the trends of the global depletion times, star-forming mass fractions, and freefall times in star-forming gas with the parameters of the star formation and feedback prescriptions used in simulations. In Section~\ref{sec:model}, we summarize the model for global star formation presented in \P1 and show that it can reproduce the trends in our simulation results both qualitatively and quantitatively. In Section~\ref{sec:observ}, we compare our simulation results and model predictions to the observed star-forming properties of real galaxies and identify a combination of the star formation efficiency and the feedback strength that satisfies all considered observational constraints. In Section~\ref{sec:disc:prev}, we compare our predictions with the results of previous recent studies and interpret their results in the context of our model. Finally, in Section~\ref{sec:summary}, we summarize our results and  conclusions.

\section{Simulations}
\label{sec:sims}

\subsection{Method Overview}
\label{sec:sims:methods}

To understand the connection between local and global star formation, we explore the effects of local star formation and feedback parameters in a suite of \Lstar-sized isolated galaxy simulations performed with the adaptive mesh refinement (AMR) gasdynamics and $N$-body code ART \citep{Kravtsov.1999,Kravtsov.etal.2002,Rudd.etal.2008,Gnedin.Kravtsov.2011}. In this section, we briefly summarize the adopted initial conditions and subgrid models, and for details we refer the reader to Section~3 of \P1. 

Our simulations start from the initial conditions of the AGORA code comparison project \citep{agora2}, in which an \Lstar-sized exponential galactic disk with a stellar bulge is embedded into a dark matter halo. The disk scale height and radius are $h_{\rm d} \approx 340\pc$ and $r_{\rm d} = 10\;h_{\rm d}$, respectively, and its total mass is $M_{\rm d} \approx 4.3 \times 10^{10} \Msun$, 20\% of which is in the gaseous disk and the rest is in the initial population of old stellar particles. The stellar bulge has a \citet{Hernquist.1990} density profile, with the scale radius of $a = h_{\rm d}$ and the total mass of $M_{\rm \star,b} \approx 4.3 \times 10^9 \Msun$. The dark matter halo is initialized with a Navarro--Frenk--White profile \citep{NFW.1996,NFW.1997}, with the characteristic circular velocity of $v_{c,200} = 150 \kms$ and the concentration of $c=10$. 

In our simulations, we adaptively resolve cells whose gas mass exceeds $\sim 8\,300 \Msun$ and reach the maximum resolution of $\Delta=40\pc$. The Poisson equation for the gravity of gas and stellar and dark matter particles is solved using a fast Fourier transform on the zeroth uniform level of the AMR grid and using the relaxation method on all refinement levels. The resolution for gravity is therefore also set by the local resolution of the AMR grid, and in the ART code it corresponds to $\sim 2$ grid cells \citep[see Figure~6 in][]{Kravtsov.etal.1997}.

For cooling and heating, we adopt the \citet{Gnedin.Hollon.2012} model assuming constant metallicity of $Z = Z_\odot$ and the background radiation field with the photodissociation rate of $10^{-10}\ \rm s^{-1}$ \citep{Stecher.Williams.1967}. Molecular gas shielding is modeled using a prescription calibrated against radiative transfer ISM simulations \citep[the ``L1a'' model in ][]{Safranek-Shrader.etal.2017}. In each computational cell, we dynamically follow unresolved turbulence using the ``shear-improved'' model of \citet{Schmidt.etal.2014}, whose implementation in the ART code is discussed in \citet{Semenov.etal.2016}. Subgrid turbulence dynamically acts on resolved gas motions, and its distribution allows us to predict velocity dispersions in star-forming regions that we account for in our star formation prescription (Section~\ref{sec:sims:sf-fb}).

Analysis of time evolution shows that all our simulations exhibit a short ($\lesssim 300 \Myr$) initial transient stage, after which the simulated galaxy settles into a quasi-equilibrium state with approximately constant global galaxy parameters, such as gas depletion time, $\tglob$ (see, e.g., Figure~\ref{fig:taudep_fid} below). Thus, in our subsequent analysis we average the equilibrium values of the parameters of interest between 300 and 600 Myr, as this time interval is sufficiently long to average out the temporal variability of such quantities, but it is also shorter than $\tglob$, and hence the galaxy maintains the approximate equilibrium over this time interval. The only exceptions are the runs without feedback and with high local star formation efficiency of $\epsff \geq 10\%$, in which $\tglob$ is very short and the total gas mass decreases appreciably between 300 and 600 Myr. The equilibrium assumption is also violated for the central region in simulations with $\epsff \leq 0.1\%$, where the central density keeps increasing owing to continuous accretion. However, outside the central 1 kpc the total gas mass and the value of $\tglob$ remain approximately constant, and therefore we exclude gas in the central 1 kpc region when computing quantities in our analysis.

To explore gas flows between different states in the ISM, we use passive gas-tracer particles that are exchanged between adjacent computational cells stochastically, with the probability proportional to the gas mass flux between the cells, as proposed by \citet{Genel.etal.2013}. These tracer particles are initialized proportionally to the local gas density after 400 Myr of disk evolution. By this point, all initial transients have dissipated away and ISM gas distribution has become stationary. Using tracer particles, we average this distribution between 400 and 600 Myr and measure at each step the instantaneous contribution of each tracer into gas fluxes as the second-order time derivatives between the previous and subsequent snapshots. To accurately track these gas fluxes, we output positions, densities, and subgrid velocity dispersion for each gas tracer every 1 Myr. To account for gas consumption, whenever a stellar particle is formed, relative weights of all tracers inside the host cell are decreased correspondingly. 

We note that the analysis presented in this paper differs from that in \P1 in our implementation of gas tracers: we now use stochastic tracer particles instead of classical velocity tracers, and also initialize these particles proportionally to gas density rather than uniformly as in \P1. Both these changes allow us to follow the evolution of gas distribution more accurately. We checked, however, that all the conclusions of \P1 remain valid after these changes.

\subsection{Star Formation and Feedback}
\label{sec:sims:sf-fb}

As our goal is to explore the effects of star formation and feedback model parameters, we adopt a usual parameterization of the local star formation rate with a star formation efficiency per freefall time, $\epsff$,
\begin{equation}
\label{eq:rhoSFR}
\rhoSFR = \epsff \frac{\rho}{\tff},
\end{equation} 
and systematically vary $\epsff$ as will be explained at the end of this section. We allow star formation to occur only in the gas that satisfies a chosen criterion. To explore the effects of such a criterion, we adopt thresholds in either the gas virial parameter, $\avirsf$, or the density, $\nsf$, and also vary the values of $\avir$ and $\nsf$. 
 
As our fiducial star formation criterion, we adopt a threshold in $\avir$ and define all gas with $\avir < \avirsf$ as star-forming. For a computational cell with a side $\Delta$, the local virial parameter is defined as for a uniform sphere of radius $R = \Delta/2$ \citep{Bertoldi.McKee.1992}:
\begin{equation}
\label{eq:avir}
\avir \equiv \frac{5 \stot^2 R}{3GM} \approx 9.35 \frac{ (\stot/10\kms)^2 }{ (n/100\cc) (\Delta/40 \pc)^2}, 
\end{equation}
where $\stot = \sqrt{ \st^2 + \cs^2 }$ is the total subgrid velocity dispersion due to turbulent and thermal motions, and subgrid turbulent velocities, $\st$, are dynamically modeled in each cell following \citet{Schmidt.etal.2014}.

The choice of the star formation threshold in $\avir$ is motivated by theoretical models of star formation in turbulent giant molecular clouds (GMCs), which generically predict an exponential increase of $\epsff$ with decreasing $\avir$ \citep[see][for a review]{Padoan.etal.2014}. We set our fiducial values of parameters to $\epsff=1\%$ and $\avirsf=10$, as supported by the observed efficiencies and virial parameters of star-forming GMCs \citep[e.g.,][]{Evans.etal.2009,Evans.etal.2014,Heiderman.etal.2010,Lada.etal.2010,Lada.etal.2012,Lee.etal.2016,Vutisalchavakul.etal.2016,MivilleDeschenes.etal.2016}, and also consistent with the results of high-resolution GMC simulations \citep[e.g.,][]{Padoan.etal.2012,Padoan.etal.2017}, which show a sharp increase of $\epsff$ below $\avir \sim 10$. Note also that the threshold in $\avir$ is equivalent to a threshold in the local Jeans length that accounts for both the thermal and turbulent pressure support: $\lambda_{\rm J} = \stot \sqrt{ \pi/G\rho } = \pi \Delta \sqrt{\avir/5}$, and thus $\avirsf=10$ implies that gas becomes star-forming when the local Jeans length is resolved by less than $\lambda_{\rm J}/\Delta \approx 4.5$ cells.

In galaxy simulations that do not track subgrid turbulence, the GMC-scale $\avir$ is not readily available owing to insufficient resolution. Instead, such simulations often adopt a star formation threshold in gas density, $n$, and define star-forming gas as the gas with $n > \nsf$. To show that our conclusions remain valid for such a threshold, we explore models with varied density-based thresholds in addition to our fiducial $\avir$-based threshold.

The feedback from young stars is implemented by injection of thermal energy and radial momentum generated during supernova (SN) remnant expansion in a nonuniform medium in the amounts calibrated against simulations by \citet{Martizzi.etal.2015}. The total number of SNe exploded in a single stellar particle is computed assuming the \citet{Chabrier.2003} initial mass function. To mimic the effects of pre-SN feedback, such as radiation pressure and winds from massive young stars, the momentum injection commences at the moment when a stellar particle is created and continues for 40 Myr. 

The explicit injection of the generated radial momentum allows one to partially resolve the overcooling problem and efficiently couple the feedback energy to the resolved dynamics of gas, which explains the growing popularity of the method \citep[e.g.,][]{Simpson.etal.2015,Grisdale.etal.2017,Hopkins.etal.2017b}. However, the injected momentum is still partially lost as a result of advection errors \citep[see, e.g.,][]{Agertz.etal.2013}, and to compensate for this loss, we boost the momentum predicted by \citet{Martizzi.etal.2015} by a factor of 5. This value is motivated by our idealized tests of a stellar particle exploding in a uniform medium with additional translational motion at velocity $200\kms$, which is comparable to the rotational velocity of the simulated galaxy. Such a fiducial boosting factor also absorbs uncertainties related to SNe clustering \citep{Gentry.etal.2017,Gentry.etal.2018} and the total energy of a single SN. To explore the effects of the feedback strength on the global depletion times, in addition to this fiducial boosting, we multiply the injected momentum by a factor $\b$, which is systematically varied. 

In the end, in our simulations, star formation and feedback are parameterized by three numbers: the star formation efficiency, $\epsff$, the star formation threshold, $\avirsf$ or $\nsf$, and the feedback boost factor, $\b$, which we vary in order to explore their effects on the global star formation. To assess the effect of the local star formation efficiency, we vary $\epsff$ from $0.01\%$ to $100\%$, i.e., by four orders of magnitude around our fiducial value of $\epsff=1\%$. To explore the effects of the star-forming gas definition, we vary $\avirsf$ between 10 and 100 and $\nsf$ between $10\cc$ and $100\cc$. We expect that such $\avir$ and $n$ are well resolved in our simulations, because they are sufficiently far from the resolution-limited values of $\avir\sim2$ and $n\sim 10^4\cc$ in a simulation with $\epsff=0.01\%$, in which gas contraction is not inhibited by stellar feedback (see the bottom left panel of Figure~\ref{fig:phases_fid} below). Finally, in order to explore the effect of the feedback strength, in addition to the fiducial case of $\b=1$, we also consider the 5 times stronger feedback ($\b = 5$), the 5 times weaker feedback ($\b = 0.2$), and the case of no feedback at all ($\b=0$). Such wide variation of model parameters allows us to explore the connection between the subgrid scale and the global star formation in the simulated galaxy.

\section{Effects of star formation and feedback parameters on global star formation }
\label{sec:results}

The analysis presented in this section focuses on the quantities that characterize the global star formation of the simulated galaxy: the global gas depletion time, 
 \begin{equation}
\label{eq:tglob_def}
\tglob \equiv \frac{\Mg}{\SFR},
\end{equation}
as well as the mass fraction of star-forming gas, $\fsf=\Msf/\Mg$, and the mean freefall time of star-forming gas, $\tauff = \langle 1/\tff \rangle_{\rm sf}^{-1}$. Here the star-forming gas mass, $\Msf$, is the total mass of all gas in the galaxy that satisfies the adopted star formation criterion. Consequently, the average freefall time is defined by analogy with Equation~(\ref{eq:rhoSFR}), $\SFR = \epsff \Msf/\tauff$, and thus $\tauff$ depends on the local $\tff$ via $\epsff/\tauff = \SFR/\Msf = \int (\epsff/\tff) \rho dV / \int \rho dV = \epsff \langle 1/\tff \rangle_{\rm sf}$, where the integrals are taken over all star-forming gas. The values of $\tglob$, $\fsf$, and $\tauff$ are closely related. For example, the global depletion time can be expressed as  
\begin{equation}
\label{eq:tglob_epsff}
\tglob \equiv \frac{\Mg}{\SFR} = \frac{\Msf}{\SFR} \frac{\Mg}{\Msf} = \frac{\tauff}{\epsff \fsf}.
\end{equation}

\begin{figure}
\centering
\includegraphics[width=\columnwidth]{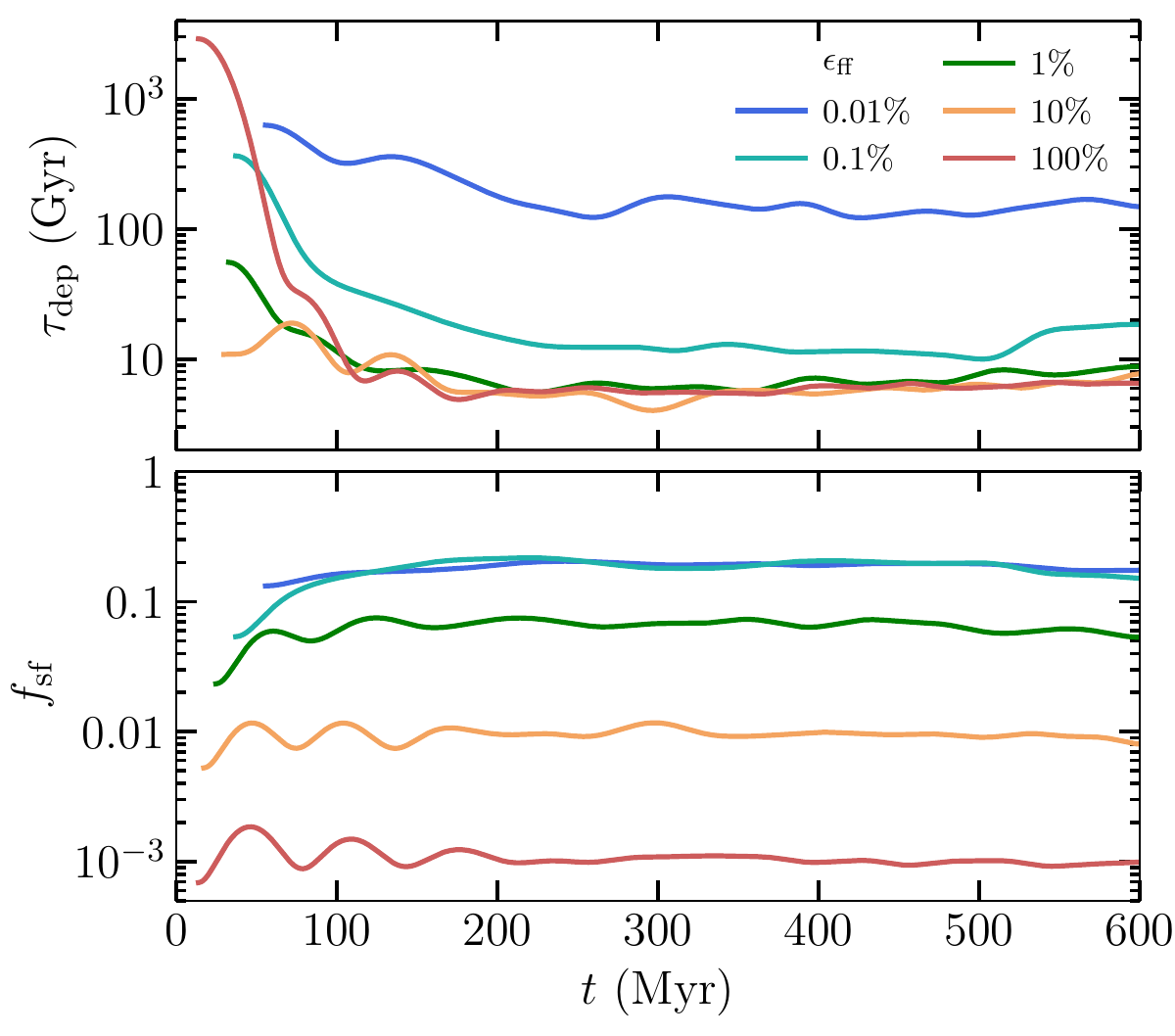}
\caption{\label{fig:taudep_fid} Evolution of the global depletion time, $\tglob$, and the star-forming mass fraction, $\fsf$, in the simulations with varying $\epsff$ at the fiducial feedback strength ($\b=1$) and star formation threshold ($\avirsf=10$). To compare different runs at the same temporal resolution, all curves are smoothed using a Gaussian filter with a width of $30\Myr$. All quantities are measured in a cylindrical volume centered at the disk center with $|z| < 2\kpc$ and $1 < R < 20\kpc$.}
\end{figure}

\begin{figure}
\centering
\includegraphics[width=\columnwidth]{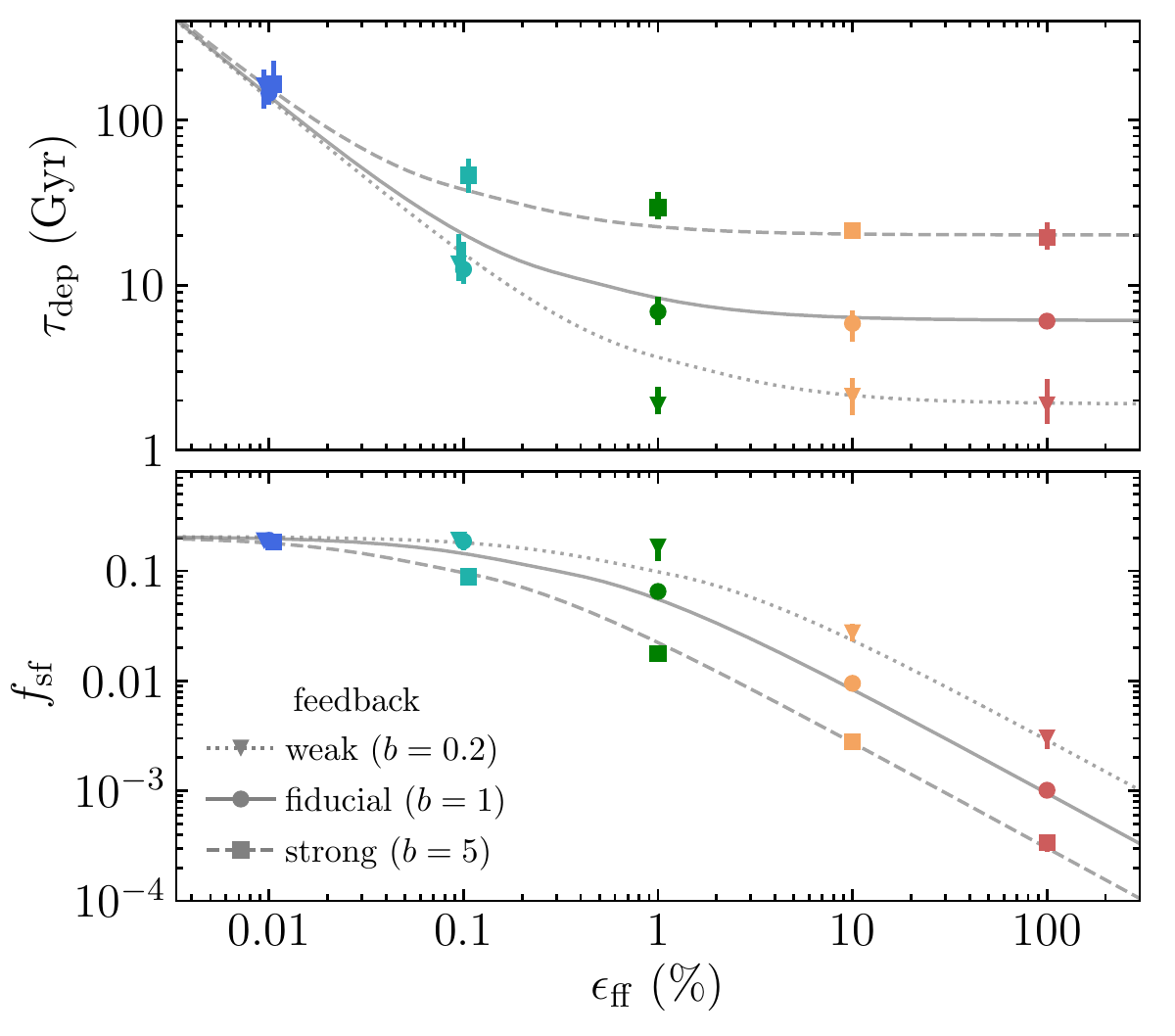}
\caption{\label{fig:taudep_epsff} Dependence of the equilibrium $\tglob$ and $\fsf$ values on the $\epsff$ value in our simulations with fiducial star formation threshold ($\avirsf=10$) and different feedback boosts: weak ($b=0.2$; triangles), fiducial ($b=1$; circles), and strong ($b=5$; squares). The values of $\tglob$ and $\fsf$ are time-averaged between 300 and 600 Myr, with error bars indicating 5$^{\rm th}$ and 95$^{\rm th}$ percentiles over this time interval. The choice of the averaging interval is explained in Section~\ref{sec:sims:methods}. Gray lines show the predictions of our model, which will be explained in Section~\ref{sec:model}. The figure illustrates qualitatively different behavior of $\tglob$ and $\fsf$ at low and high $\epsff$. }
\end{figure}

\begin{figure*}
\centering
\includegraphics[width=0.36\textwidth]{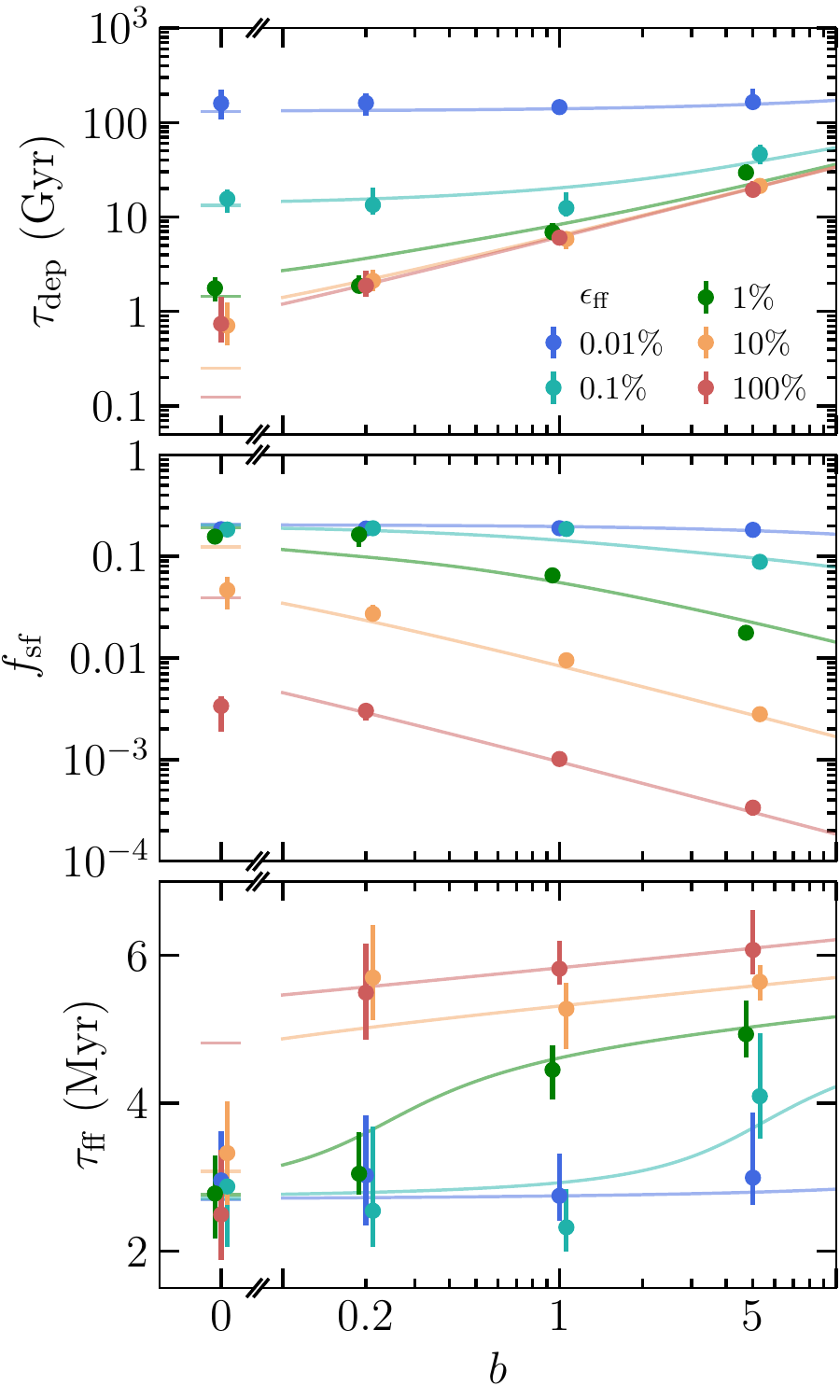}%
\includegraphics[width=0.34\textwidth]{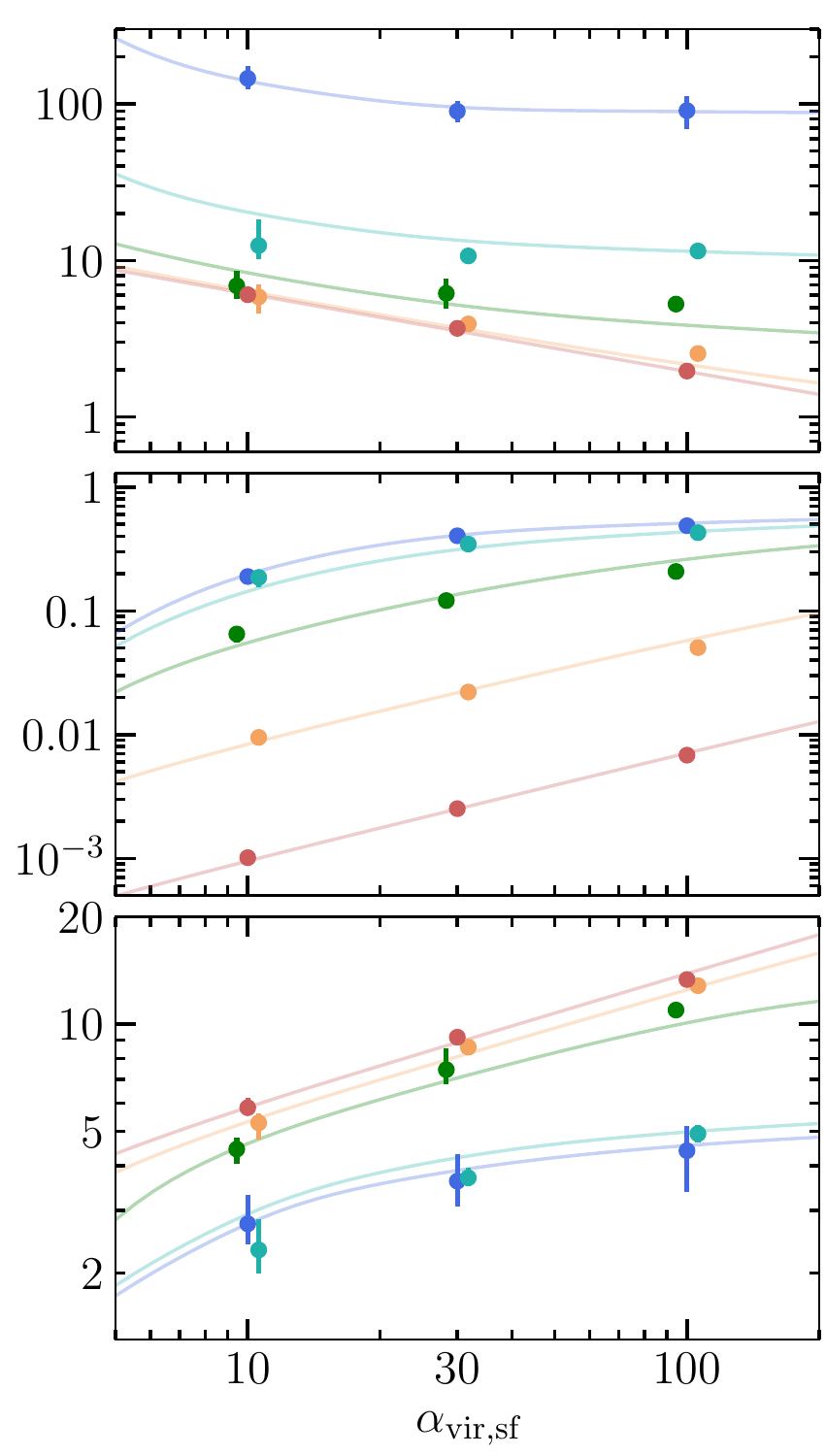}%
\includegraphics[width=0.3\textwidth]{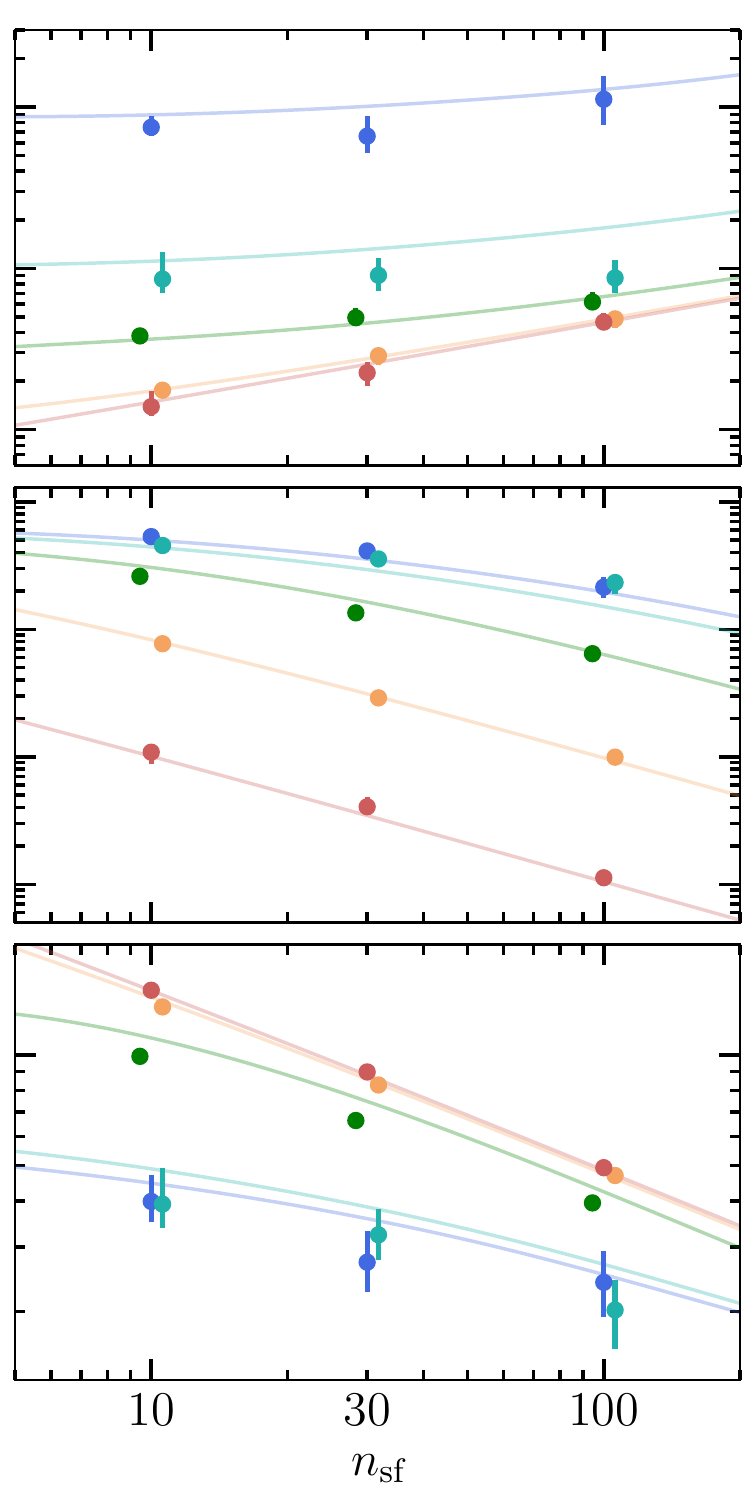}
\caption{\label{fig:fb} Equilibrium values of $\tglob$, $\fsf$, and $\tauff$, in simulations with varying $\epsff$ and different feedback strengths ($\b$; left column) and star formation thresholds set in virial parameter ($\avir < \avirsf$; middle columns) and gas density ($n>\nsf$; right column). The feedback strength is varied at the fiducial threshold value ($\avirsf=10$), whereas the threshold is varied at the fiducial feedback strength ($\b=1$). Points indicate the values of $\tglob$, $\fsf$, and $\tauff$ time-averaged between 300 and 600 Myr, with error bars indicating 5$^{\rm th}$ and 95$^{\rm th}$ percentiles over this time interval. To avoid overlap, points for $\epsff = 0.1\%$, 1\%, and 10\% are slightly shifted horizontally around the actually used values of $\b$, $\avirsf$, and $\nsf$. Lines show our analytical model detailed in Section~\ref{sec:model} and summarized in Appendix~\ref{app:model}.}
\end{figure*}

Below, we describe the trends of $\tglob$, $\fsf$, and $\tauff$ with the main parameters of the star formation and feedback prescriptions  in our \Lstar-sized galaxy simulations: the efficiency $\epsff$, the feedback strength parameter $\b$, and the star formation threshold $\avirsf$ or  $\nsf$. The efficiency $\epsff$ affects local star formation in the most direct way, while the feedback strength $\b$ affects the {\it integral} local star formation efficiency by controlling the time that gas spends in the star-forming state. The interplay between star formation and feedback also affects the overall distribution of gas in a galaxy. For a given distribution, the star formation thresholds control the mass fraction, $\fsf$, and the mean density of star-forming gas, and thus its mean freefall time, $\tauff$. 

Figure~\ref{fig:taudep_fid} shows the evolution of $\tglob$ and $\fsf$ in simulations with varying $\epsff$ at the fixed fiducial feedback strength ($\b = 1$) and the star formation threshold ($\avirsf = 10$). After the initial transient stage, $\tglob$ and $\fsf$ become approximately constant in time at values that depend on the choice of $\epsff$. To explore this dependence on $\epsff$, we average the equilibrium values of $\tglob$ and $\fsf$ between 300 and 600 Myr\footnote{The choice of this time interval is explained in Section~\ref{sec:sims:methods}.} and show them in Figure~\ref{fig:taudep_epsff} with error bars indicating temporal variability around the average. In addition to simulations with fiducial feedback (circles), the figure also shows the results for 5 times weaker (triangles) and 5 times stronger feedback (squares). Star formation histories in these and all our other simulations are qualitatively similar to those shown above, and thus for quantitative comparison from now on we will consider only the equilibrium values of $\tglob$ and $\fsf$. Gray lines in this figure show the predictions of our model that will be described and discussed in Section~\ref{sec:model}. 

Figure~\ref{fig:taudep_epsff} clearly shows that the dependence of $\tglob$ and $\fsf$ on $\epsff$ is qualitatively different when $\epsff$ is low and when it is high. When $\epsff$ is low, $\leq 0.01\%$, $\tglob$ scales as $\epsff^{-1}$, whereas the star-forming mass fraction remains independent of $\epsff$. When $\epsff$ is high, $\epsff \geq 1\%$, the trends are reversed: $\tglob$ is independent of $\epsff$, whereas $\fsf$ scales as $\epsff^{-1}$. Such independence of $\tglob$ from $\epsff$ has been referred to as {\it self-regulation} in the literature.

The figure also shows that this dependence on $\epsff$ remains qualitatively similar at different feedback strengths, and the limiting regimes of low and high $\epsff$ exist at all $\b$. However, for stronger feedback, the transition to the self-regulation regime occurs at smaller $\epsff$ and depletion time at high $\epsff$ increases.

This increase of $\tglob$ with feedback strength at high $\epsff$ is easier to quantify in the top left panel of Figure~\ref{fig:fb}, which shows $\tglob$ as a function of feedback boost $b$ at different $\epsff$. As before, the error bars indicate temporal variability around the average, and lines show the predictions of our model that will be detailed in Section~\ref{sec:model}. From the figure, depletion time at high $\epsff$ increases almost linearly with $\b$: $\tglob \sim (6 \Gyr) \;\b^{0.75}$. The middle left panel shows that $\fsf$ exhibits the opposite trend with $b$. The bottom left panel also shows that despite wide variation of $\epsff$ and $\b$, the average freefall time in star-forming gas varies only mildly, from $\tauff \sim 3\Myr$ at low $\epsff$ to $\tauff \sim 5\text{--}6\Myr$ at high $\epsff$.

The middle column of panels in Figure~\ref{fig:fb} shows the variation of $\tglob$, $\fsf$, and $\tauff$ in the runs with different $\epsff$ and values of the adopted star formation threshold: $\avirsf = 10$, 30, and 100. Again, for every value of $\avirsf$, the dependence on $\epsff$ is qualitatively similar to the fiducial case. In the high-$\epsff$ regime, $\tglob$ decreases at higher $\avirsf$, i.e., when the threshold becomes less stringent and makes more gas eligible to star formation. At a less stringent threshold, $\fsf$ and $\tauff$ both increase, and this increase is stronger in the high-$\epsff$ regime. In the right panels of Figure~\ref{fig:fb}, the star formation threshold is set in the gas density rather than in $\avir$, and the behavior of $\tglob$, $\fsf$, and $\tauff$ remains qualitatively the same, but the direction of all trends is opposite since the density-based threshold becomes less stringent at smaller $\nsf$.

The presented results show that the key global star formation properties of our simulated galaxies change systematically with changing parameters of the local star formation and feedback. The trends are well defined and exhibit distinct behavior in the low-$\epsff$ and high-$\epsff$ regimes. In the latter, the global star formation rate and the gas depletion time become insensitive to the variation of $\epsff$, while the mass fraction of the star-forming gas, $\fsf$, is inversely proportional to $\epsff$. In the low-$\epsff$ regime, the trends are reversed: $\tglob$ scales inversely with $\epsff$, while $\fsf$ is almost insensitive to it. The dependence of $\tglob$ on the feedback strength parameter $\b$ is the opposite to the dependence on $\epsff$: in the low-$\epsff$ regime, $\tglob$ is insensitive to $\b$, while in the high-$\epsff$ regime $\tglob$ exhibits a close-to-linear scaling with $\b$.

\section{Analytic model for global star formation in galaxies}
\label{sec:model}

As solid lines in Figures~\ref{fig:taudep_epsff} and \ref{fig:fb} show, the trends of $\tglob$, $\fsf$, and $\tauff$ are well described  by a physical model of gas cycling in the interstellar medium formulated in \P1. This model is based on the basic mass conservation between different parts of the interstellar gas. In this section, we summarize the main equations of our model and its predictions for the global gas depletion time and the star-forming gas mass fraction. We then discuss the qualitative predictions of the model for the trends of $\tglob$, $\fsf$, and $\tauff$ in simulations and provide a physical interpretation of these trends. We then show that with a minimal calibration, our model can reproduce these trends {\it quantitatively}. For convenience, the meanings of quantities used in our model are summarized in Table~\ref{tab:defs} in Appendix~\ref{app:model}. 

\subsection{Description of the Model}
\label{sec:model:summary}

\begin{figure}
\centering
\includegraphics[width=\columnwidth]{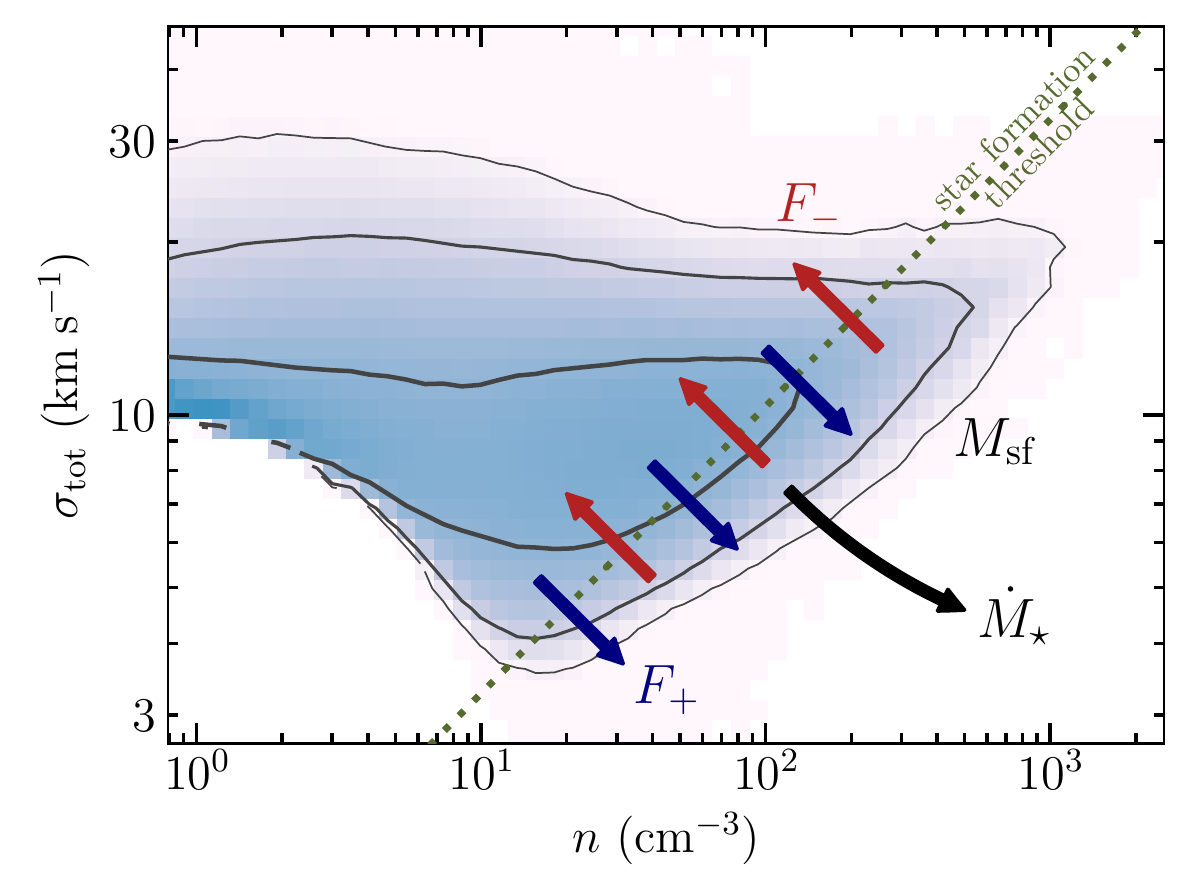}
\caption{\label{fig:model} Illustration of the analytical model of \P1 using the gas distribution from our fiducial simulation in the plane of the gas density, $n$, and the total subgrid velocity dispersion, $\stot=\sqrt{c_s^2+\sigma_{\rm turb}^2}$. The contours enclose 68\%, 95\%, and 99\% of the ISM gas mass. The star-forming gas in this diagram resides below the star formation threshold, $\avir < \avirsf = 10$ (Equation~\ref{eq:avir}), shown by the dotted line. Thick blue and red arrows illustrate the total positive and negative gas mass fluxes through the star formation threshold, while the thick black arrow illustrates gas conversion into stars at a rate $\SFR$.}
\end{figure}

Figure~\ref{fig:model} illustrates our model for $\tglob$, $\fsf$, and $\tauff$ using the distribution of gas in our fiducial simulation in the plane of gas density, $n$, and total velocity dispersion that includes both the thermal and subgrid turbulent motions, $\stot = \sqrt{\cs^2+\st^2}$. The values of $\tglob$, $\fsf$, and $\tauff$ are defined by the distribution of star-forming gas, which resides below the adopted star formation threshold, $\avir < \avirsf = 10$, shown with the dotted line in the figure. This distribution of star-forming gas is shaped by gas motions in the \ns plane, and its total mass, $\Msf = \fsf \Mg$, changes as a result of the gas consumption at a rate $\SFR$ and the net gas flux through the star formation threshold, which in general can be decomposed into a positive and a negative component, $\Fp$ and $\Fm$:
\begin{equation}
\label{eq:masscons}
\dotMsf = \Fp - \Fm - \SFR.
\end{equation}

As the local dynamical time scales of processes controlling $\Fp$ and $\Fm$ are short compared to the global time scales, such as rotation period or gas consumption time, isolated galaxies settle into a quasi-equilibrium state.\footnote{We stress that an assumption of the quasi-equilibrium is not required in general and is made here only to simplify notation. As was shown in \P1, the out-of-equilibrium state of a galaxy (or a given ISM patch) results in an extra term in the final expression for $\tglob$, which contributes to the scatter of the depletion time. For normal star-forming galaxies, this term is small and can become significant only if the global dynamical properties of the galaxy change on a timescale much shorter than the local depletion time. Thus, in case of, e.g., starburst mergers, a more general Equation (10) from \P1 should be used instead of Equation~(\ref{eq:Nc}) below. } In this state, $\dotMsf \approx 0$ over a suitably short time interval, and, therefore, the global SFR is balanced by the net inflow of the star-forming gas, $\SFR \approx \Fp-\Fm$. As was shown in \P1, in normal star-forming galaxies the small net flux required by the observed small SFRs results from the near cancellation of $\Fp$ and $\Fm$, both of which are much larger than the resulting net mass  flux. The total positive flux, $\Fp$, results from a combined effect of gravity, cooling, compression in ISM turbulence, etc., while the negative flux is due to the dispersal of star-forming regions by stellar feedback, $\Fmfb$, and dynamical processes like the turbulent shear, the differential rotation, and the expansion behind spiral arms, $\Fmdyn$: $\Fm = \Fmfb + \Fmdyn$.

The global depletion time and the star-forming mass fraction can be related to the parameters of star formation and feedback if the terms in the above equation are parameterized as
 \begin{align}
\label{eq:Fp} \Fp &\equiv \frac{(1-\fsf) \Mg}{\tp},\\
\label{eq:Fmfb} \Fmfb &\equiv \frac{\fsf \Mg}{\tmfb} = \xi \SFR,\\
\label{eq:Fmdyn} \Fmdyn &\equiv \frac{\fsf \Mg}{\tmd},\\
\label{eq:SFR} \SFR &\equiv \frac{\fsf \Mg}{\taust}.
\end{align}
In these expressions, the adopted star formation prescription determines the average consumption time in the star-forming gas, $\taust$, which in our case is equal to $\tauff/\epsff$, and the strength of the stellar feedback is reflected in the parameter $\xi$, which is analogous to the usual feedback mass-loading factor, but is defined on the scale of star-forming regions. The final expressions for the global depletion time and the star-forming fraction follow from Equation~(\ref{eq:masscons}) after the substitution of Equations~(\ref{eq:Fp}--\ref{eq:SFR}):
\begin{align}
\label{eq:tglob} \tglob &= \Nc \tp + \taust, \\
\label{eq:fsf}   \fsf &= \frac{\taust}{\tglob} = \left( \Nc \frac{\tp}{\taust} + 1 \right)^{-1},
\end{align}
where $\Nc$ in the steady state with $\dot{M}_{\rm sf} \approx 0$ is given by
\begin{equation}
\label{eq:Nc} 
\Nc \approx 1 + \xi + \frac{\taust}{\tmd}. 
\end{equation}

Equations~(\ref{eq:tglob}--\ref{eq:Nc}) explicitly connect $\tglob$ and $\fsf$ to the parameters of subgrid models for star formation (via $\taust$) and feedback (via $\xi$), and their physical interpretation is clear. In the ISM, gas is gradually converted into stars as individual gas parcels frequently cycle between the non-star-forming and actively star-forming states. On average, a given gas parcel transits from the non-star-forming to the star-forming state on a dynamical time scale, $\tp$, determined by a mix of processes such as ISM turbulence, gravity, cooling, etc. 

\begin{figure*}
\centering
\includegraphics[width=\textwidth]{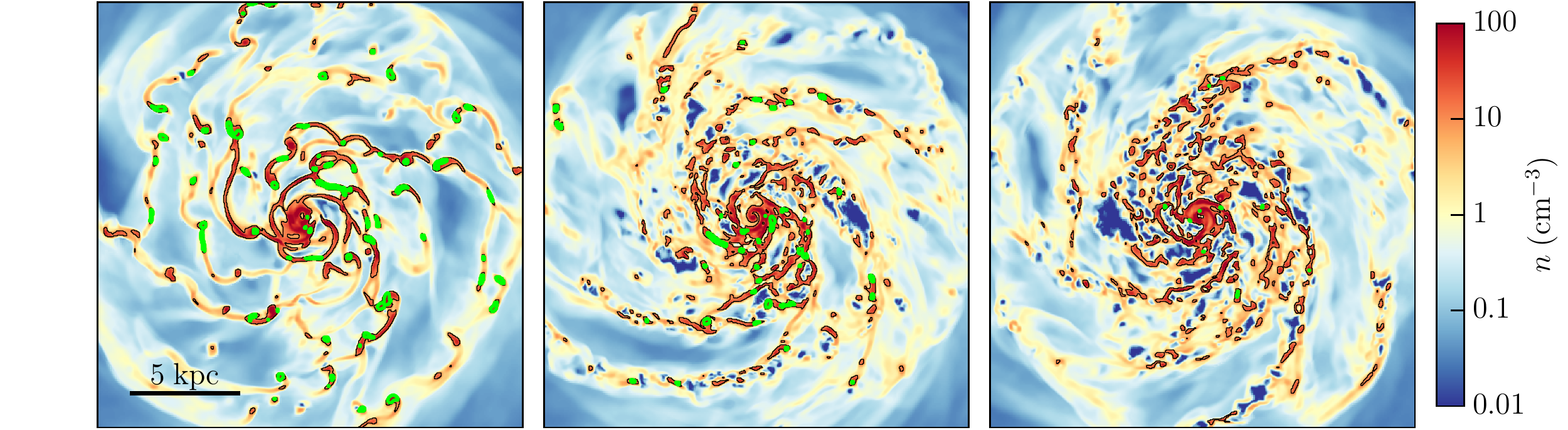}\\
\includegraphics[width=\textwidth]{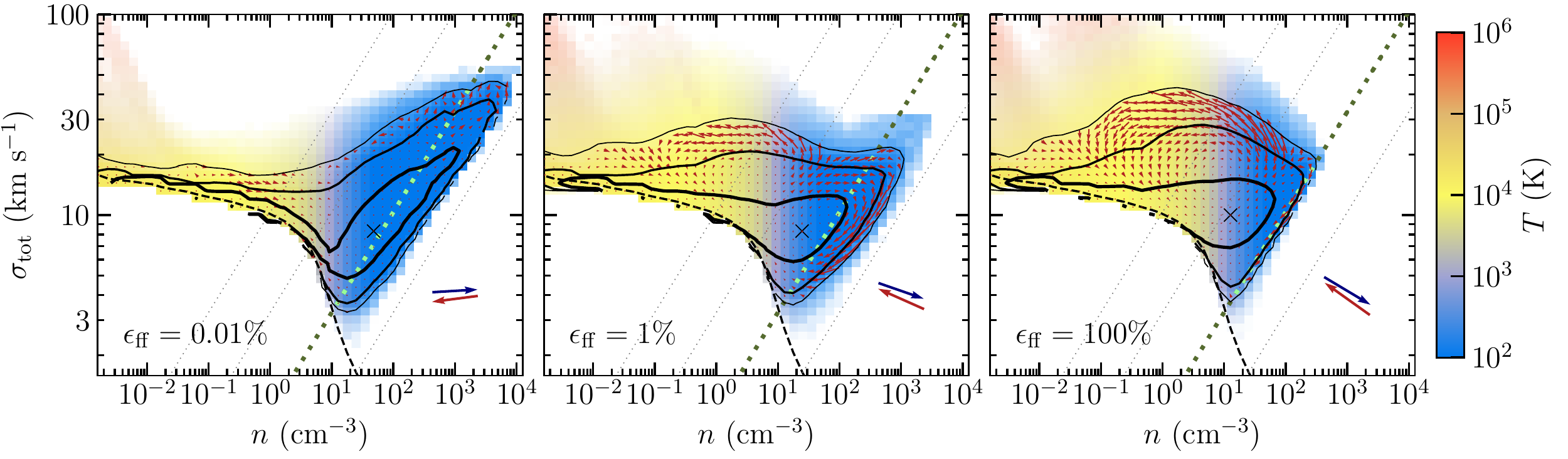}
\caption{\label{fig:phases_fid} Effect of $\epsff$ on the spatial gas distribution and the gas distribution in the phase space of the gas density, $n$, and the total subgrid velocity dispersion, $\stot$. The adopted value for $\epsff$ changes from left to right: $0.01\%$, $1\%$, and $100\%$. The top row of panels shows the midplane density slices at $t=500\Myr$, with black contour indicating cold dense gas, $n > 10\cc$, and green contour indicating star-forming regions, $\avir < \avirsf = 10$. The bottom row of panels shows \ns diagrams colored with the mass-weighted temperature in each bin. The distribution is time averaged between 400 and 600 Myr using gas-tracer particles at $R>1\kpc$ (see Section~\ref{sec:sims:methods}). Black contours indicate 68\%, 95\%, and 99\% of resulting gas tracers' PDF. Thin red arrows throughout the diagram show the average net flux of gas tracers, while the thick blue and red arrows in the corner of each panel illustrate magnitudes and directions of the average positive and negative gas fluxes, respectively, measured at the point indicated with a cross. All arrows can be directly compared to each other because their normalization is the same: the arrow extent corresponds to the distance that a tracer traverses over 5 Myr. Star-forming gas in each \ns diagram resides below the thick dotted line, which corresponds to the star formation threshold of $\avirsf=10$. Thin dotted lines parallel to the star formation threshold show constant values of $\avir=1000$, 100, and 1 (from left to right).}
\end{figure*}

In order to be converted into stars, a gas parcel needs to spend one average depletion time \textit{of star-forming gas} in the star-forming state: $\taust \equiv \Msf/\SFR = \langle 1/\tdep \rangle_{\rm sf}^{-1}$. However, before the gas parcel is converted into stars, it can be removed back into the non-star-forming state by efficient feedback or dynamical processes, and then this gas parcel has to start the cycle from the beginning. Overall, if star-forming stages on average last for $\tsf$, then $\Nc = \taust/\tsf$ such replenishment-expulsion cycles are required to convert all gas into stars. The global depletion time can be expressed by Equation~(\ref{eq:tglob}), where the first and the second terms in the sum correspond to the total times in the non-star-forming and the star-forming states, respectively. The star-forming mass fraction is then given by the ratio of the time spent in the star-forming state to the total depletion time, as expressed by Equation~(\ref{eq:fsf}). In a steady state with the constant total star-forming gas mass, the number of transitions, $\Nc$, is controlled by the stellar feedback and the dynamical processes that destroy star-forming regions and thereby define the average duration of star-forming stages (Equation~\ref{eq:Nc}).

As was shown in Figures~\ref{fig:taudep_epsff} and \ref{fig:fb}  and as we will discuss in more detail below, Equations~(\ref{eq:tglob}--\ref{eq:Nc}) can predict the trends of $\tglob$ and $\fsf$ observed in our simulations (Section~\ref{sec:results}) with varied star formation efficiency $\epsff$, star formation threshold, and feedback strength $\b$. We note that the latter is closely related to the $\xi$ parameter of the model. Both these parameters reflect the strength of feedback per unit stellar mass formed and its efficacy in dispersing star-forming regions. However, these parameters are not identical: $\b$ is a relative strength of the momentum injection in our implementation of feedback, while $\xi=\Fmfb/\SFR$ is an average ``mass-loading factor'' that characterizes the efficacy of gas removal from star-forming regions by feedback. We also note that in equations for $\tglob$ and $\fsf$ the average freefall time in the star-forming gas, $\tauff$, is a model parameter, but, as we will show in Section~\ref{sec:predict:ebdep} and Appendix~\ref{app:model}, its trends with simulation parameters discussed in Section~\ref{sec:results} can also be understood using our model predictions. 

For our subsequent discussion, it is convenient to combine Equations~(\ref{eq:tglob}) and (\ref{eq:Nc}) and rearrange terms in the resulting equation as
\begin{equation}
\label{eq:tglob2}
\tglob = (1 + \xi) \tp + \left( 1 + \frac{\tp}{\tmd} \right) \frac{\tauff}{\epsff},
\end{equation}
where we have substituted $\taust = \tauff/\epsff$. Similarly, using Equation~(\ref{eq:fsf}), the star-forming mass fraction can be expressed as
\begin{equation}
\label{eq:fsf2}
\fsf =  \left[ (1 + \xi) \tp\frac{\epsff}{\tauff} + 1 + \frac{\tp}{\tmd} \right]^{-1}.
\end{equation}

Equation~(\ref{eq:tglob2}) readily shows that the global depletion time is a sum of two terms, one of which may dominate depending on the parameters. For example, the first term, $(1+\xi)\tp$, will dominate when feedback is sufficiently strong, i.e. $\xi$ is large, or star formation efficiency $\epsff$ is sufficiently high so that the second term, $(1+\tp/\tmd)\tauff/\epsff$, is subdominant. Conversely, the second term may dominate if feedback is inefficient or $\epsff$ is low. In these two regimes, the dependence of depletion time on the parameters of star formation and feedback will be qualitatively different. Specifically, when the first term in the equation dominates, $\tglob$ is insensitive to $\epsff$ and scales with feedback strength $\xi$. Conversely, when the second term dominates, $\tglob$ scales as $\epsff^{-1}$ and is independent of feedback strength.

Physically, these two regimes reflect the dominance of different negative terms in Equation~(\ref{eq:masscons}) and thus different mechanisms that limit lifetimes of star-forming regions. In the first regime, $\tglob \approx (1+\xi)\tp$ and the lifetime of gas in the star-forming state is limited by feedback and star formation itself. We therefore will refer to this case as the {\it ``self-regulation regime''} because this was the term used to indicate insensitivity of $\tglob$ to $\epsff$ in previous studies. In the second regime, $\tglob \approx (1+\tp/\tmd)\tauff/\epsff$ and star-forming gas lifetime is limited by dynamical processes dispersing star-forming regions, such as turbulent shear, differential rotation, and expansion behind spiral arms, operating on timescale $\tmd$. We will refer to this case as the {\it ``dynamics-regulation regime,''} as star formation passively reflects the distribution of ISM gas regulated by these dynamical processes, rather than actively shaping it by gas consumption and associated feedback. 

In the next section, we will consider dynamics- and self-regulation regimes in more detail. We will illustrate these regimes using our simulations with the fiducial feedback strength and star formation threshold but varying $\epsff$ from a low value of $0.01\%$, corresponding to the dynamics-regulation regime, to a high value of $100\%$, corresponding to the self-regulation regime. As Figure~\ref{fig:phases_fid} shows, in different regimes the quasi-equilibrium ISM gas distribution is qualitatively different. The figure shows the midplane density slices and \ns diagrams (like the one in Figure~\ref{fig:model}) colored according to the average gas temperature, with arrows indicating average gas fluxes. In all cases, small net fluxes result from the near cancellation of strong positive and negative fluxes, $\Fp$ and $\Fm$, whose typical magnitudes are shown with the thick blue and red arrows, respectively, in the lower right corner of each diagram. Depending on the $\epsff$ value, the negative flux can be dominated by either $\Fmdyn$ or $\Fmfb$, which in turn results in qualitatively different behavior of Equation~(\ref{eq:tglob2}).

\subsection{Predictions for Trends of $\tglob$, $\fsf$, and $\tauff$}
\label{sec:model:predict}

\subsubsection{Interpretation of Scalings in the Dynamics-regulation Regime}
\label{sec:model:lowepsff}

As discussed above, dynamics-regulation occurs when $\epsff$ or $\xi$ are small, so that the second term on the right-hand side of Equation~(\ref{eq:tglob2}) dominates. In this case, $\tglob$ scales inversely with $\epsff$:
\begin{equation}
\label{eq:tglob_lowepsff} \tglob \approx \left( 1 + \frac{\tp}{\tmd} \right) \frac{\tauff}{\epsff}.
\end{equation}
The star-forming mass fraction, on the other hand, remains independent of $\epsff$ because, according to Equation~(\ref{eq:fsf2}),
\begin{equation}
\label{eq:fsf_lowepsff} \fsf \approx \left( 1 + \frac{\tp}{\tmd} \right)^{-1}. 
\end{equation}
Such scalings, $\tglob \propto \epsff^{-1}$ and $\fsf \approx \const$, indeed persist in our simulations with low $\epsff$ values (see $\epsff=0.01\%$ and $0.1\%$ in Figures~\ref{fig:taudep_fid}--\ref{fig:fb}). 

Physically, these scalings arise because at low $\epsff$ and $\xi$ the contributions of star formation ($\SFR$) and feedback ($\Fmfb$) terms to the overall mass flux balance in Equation~(\ref{eq:masscons}) become small. As a result, the steady state is established with $\Fmdyn \approx \Fp$, which yields Equations~(\ref{eq:tglob_lowepsff}) and (\ref{eq:fsf_lowepsff}). In our simulated galaxy, such a state is established as gas is compressed into new star-forming clumps at the same rate at which old clumps are dispersed by differential rotation and tidal torques, and neither of these processes depends on $\epsff$. The interplay between compression and dynamical dispersal determines the steady-state distribution of gas in the \ns diagram (the bottom left panel of Figure~\ref{fig:phases_fid}), which is also insensitive to $\epsff$. As a consequence, the star-forming mass fraction, $\fsf$, and the mean freefall time in star-forming gas, $\tauff$, also do not depend on $\epsff$ and are determined solely by the definition of the star-forming gas. The global depletion time, however, does depend on $\epsff$ as is evident from Equation~(\ref{eq:tglob_lowepsff}).

As $\Fmfb$ is subdominant in this regime, $\tglob$, $\fsf$, and $\tauff$ are also insensitive to the feedback strength, but they do depend on the star formation threshold. Indeed, as blue lines in the left column of Figure~\ref{fig:fb} show, $\tglob$, $\fsf$, and $\tauff$ remain approximately constant when feedback boost factor, $\b$, is varied from 0 to 5. At the same time, when star formation threshold is varied such that more gas is included in the star-forming state, both $\fsf$ and $\tauff$ increase because more low-density gas is added, while $\tglob$ decreases as additional star-forming gas increases SFR. It is worth noting that these dependencies on star formation threshold are rather weak when the threshold encompasses significant fraction of the ISM gas, but they become stronger when the threshold selects gas only from the high-density tail of distribution, because it is this high-density gas that mostly determines $\tglob$, $\fsf$, and $\tauff$.

Finally, it is also worth noting that for some galaxies, or certain regions within galaxies, equilibrium may not be achievable, so that $\Fp > \Fmdyn$ or $\Fp < \Fmdyn$. In this case distribution of gas evolves, and thus $\tglob$, $\fsf$, and $\tauff$ also change with time. This occurs in the central regions of galaxies in simulations with $\epsff = 0.1\%$ and $0.01\%$, where the central gas concentration grows owing to accretion, and which we thus exclude from our analysis (see Section~\ref{sec:sims:methods}).

\subsubsection{Interpretation of Scalings in the Self-regulation Regime}
\label{sec:model:highepsff}

Self-regulation occurs when $\epsff$ or $\xi$ are sufficiently large, so that the first term on the right-hand side of Equation~(\ref{eq:tglob2}) dominates and depletion time is given by
\begin{equation}
\label{eq:tglob_highepsff}
\tglob \approx (1+\xi) \tp,
\end{equation}
and is thus independent of $\epsff$, but scales almost linearly with $\xi$. In this regime, the star-forming mass fraction scales inversely with $\epsff$ (see Equation~\ref{eq:fsf2}):
\begin{equation}
\label{eq:fsf_highepsff}
\fsf \approx \frac{1}{(1+\xi)\epsff} \frac{\tauff}{\tp},
\end{equation}
which also implies $\fsf \ll 1$ because $\tauff/\epsff \ll (1+\xi)\tp$ is required for the subdominance of the terms proportional to $\epsff^{-1}$ in Equation~(\ref{eq:tglob2}). 

The scalings of Equations~(\ref{eq:tglob_highepsff}) and (\ref{eq:fsf_highepsff}) are consistent with the results of our simulations with large $\epsff$ values (Figures~\ref{fig:taudep_fid}--\ref{fig:fb}). The insensitivity of $\tglob$ to $\epsff$ and its scaling with feedback strength have also been observed in other simulations with high $\epsff$ and efficient feedback \citep[e.g.,][]{Agertz.Kravtsov.2015,Hopkins.etal.2017,Orr.etal.2017}. In the literature, these phenomena are also usually referred to as ``self-regulation.''

As detailed in \P1, self-regulation occurs when gas spends most of the time in non-star-forming stages, $\fsf \ll 1$, and the rate of star-forming gas supply, $\Fp$ in Equation~(\ref{eq:masscons}), is balanced by rapid gas consumption and strong feedback-induced gas dispersal: $\Fp \approx \SFR+\Fmfb$. In this case, global depletion time is given by $\tglob \approx \Nc \tp$, where $\Nc$ is the total number of cycles between non-star-forming and star-forming states (see Section~\ref{sec:model:summary}). Due to  large $\SFR+\Fmfb \propto (1+\xi)\epsff$, the duration of star-forming stages, $\tsf$, is regulated by star formation and feedback: when $\epsff$ or $\xi$ are increased, the lifetime of gas in the star-forming state shortens as $\tsf \propto [(1+\xi)\epsff]^{-1}$. However, the {\it total} time spent in the star-forming state before complete depletion depends on $\epsff$ but not on $\xi$: $\taust \propto \epsff^{-1}$. The dependence on $\epsff$ thus cancels out in $\Nc = \taust/\tsf$ and global depletion time becomes independent of $\epsff$ but maintains scaling with $\xi$. 

Therefore, in the self-regulation regime, star formation regulates itself by controlling the timescale on which feedback disperses star-forming regions and by conversion of gas into stars in these regions. The relative importance of these processes is determined by the feedback strength per unit of formed stars, i.e. the $\xi$ value. 

When feedback is efficient, $\xi \gg 1$, as is the case in our simulations\footnote{Our results in Section~\ref{sec:predict:ebdep} and Appendix~\ref{app:model} suggest that $\xi \sim 60$ in our simulations with fiducial feedback and star formation threshold.} shown in Figure~\ref{fig:phases_fid}, the ISM gas distribution at high $\epsff$ is shaped by feedback-induced gas motions, $\Fp \approx \Fmfb$. Specifically, as the top panels show, at $\epsff=1\%$ and $100\%$, efficient feedback makes ISM structure flocculent and devoid of dense star-forming clumps, which are typical in the $\epsff = 0.01\%$ simulation. The bottom panels show that at high $\epsff$ efficient feedback keeps most of the dense gas above the star formation threshold or close to it. This results in a significant decrease of $\fsf$ and increase of $\tauff$ in this regime, compared to the dynamics-regulated regime.

When feedback is inefficient, $\xi \ll 1$, or even completely absent, $\xi = 0$, the gas consumption dominates at high $\epsff$, $\Fp \approx \SFR$. In this regime, all available star-forming gas is rapidly converted into stars and the global depletion time is determined by the timescale on which new star-forming gas is supplied, i.e. $\tglob \sim \tp$. Thus, this regime is analogous to the ``bottleneck'' scenario envisioned by \citet{Saitoh.etal.2008}. Our simulations with $\b=0$ and $\epsff \geq 10\%$ operate in this regime, and because $\tp$ is short, $\tglob$ is also short, so that gas is rapidly consumed and the simulated galaxy cannot settle into an equilibrium state.

Dependence of $\tglob$, $\fsf$, and $\tauff$ on the choice of the star formation threshold can also be understood as follows. As $\epsff$ and $\xi$ increase, the average density of the star-forming gas decreases, which increases $\tauff$. For the density-based threshold, the value of $\tauff$ becomes independent of $\epsff$ and $\xi$ as the star-forming gas is kept at the density close to the threshold, $n\sim\nsf$. Larger $\avirsf$ (or smaller $\nsf$) in Figure~\ref{fig:fb} results in shorter $\tglob \propto \tp$, because $\tp$ decreases as it takes less time for gas to evolve from the typical ISM density and $\avir$ to the values of the star-forming gas. As typical densities of the star-forming gas decrease, $\tauff$ increases and thus $\fsf \propto \tauff/\tp$ (Equation~\ref{eq:fsf_highepsff}) also increases because of both longer $\tauff$ and shorter $\tp$.

In the above discussion, the dynamical time $\tp$ was assumed to be independent of $\epsff$ and the feedback strength. This is certainly a simplification,  as $\tp$ can be determined by feedback, which can limit the lifetime of star-forming regions, drive large-scale turbulence in the ISM, inflate low density hot bubbles, launch fountain-like outflows, and sweep gas into new star-forming regions. These processes are reflected in the complicated pattern of the net gas flux in the \ns plane in the bottom middle panel of Figure~\ref{fig:phases_fid}, which shows a prominent clockwise whirl near the star formation threshold and a counterclockwise whirl in the lower-density gas. The clockwise whirl originates from the ISM gas being swept by SN shells, while the counterclockwise whirl is shaped by the gas in freely expanding shells (see Section 4.2 of \P1 for a more detailed discussion). Nevertheless, we find that  the dependence of $\tp$ on the feedback strength variation is much weaker than the linear scalings of $\tglob$ and $\fsf$ with $\xi$ and $\epsff$ (see Section~\ref{sec:predict:ebdep}), and thus our simplification is warranted.

\subsubsection{Transition between the Regimes}
\label{sec:model:transition}

Self-regulation or dynamics-regulation regimes occur when the first or second term in Equation~(\ref{eq:tglob2}) dominates. In Section~\ref{sec:results}, we illustrated these regimes using simulations in which $\epsff$, feedback strength, and star formation threshold are varied in a wide range. The transition between the two regimes depends on all of these parameters. For example, the dependence of transition on the feedback strength is evident from Figure~\ref{fig:taudep_epsff}: at stronger feedback, the transition occurs at smaller $\epsff$. As a result, the run with $\epsff=1\%$ and weak feedback, $\b=0.2$, exhibits behavior of the dynamics-regulation regime, while the galaxy in the run with the same $\epsff$ but with much stronger feedback, $\b=5$, is in the self-regulation regime. Similarly, from the middle and right panels of Figure~\ref{fig:fb}, when $\epsff=1\%$ and threshold defines a significant fraction of gas as star-forming (e.g., $\avirsf=100$ or $\nsf=10\cc$), simulated galaxies are in the dynamics-regulation regime. On the other hand, when threshold defines only a small fraction of gas as star-forming (e.g., $\avirsf=10$ or $\nsf=100\cc$), galaxies are in the self-regulation regime. 

Note, however, that achieving self-regulation with the threshold variation is not always possible, because the threshold affects both terms in Equation~(\ref{eq:tglob2}), and thus the value of the threshold at which the first term dominates does not always exist. For example, in the top middle panel of Figure~\ref{fig:fb}, when $\epsff < 1\%$, depletion time bends upward at $\avirsf<10$ and remains inversely proportional to $\epsff$ and therefore never reaches the self-regulation regime.

In the transition between dynamics-regulated and self-regulated regimes, the relation between our model parameters follows from the condition that the terms in Equation~(\ref{eq:tglob2}) are comparable:
\begin{equation}
\label{eq:epsffcr}
(1 + \xi)\epsff \sim \left( 1 +  \frac{\tp}{\tmd} \right) \frac{\tauff}{\tp}.
\end{equation}
Notably, in this case a given galaxy has the same star-forming mass fraction independent of $\epsff$ or the feedback strength. Indeed, after substituting condition (\ref{eq:epsffcr}) into Equation~(\ref{eq:fsf2}), we get
\begin{equation}
\label{eq:fsfcr}
\fsf \sim \frac{1}{2} \left( 1 + \frac{\tp}{\tmd} \right)^{-1},
\end{equation}
i.e., the star-forming mass fraction at the transition is half of that in the dynamics-regulation regime (Equation~\ref{eq:fsf_lowepsff}).

\subsubsection{Quantitative Predictions as a Function of $\epsff$ and Feedback Strength}
\label{sec:predict:ebdep}

So far, we described how the model presented above can explain the trends and regimes revealed by our simulations qualitatively. Here we will show that the model can also describe the simulation results {\it quantitatively.} 

To predict $\tglob$ and $\fsf$ in the simulations using Equations~(\ref{eq:tglob2}) and (\ref{eq:fsf2}), we note that the unknown parameters enter these equations only in three different combinations: $(1+\xi)\tp$, $\tp/\tmd$, and $\tauff/\epsff$. These can be calibrated against a small subset of the simulations in the dynamics- and self-regulation regimes using scalings discussed in Sections~\ref{sec:model:lowepsff} and \ref{sec:model:highepsff} as a guide. Quantitative predictions of the model with calibrated parameters for the trends of $\tglob$ and $\fsf$ can then be compared with the results of other simulations, not used in the calibration. 

Specifically, using two runs in the self-regulated regime with $\epsff=100\%$, we measure the normalization of $(1+\xi) \tp$ and its scaling with the feedback boosting factor $\b$. Equation~(\ref{eq:tglob_highepsff}) gives the normalization of the global depletion time in the high-$\epsff$ run with $\b=1$: $[(1+\xi)\tp]_0 \approx \tglob(\b=1) \sim 6\Gyr$. Adopting $(1+\xi)\tp \propto \b^\beta$ for the scaling with $\b$, the slope $\beta = \Delta \log \tglob / \Delta \log \b \approx 0.75$ is measured using the second run with $\b=5$, and thus the final relation is
\begin{equation}
\label{eq:xitp}
(1+\xi) \tp \approx 6\, \b^{0.75}\ \Gyr,
\end{equation} 
i.e. $(1+\xi)\tp$ is long and increases almost linearly with $\b$.

\begin{figure}
\centering
\includegraphics[width=\columnwidth]{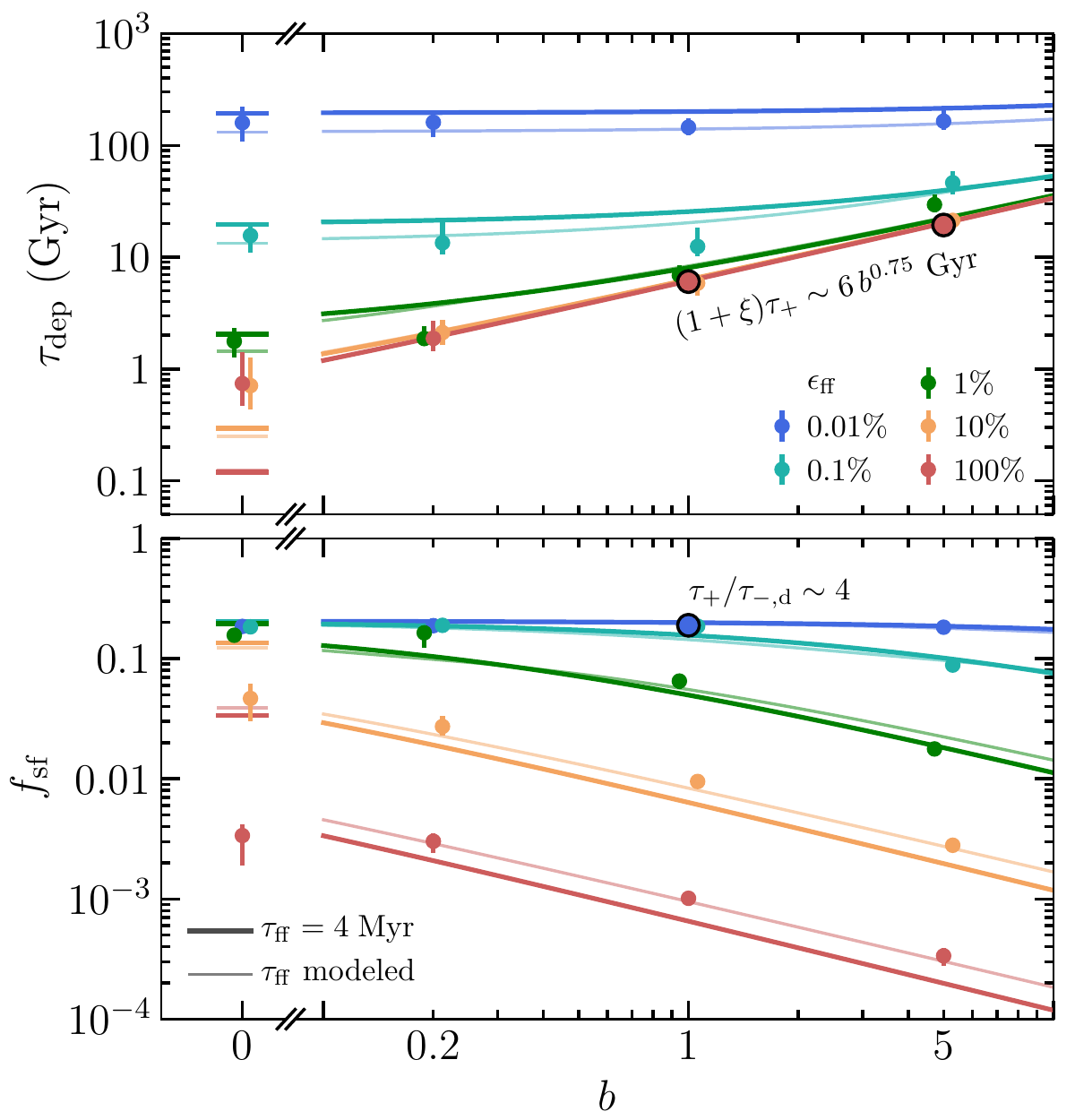}
\caption{\label{fig:predict} Comparison of our model predictions (shown with lines) for the global depletion time ($\tglob$; top panel) and the star-forming mass fraction ($\fsf$; bottom panel) with the results of our simulations with varying $\epsff$ and the feedback boost factor, $\b$, assuming the fiducial star formation threshold, $\avirsf=10$ (notation repeats that of Figure~\ref{fig:fb}). To fix the model parameters, we use $\tglob$ in two high-$\epsff$ runs (red circled points in the top panel), which give $(1+\xi)\tp \sim (6\Gyr)\;\b^{0.75}$, and $\fsf$ from a low-$\epsff$ run (blue circled point in the bottom panel), which gives $\tp/\tmd \sim 4$. As thick lines show, if we neglect variation of $\tauff$ and assume the average $\tauff=4\Myr$, our model correctly predicts the overall behavior of $\tglob$ and $\fsf$. As thin lines show, predictions of our model are improved if the variation of $\tauff$ is also modeled as explained in Appendix~\ref{app:model}. To avoid clutter, simulation points for $\epsff = 0.1\%$, 1\%, and 10\% are slightly shifted horizontally around the actually used values of $\b=0$, 0.2, 1, and 5.}
\end{figure}

Using a simulation with $\epsff=0.01\%$ (i.e., the dynamics-regulation regime) and Equation~(\ref{eq:fsf_lowepsff}), we estimate the ratio of dynamical times $\tp/\tmd$ from the value of star-forming mass fraction, $\fsf \approx 0.2$, measured in this simulation:
\begin{equation}
\label{eq:tptmd}
\frac{\tp}{\tmd} \approx \frac{1}{\fsf} - 1 \sim 4,
\end{equation}
which implies that in the absence of feedback the star-forming gas is supplied 4 times more slowly than it is dispersed by dynamical effects.

Finally, the last unknown parameter is the average freefall time in the star-forming gas, $\tauff$. In our simulations, $\tauff$ varies only mildly, from $\tau_{\rm ff} \approx 2\text{--}3\Myr$ in the dynamics-regulation regime to  $\tau_{\rm ff} \approx 5\text{--}6\Myr$ in the self-regulation regime. In the simplest case, we can make predictions assuming a constant $\tauff = 4\,\Myr$, which is representative of the freefall time in star-forming regions both in our simulations and in observations. 

Figure~\ref{fig:predict} compares the simulation results for $\tglob$ and $\fsf$ as a function of the feedback strength, $\b$, with the predictions of our model with constant $\tauff=4\,\Myr$ (thick lines). Of the 20 simulation results shown by points in the figure, only three were used to calibrate the four model parameters,  $[(1+\xi)\tp]_0$, $\beta$, $\tp/\tmd$, and $\tauff$, as described above; these simulations are shown by the large circled points.  For the other 17 simulations, the lines show {\it predictions} of the model. Figure~\ref{fig:predict} shows that the model correctly predicts a wide variation of $\tglob$ and $\fsf$ with $\epsff$ and the feedback strength $\b$ in the entire suite of simulations. 

Moreover, $\tglob$ and $\fsf$ involve two independent quantities, $\SFR$ and $\Msf$, measured in the simulations. Thus, our four-parameter model calibrated using three simulations describes well $17\times 2=34$ independent data points. The fact that our model closely agrees with the simulations when we treat $\tauff$ as a fixed parameter and $\tp$ as independent of $\xi$ and $\epsff$ indicates that most of the variation of $\tglob$ and $\fsf$ is driven by their explicit dependence on $\epsff$ and $\xi$ in Equations~(\ref{eq:tglob2}) and (\ref{eq:fsf2}), whereas any variation of $\tauff$ and $\tp$ with $\epsff$ and $\xi$ is secondary. 

Nevertheless, accounting for $\tauff$ variations can somewhat improve the accuracy of our model. Thin lines in Figure~\ref{fig:predict} and in the left panels of Figure~\ref{fig:fb} show our model predictions incorporating $\tauff$ variation with $\epsff$ and $\xi$ values. To model this variation, we note that the increase of $\tauff$ during the transition from the dynamics-regulation regime to the self-regulation regime is controlled by the total rate of the star-forming gas removal by gas consumption and feedback: $\SFR+\Fmfb \propto (1+\xi)\epsff$. Thus, we calibrate the values of $\tauff$ in these regimes using the same three simulations as before, and we interpolate $\tauff$ as a function of $(1+\xi)\epsff$ for all other simulations. The details of this calibration and the adopted interpolation function are presented in Appendix~\ref{app:model}.

\subsubsection{Quantitative Predictions as a Function of the Star Formation Threshold}
\label{sec:predict:thresholds}

To predict how $\tglob$, $\fsf$, and $\tauff$ depend on the star formation threshold, $\avirsf$, we need to calibrate model parameters as a function of $\avirsf$. Analogously to the previous section, we constrain these dependencies using runs in the limiting regimes and use our model to predict $\tglob$, $\fsf$, and $\tauff$ in the other simulations. Our model predictions are shown with lines in the middle column of panels in Figure~\ref{fig:fb} using calibrations done as follows.

First, the dependence of $(1+\xi)\tp$ and $\tauff$ in the self-regulation regime on $\avirsf$ can be assessed using a run with $\epsff=100\%$ and fiducial $\avirsf=10$ and an additional run with $\avirsf=100$ to obtain the following scalings:
\begin{align}
\label{eq:xitp-avirsf}
(1+\xi)\tp &\propto \avirsf^{-0.5},\\
\label{eq:tauffhi-avirsf}
\tauff &\propto \avirsf^{0.4}.
\end{align}
The scaling of $(1+\xi)\tp$ is measured as the slope of $\tglob$ in the top middle panel of the figure. For the typical density of the star-forming gas $\bar{n}$, the freefall time is $\tauff \propto \bar{n}^{-0.5}$ and the slope of 0.4 in Equation~(\ref{eq:tauffhi-avirsf}) thus indicates that $\bar{n} \propto \avirsf^{-0.8}$. Given that $\avir \propto \sigma_{\rm t}^2/n$, this means that the typical velocity dispersion in the star-forming gas scales as $\bar{\sigma}_{\rm t} \propto \avirsf^{0.1}$.

Second, we note that to constrain the behavior of $\tp/\tmd$ and $\tauff$ in the dynamics-regulated regime, no extra runs are needed, and all the required information can be obtained directly from the simulation with $\epsff=0.01\%$ and $\b=1$, which has been already used in the previous section. This is because in the dynamics-regulated regime the gas distribution in the \ns plane is not affected by star formation and feedback, and thus we expect it to be the same as in the bottom left panel of Figure~\ref{fig:phases_fid}. Therefore, $\fsf$---which yields $\tp/\tmd$ from Equation~(\ref{eq:fsf_lowepsff})---and $\tauff$ as a function of the star formation threshold can be directly measured from this distribution. We spline $\fsf(\avirsf)$ and $\tauff(\avirsf)$ in the low-$\epsff$ simulation with fiducial $\avirsf$ and show these functions with blue lines in the bottom two panels of the middle column in Figure~\ref{fig:fb}. 

These two steps fix the dependencies of $(1+\xi)\tp$, $\tp/\tmd$, and $\tauff$ on the star formation threshold, and thus we can predict how $\tglob$, $\fsf$, and $\tauff$ depend on the threshold at different $\epsff$ and our predictions closely agree with the results of simulations, as shown in the middle column of panels in Figure~\ref{fig:fb}. To test our model, we repeated the above steps for the simulations with the star formation threshold in the gas density rather than in $\avir$. As the right column of Figure~\ref{fig:fb} shows, our predictions again closely agree with the results of the simulations, although the values of the parameters are of course different (see Appendix \ref{app:model}).

\subsection{Generic Approach to Calibrating the Star Formation and Feedback Parameters in Simulations}

Galaxy simulations can differ significantly in numerical methods used to handle hydrodynamics and in specific details of the implementation of star formation and feedback processes. The implementations can also be applied at different resolutions, so that the values and sometimes even the physical meaning of the parameters change. Thus, the parameter values of our model that we calibrated above should be used with caution and applied only when similar numerical techniques, resolutions, and implementations of star formation and feedback are used.  

Nevertheless, the overall calibration approach can still be used in all cases to choose the values of the star formation and feedback parameters. For example,  one can calibrate $\tglob$ and $\fsf$ dependence on the parameters in the dynamics-regulation regime using one simulation with a very low (or even zero) value of $\epsff$, as was done in Sections~\ref{sec:predict:ebdep} and \ref{sec:predict:thresholds}. Then, the $\tglob$ and $\fsf$ behavior in the self-regulation regime can be anchored using several simulations with varying feedback strength and star formation threshold at sufficiently high $\epsff$. The value of $\epsff$ appropriate for this second step can be chosen from the condition that the local depletion time at typical densities of the star-forming gas must be much shorter than the global depletion time, which thus implies $\epsff \gg \tauff/\tglob$. The appropriately high value of $\epsff$ will also result in $\fsf$ much smaller than the $\fsf$ in the simulation with low $\epsff$.

\section{Comparisons with observations}
\label{sec:observ}

Results presented in the previous section demonstrate that our general theoretical framework for star formation in galaxies can describe and explain the results of galaxy simulations both qualitatively and quantitatively. The model can thus be also used to interpret and explain observational results, in particular the observed long gas depletion times in galaxies, as we showed in \P1. In this section, we use the observations to constrain the parameters of our model, in particular, the efficiency of star formation per freefall time, $\epsff$. We also use the model to infer whether observed galaxies are in the dynamics- or self-regulation regime.

Specifically, we use the observed values of the depletion time of atomic$+$molecular and just molecular gas at different scales---from global galactic values to the scales comparable to our resolution limit of $\sim 40\pc$---as well as the mass fraction of gas in star-forming regions and in the molecular phase. Comparisons and inferences from observations on different scales are presented in separate sections below. In most of the comparisons, we use observations in the Milky Way, where star formation is studied most extensively. However, whenever possible, we also use recent observations of other nearby galaxies. Note that we focus here on the inferences specific to $\sim$\Lstar-sized galaxies, as our simulated galaxy model has structural parameters typical for such galaxies. 

In what follows, we use the star formation rates in simulations computed differently on different scales, in ways that approximate how corresponding rates are estimated in observations. 
We compute the local SFR using the total mass of stellar particles younger than some age $t_{\rm sf}$ in the cell: $\dot M^{\rm cell}_\star \equiv M^{\rm cell}_{\star}(<t_{\rm sf})/t_{\rm sf}$, where the choice of $t_{\rm sf}$ is motivated by star formation indicators used in observations. In Sections~\ref{sec:observ:KSR:kpc} and \ref{sec:observ:KSR:bias}, we compare our results with extragalactic studies that use H$\alpha$ and far IR indicators sensitive to the presence of massive young stars, and we thus adopt $t_{\rm sf} = 10\Myr$ \citep[see, e.g., Table 1 in][]{Kennicutt.Evans.2012}. In Section~\ref{sec:observ:KSR:local} we compare with observations of individual star-forming regions, where SFR is estimated by direct counting of pre-main-sequence young stellar objects, and thus we adopt $t_{\rm sf} = 1\Myr$ in this case.

To compare our results with the observed distribution of molecular gas, in each computational cell we estimate the molecular mass as $\rho_{\rm H_2}\Delta^3 = f^{\rm cell}_{\rm H_2} \rho (1-Y_{\rm He}) \Delta^3$, assuming the helium mass fraction of $Y_{\rm He}=0.25$ and computing $f^{\rm cell}_{\rm H_2}$ using the model of \citet{KMT1,KMT2} and \citet{McKee.Krumholz.2010}: $f^{\rm cell}_{\rm H_2} = \max[0,(1-0.75 s/(1+0.25 s))]$ with $s\approx 1.8/\tau_{\rm c}$ and $\tau_{\rm c} = 320 (\rho\Delta / {\rm g\ cm^{-2}})$ at solar metallicity.

\subsection{Global Star Formation}
\label{sec:observ:global}

\subsubsection{Comparison with Observed $\tglob$ and $\fsf$}
\label{sec:observ:tglob-fsf}

\begin{figure}
\centering
\includegraphics[width=\columnwidth]{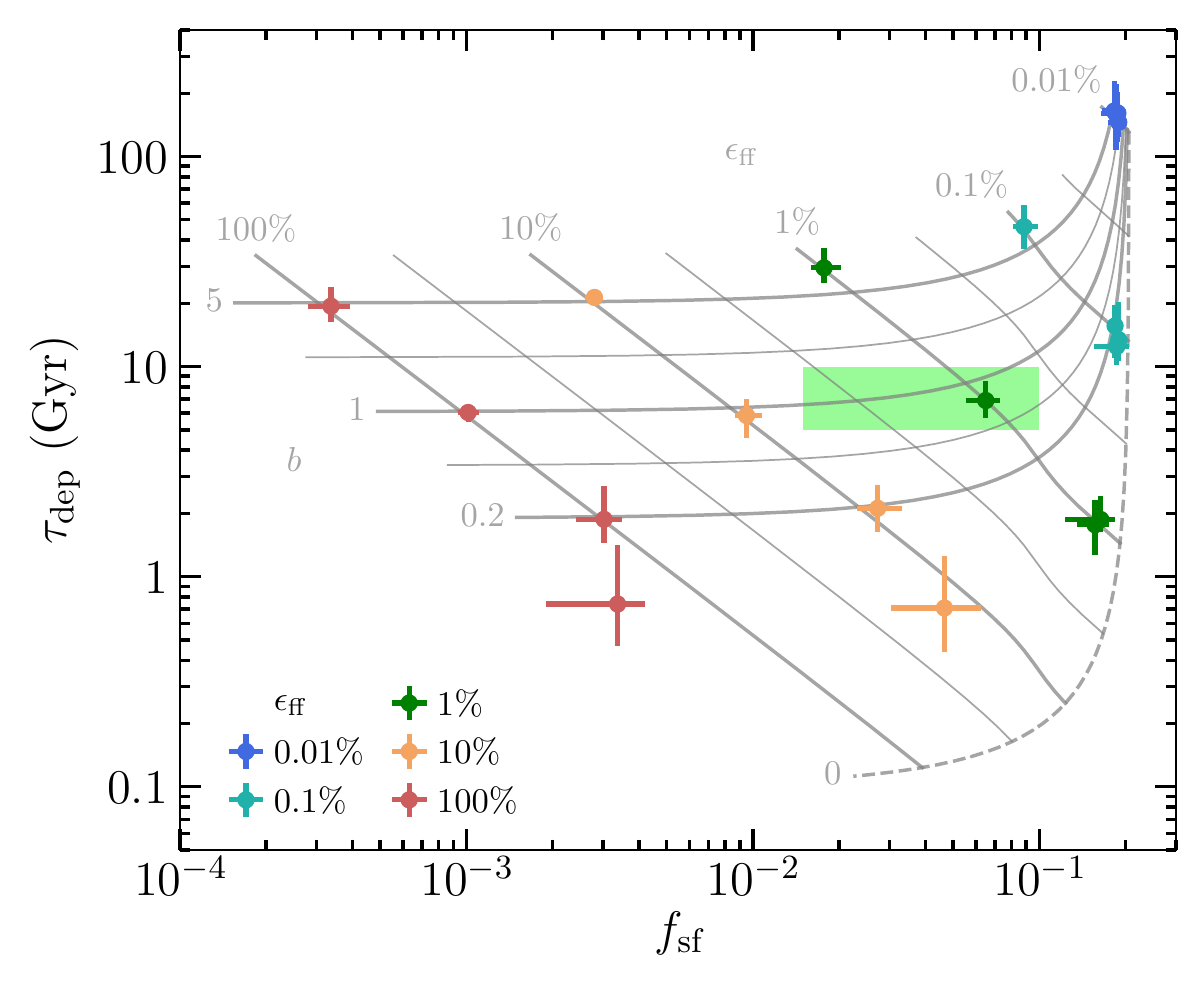}
\caption{\label{fig:fsf-taudep} Comparison of our simulation results (points) and our model predictions (gray lines) for the star-forming mass fraction, $\fsf$, and the global depletion time, $\tglob$, with their values in the Milky Way (green rectangle). Notation of points repeats that of Figure~\ref{fig:fb}, with color indicating $\epsff$ and the feedback boost factor, $\b$, increasing upward: 0, 0.2, 1, and 5. Solid gray lines show the predictions of our model calibrated in Section~\ref{sec:predict:ebdep} for the constant values of $0.2<\b<5$ and $0.01\%<\epsff<100\%$, with thicker lines corresponding to the values used in the simulations and thinner lines showing intermediate values: $\b\approx0.45$ and 2.2 and $\epsff \approx 0.032\%$, 0.32\%, 3.2\%, and 32\%. The dashed line indicates model predictions for runs without feedback ($\b=0$), assuming $\tp=100\Myr$, as motivated by the results of \P1. The green rectangle indicates the range estimated for the Milky Way, $\fsf \sim 1.5\% - 10\%$ and $\tglob \sim 5\text{--}10\Gyr$, as explained in the text.}
\end{figure}

We start our comparisons with observations by comparing our model and simulation predictions as a function of $\epsff$ and the feedback strength $\b$ with the global values of the depletion time, $\tglob$, and the mass fraction of star-forming gas, $\fsf$. To make a fair comparison, $\tglob$ and $\fsf$ in observations must be defined consistently with their definition in the simulations. While $\tglob$ can be compared directly using the total gas mass and SFR, the comparison of $\fsf$ is more nuanced, because one needs to choose which gas in real galaxies corresponds to the star-forming gas in simulations. Our fiducial star formation criterion, $\avir<\avirsf=10$, is motivated by $\avir$ in observed GMCs, and it selects molecular gas with the lowest turbulent velocity dispersions on the scale of our resolution, $\Delta=40\pc$. Such a criterion also results in the average freefall time in star-forming regions of $\tauff \approx 3\text{--}6$ Myr, which is consistent with typical $\tauff$ values estimated for observed GMCs \citep[see, e.g., Figure~1 in][]{Agertz.Kravtsov.2015}. In simulations with larger $\avirsf$, $\tauff$ becomes several times longer than observed in GMCs (see the bottom middle panel of Figure~\ref{fig:fb}). Thus, we argue that our fiducial value of $\avir=10$ corresponds to the definition of the star-forming regions in observations most closely, and we will use the simulations with this value to constrain $\epsff$. We will, however, discuss the dependence on the assumed threshold below, whenever it is relevant.  

To compare our model results, we use the global depletion time and the mass fraction of the star-forming gas in the Milky Way, $\tglob \sim 5\text{--}10 \Gyr$ and $\fsf \sim 1.5\%\text{--}10\%$ estimated as follows. The range of $\tglob$ follows from $\Mg \sim 10^{10}\Msun$ \citep[e.g.,][]{Kalberla.Kerp.2009} and $\SFR \sim 1\text{--}2\Msunyr$ \citep[e.g.,][]{Licquia.Newman.2015}. The upper limit on the star-forming mass fraction follows from the assumption that all molecular gas in the Milky Way is star-forming, and thus $\fsf < \fH2 = \MH2/\Mg \sim (10^{9}\Msun)/(10^{10}\Msun) \sim 10\%$ \citep{Heyer.Dame.2015}. A conservative lower limit on $\fsf$ can be estimated using the total mass in the largest star-forming GMCs in the Milky Way from \citet{Murray.2011}, with sizes  comparable to our resolution of $40\pc$. These massive GMCs account for 33\% of total SFR in the Milky Way but have a total mass of $\approx 5\times 10^7\, \Msun$. If the rest of star formation in the Milky Way were proceeding in clouds with local depletion times similar to those in the \citet{Murray.2011} sample, then the total mass of the star-forming gas would be 3 times larger, or $\approx 1.5\times 10^8\,\Msun$, which would mean $\fsf \sim 1.5\%$. However, this estimate is a conservative lower limit because the rest of the star-forming gas probably forms stars with lower efficiency, as it does not host bright radio sources associated with \ion{H}{2} regions, used by \citet{Murray.2011} to identify the star-forming GMCs. 

In Figure~\ref{fig:fsf-taudep}, the above constraints on $\tglob$ and $\fsf$ in the Milky Way (green rectangle) are compared to the results of our simulations (points with error bars) and the predictions of our analytical model (gray lines). The figure shows that only $\epsff \sim 0.5\%\text{--}5\%$ and $\b \sim 0.3\text{--}2$ can satisfy the constraints on both $\tglob$ and $\fsf$ simultaneously. It is important to note that this constraint on $\epsff$ is rather generous, due to the rather conservative lower limit estimate of $\fsf$ we use for the Milky Way. 

This conclusion would not change if we adopted a different star formation threshold. Figure~\ref{fig:fb} shows that $\avirsf$ values smaller than our fiducial $\avirsf=10$ would result in even smaller $\fsf$, while even values as large as  $\avirsf=100$ for $\epsff=100\%$ would only increase the star-forming gas mass fraction to $\fsf\approx 0.7\%$, while decreasing the depletion time to $\tglob\approx 2\Gyr$, which is still far outside the range we estimate for the Milky Way. 

Note that the figure shows that $\tglob$ and $\fsf$ in the Milky Way have values close to the transition between self-regulation and dynamics-regulation regimes. Indeed, the self-regulation regime corresponds to small $\fsf<0.01$ at which gray lines of constant $\b$ are horizontal, the dynamics-regulation regime is manifested by the convergence of these lines to $\fsf \sim 0.2$, and $\fsf$ in the Milky Way lie in between these two regimes. The conclusion that the Milky Way is in the regime intermediate between dynamics- and self-regulation regimes is also directly supported by the estimate for the second term in Equation~(\ref{eq:tglob2}), $(1+\tp/\tmd)\taust$. Indeed, observed local depletion times in the Milky Way's GMCs are $\taust \sim 100\text{--}500\Myr$ \citep[e.g.,][]{Evans.etal.2009,Evans.etal.2014,Heiderman.etal.2010,Lada.etal.2010,Lada.etal.2012,Gutermuth.etal.2011,Schruba.etal.2017}, and the prefactor in front of $\taust$ is likely similar to that obtained in our simulations, $1+\tp/\tmd \sim 5$ (Equation~\ref{eq:tptmd}), because we expect that our simulations capture dynamical time scales of star-forming gas supply and dispersal. As a result, $(1+\tp/\tmd)\taust \sim 0.5\text{--}2.5\Gyr$ contributes a sizable fraction to the observed global depletion time in the Milky Way, $\tau_{\rm dep,MW} \sim 5\text{--}10\Gyr$, and thus the Milky Way is in the intermediate regime.

\subsubsection{Comparison with the Global Mass Fraction and the Depletion Time of Molecular Gas}
\label{sec:observ:tH2-fH2}

\begin{figure}
\centering
\includegraphics[width=\columnwidth]{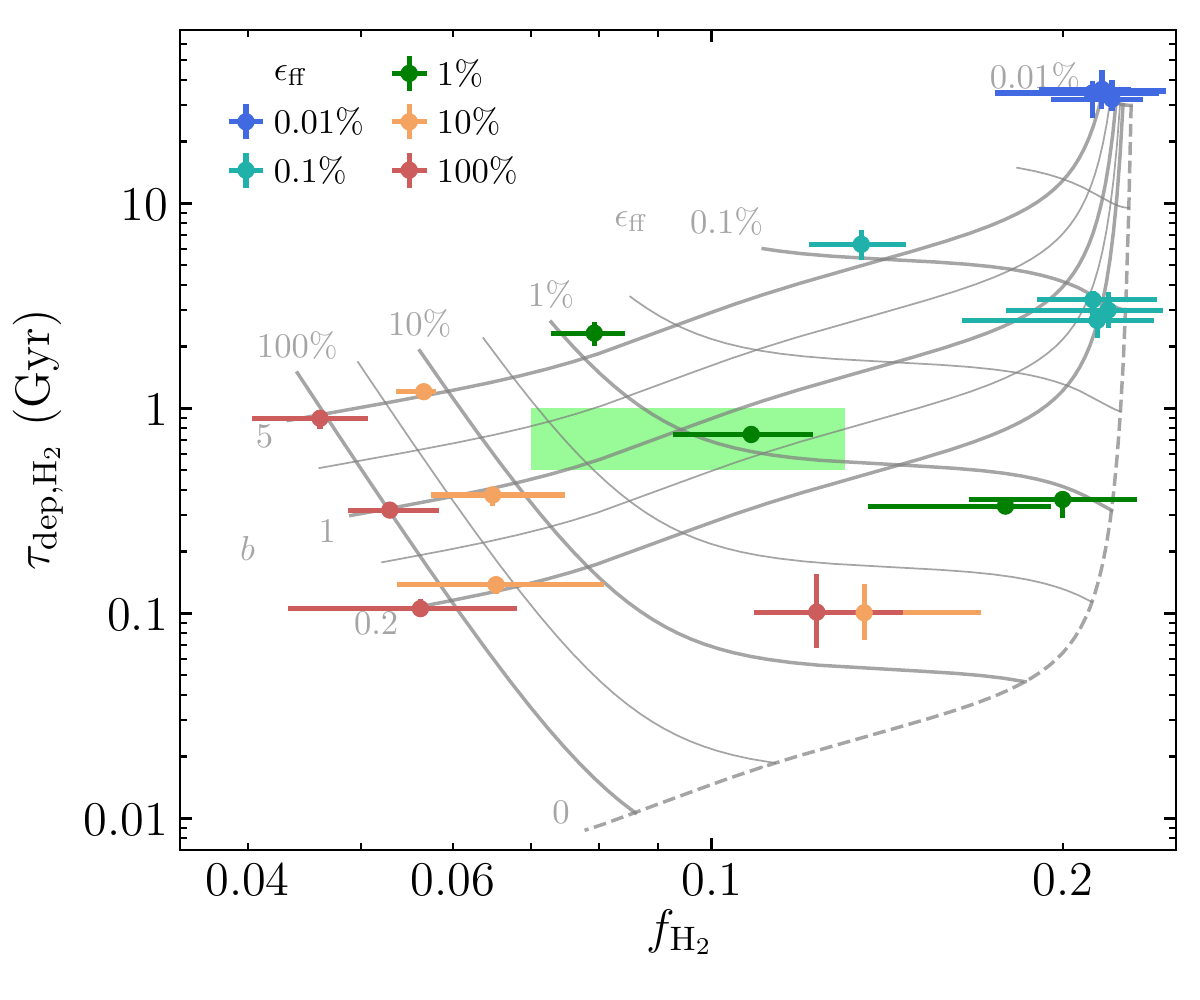}
\caption{\label{fig:fH2-taudepH2} Comparison of the simulation results (points) and our model predictions (gray lines) with the total molecular mass fraction, $\fH2$, and the global depletion time of molecular gas, $\tH2 = \fH2\tglob$, in the Milky Way (green rectangle). The symbols and lines are the same as in Figure~\ref{fig:fsf-taudep}. Our model for $\fH2$ is explained in Appendix~\ref{app:H2}. For the Milky Way, we adopt $\fH2 = (1.0 \pm 0.3)\times10^9\Msun$ \citep{Heyer.Dame.2015} and $\tH2 = (10^9 \Msun)/(1\text{--}2\Msunyr) \sim 0.5\text{--}1\Gyr$ \citep[e.g.,][]{Licquia.Newman.2015}. }
\end{figure}

Figure~\ref{fig:fH2-taudepH2} compares the global molecular gas mass fraction, $\fH2$, and its depletion time, $\tH2$, estimated for the Milky Way (green rectangle) with their values measured in our simulations (points with error bars) and predicted by our model (gray lines). For the Milky Way, we used $\fH2 = (1.0 \pm 0.3)\times10^9\Msun$ from \citet{Heyer.Dame.2015} and estimated $\tH2 = (10^9 \Msun)/(1\text{--}2\Msunyr) \sim 0.5\text{--}1\Gyr$. In the simulations, the total molecular mass, $\MH2$, required to compute $\fH2$ and $\tH2$ is derived as a sum of the molecular mass in each cell, computed as explained at the beginning of Section~\ref{sec:observ}. The model predictions are obtained using the dependence of $\fH2$ on $\epsff$ and the feedback strength, calibrated at the end of Appendix~\ref{app:H2}. The definition of the molecular gas depletion time is $\tH2 \equiv \MH2/\SFR = (\MH2/\Mg)(\Mg/\SFR) = \fH2 \tglob$, with $\tglob$ given by Equation~(\ref{eq:tglob2}).

The figure shows that $\fH2$ and $\tH2$ within the observed range can be obtained only in the simulations with $\epsff \sim 0.5\%\text{--}5\%$ and $\b \sim 0.2\text{--}3$. Note that this range of parameters is similar to the range constrained by the observed $\fsf$ and $\tglob$ in the previous section. This consistency between different constraints indicates that in our simulations with $\epsff\sim1\%$ and $\b\sim1$ the overall distribution of the ISM gas in different phases is captured correctly.

Typical values of $\fH2$ estimated in other \Lstar-sized galaxies are usually even larger than the Milky Way value \citep[e.g., $\sim10\%\text{--}30\%$ in][]{Leroy.etal.2008}. According to Figure~\ref{fig:fH2-taudepH2}, such $\fH2$, together with somewhat longer depletion times \citep[$\tH2\sim1\text{--}3\Gyr$ in][]{Bigiel.etal.2008,Bigiel.etal.2011,Leroy.etal.2013,Utomo.etal.2017}, favors small values of $\epsff$. Our model, calibrated on a specific simulation of an \Lstar-sized galaxy, does not predict values $\fH2>20\%$. However, according to our model, the values of $\fH2 > 20\%$ observed in molecular-rich galaxies can be due to a smaller ratio of dynamical time scales $\tp/\tmd$ in such galaxies as compared to the value of $\tp/\tmd \sim 4$ in our simulated galaxy, which sets the upper limit of $\fH2\sim 20\%$ in the dynamics-regulation regime (Equation~\ref{eq:fsf_lowepsff}).

Figure~\ref{fig:fH2-taudepH2} also illustrates three interesting differences in the behavior of $\fH2$ and $\tH2$ as compared to that of $\fsf$ and $\tglob$ in the previous section: (1) the range of $\fH2$ variation is substantially narrower than that of $\fsf$; (2) in contrast to $\tglob$, $\tH2$ does depend on $\epsff$ even in the self-regulation regime; and (3) the temporal variation of $\tH2$ (shown with vertical error bars) is much smaller than that of $\tglob$. The range of $\fH2$ variation is narrow because even at high $\epsff$ and $\b$ feedback cannot efficiently clear the non-star-forming molecular gas that piles up above the star formation threshold. When $\tglob$ is independent of $\epsff$, the sensitivity of $\tH2$ to $\epsff$ originates from the weak sensitivity of $\fH2$ to $\epsff$, $\tH2=\fH2\tglob$, and its temporal variation is small because $\fH2$ anticorrelates with $\tauff$, as both respond to the dispersal of the dense gas by feedback, and this anticorrelation mitigates the variation of $\tH2 \propto \fH2 \tauff$. Note that all these effects are due to the definition of the star-forming gas being different from the molecular gas and its corollary of the existence of the non-star-forming molecular gas.

\subsection{The Depletion Times of the Molecular Gas on Subgalactic Scales}
\label{sec:observ:subgal}

\subsubsection{$\tH2$ on Kiloparsec Scales}
\label{sec:observ:KSR:kpc}

\begin{figure}
\centering
\includegraphics[width=\columnwidth]{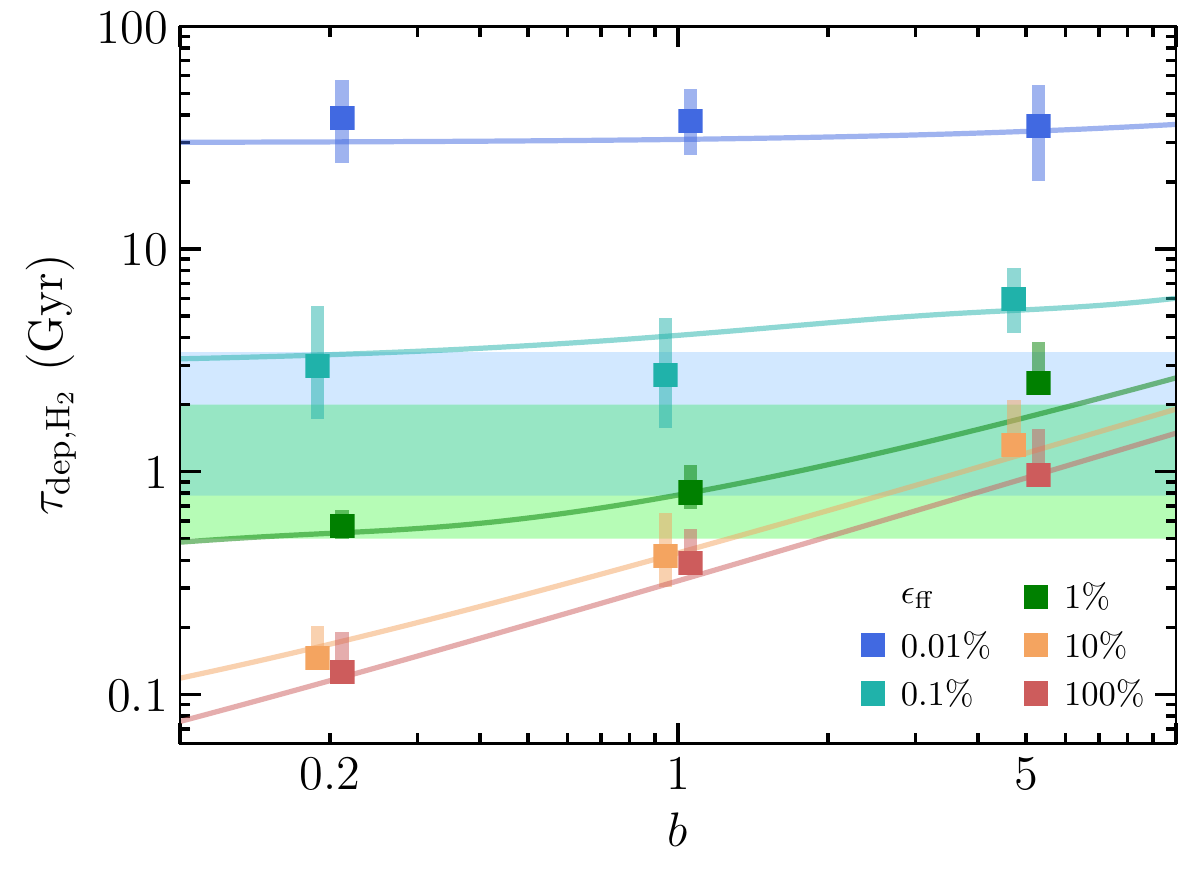}
\caption{\label{fig:b-taudepH2} Comparison of the molecular gas depletion time, $\tH2$, averaged on kiloparsec scale in our simulations (squares with vertical bands), with the observed range shown with horizontal color bands. The blue band indicates the range of $\tH2 \sim 1.6\Gyr$ (excluding correction for helium) with a factor of 2 scatter, which was derived in a number of studies \citep{Bigiel.etal.2008,Bigiel.etal.2011,Leroy.etal.2013,Bolatto.etal.2017,Utomo.etal.2017,Colombo.etal.2018}. The green band indicates the range of kiloparsec scale $\tH2$ in the Milky Way, estimated from the profiles of $\SH2$ and $\SH2$ in Figure~7 in \citet{Kennicutt.Evans.2012}. In simulations, $\tH2$ is averaged using 10 simulation snapshots between 410 and 500 Myr. Squares indicate the mass-weighted averages $\langle 1/\tH2 \rangle^{-1}$, and vertical stripes show the range of the running median for gas with $\SH2>1\Msunpc2$. For presentation purposes, the simulation points are slightly shifted horizontally around the actually used values of $\b=0.2$, 1, and 5. Colored lines show the predictions of our model for the global depletion time of the molecular gas (see Sections~\ref{sec:observ:tH2-fH2}). }
\end{figure}

\begin{figure*}
\centering
\includegraphics[width=\textwidth]{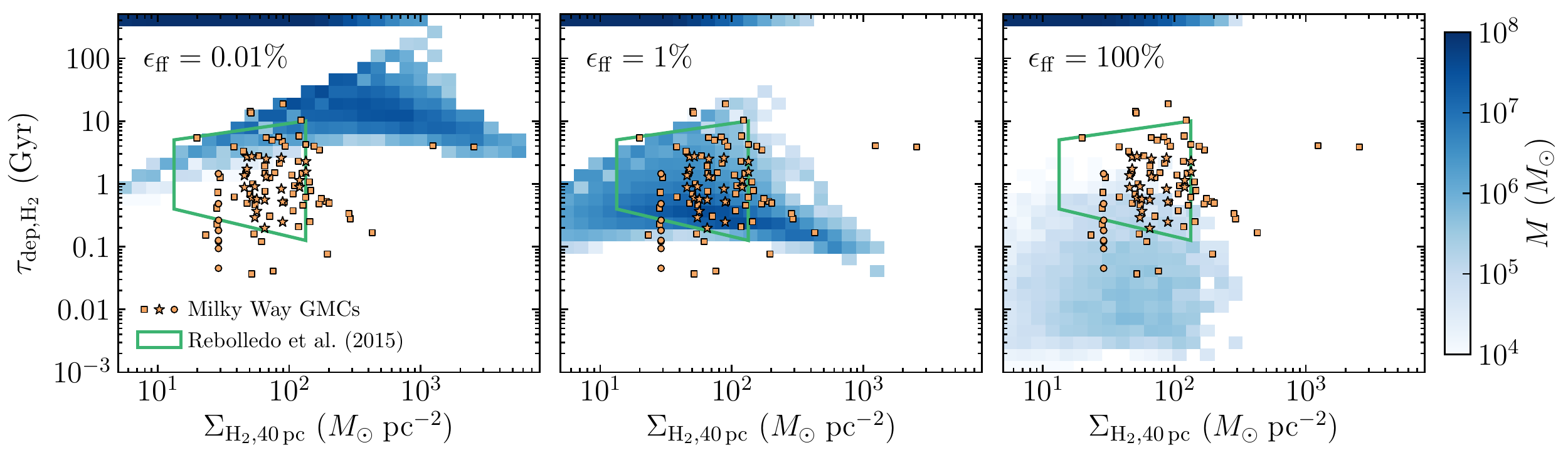}
\caption{\label{fig:tH2-local} Comparison of the molecular gas depletion time and the gas surface densities on GMC scales with their distribution on the resolution scale in our simulations, $\Delta = 40\pc$. Adopted star formation efficiency increases from the left to the right: $\epsff = 0.01\%$, $1\%$, and $100\%$. The color map shows the mass-weighted distribution of computational cells for which we define $\Sigma_{\rm H_2,40\pc} = M_{\rm H_2}^{\rm cell}/\Delta^2$ with molecular mass in a cell, $M_{\rm H_2}^{\rm cell}$, computed using the model of \citet[][see the beginning of Section~\ref{sec:observ} above]{KMT2} and $\tH2 = M_{\rm H_2}^{\rm cell}/\dot M_{\rm \star}^{\rm cell} = M_{\rm H_2}^{\rm cell}/(M_{\star}^{\rm cell}(<1\Myr)/(1\Myr))$, where in each cell $M_{\star}^{\rm cell}(<1\Myr)$ is the total mass of stars younger than $1\Myr$. Cells containing only a single stellar particle form the diagonal upper boundary of $\tH2$ distribution. Cells without young stellar particles are indicated by blue horizontal stripes on top of each axis. Orange points show the observed $\tH2$ in the Milky Way GMCs from \citet[][circles]{Lada.etal.2010}, \citet[][stars]{Heiderman.etal.2010}, and \citet[][squares]{Vutisalchavakul.etal.2016}. A green polygon indicates the range of $\SH2$ and $\tH2$ observed in three nearby spiral galaxies by \citet{Rebolledo.etal.2015}.}
\end{figure*}

Over the past two decades, star formation, the distribution of the molecular gas, and its depletion time $\tH2=\SH2/\SSFR$ have been studied observationally down to kiloparsec scales in dozens of nearby galaxies \citep[e.g.,][]{wong_blitz02,Bigiel.etal.2008,Bigiel.etal.2011,Leroy.etal.2013,Bolatto.etal.2017,Utomo.etal.2017}. These observational studies show that typical observed values of $\tH2 \sim 2\Gyr$ have a factor of $\sim 2$ scatter and are independent of the local kiloparsec-scale molecular gas surface density, $\SH2$. In the Milky Way, values of kiloparsec-scale $\tH2$ are somewhat shorter and span a range of $\tH2 \sim 0.5\text{--}2\Gyr$ \citep[estimated from Figure~7 in][]{Kennicutt.Evans.2012}. 

In Figure~\ref{fig:b-taudepH2}, we compare these values of $\tH2$ (colored bands) with the results of our simulations (squares with vertical stripes) and our model predictions (thin lines). As the figure shows, the results of our fiducial simulation with $\epsff=1\%$ and $\b=1$ agree well with the typical values of $\tH2$ inferred in observations. However, the simulations with, e.g., $\epsff\sim100\%$ and $\b\sim 5$ also agree with the observed range of $\tH2$ because the dependence of $\tH2$ on these parameters (and especially on $\epsff$) is relatively weak. Similarly to the global star-forming gas and molecular gas mass fractions considered above in Section~\ref{sec:observ:tH2-fH2}, the parameters will be constrained much better when estimates of the molecular gas fraction become available on subgalactic scales in more and more galaxies \citep[e.g.,][]{Wong.etal.2013,Leroy.etal.2016,Leroy.etal.2017}. 

To make the comparison presented in Figure~\ref{fig:b-taudepH2}, in the simulations we compute $\tH2 = \SH2/\SSFR$, where $\SH2$ and $\SSFR$ are measured by first projecting the local densities of the molecular gas and SFR perpendicular to the disk plane and then smoothing the resulting surface densities using a Gaussian filter with a width of $1\kpc$. Squares in Figure~\ref{fig:b-taudepH2} show the mass-weighted averages $\langle 1/\tH2 \rangle^{-1}$ on a kiloparsec scale, which are equivalent to the global depletion times of the molecular gas,\footnote{By definition, $\langle 1/\tH2 \rangle^{-1} \equiv [ \int dA\;(\SH2/\tH2) / \int dA\;\SH2 ]^{-1} = \int dA\;\SH2/\int dA\;\SSFR = \MH2/\SFR$.} and these averages are well approximated by our model (colored lines). A vertical band around each square indicates variation of the running median of $\tH2$ in bins of $\SH2$ at surface densities of $\SH2>1\Msunpc2$. This variation is rather small because our simulations produce constant $\tH2$, even though a density-dependent depletion time is adopted on subgrid scale: $\taust \propto \tauff \propto \rho^{-0.5}$. Such independence of $\tH2$ from $\SH2$ agrees with the observed constant $\tH2$, and its origin in our simulations is discussed in Section~4.4 of \P1. We also find that the scatter around the running median (not shown) is consistent with observations as well (see Figure~3 in \P1).

\subsubsection{$\tH2$ on Tens of Parsec Scales}
\label{sec:observ:KSR:local}

Although current observations in most galaxies  probe star formation and molecular gas only on scales $\gtrsim 1$ kpc, observations of star-forming regions in the Milky Way allow us to examine these quantities on smaller scales. Furthermore, scales of $\lesssim 100$ pc are increasingly probed in nearby galaxies \citep{Bolatto.etal.2011,Rebolledo.etal.2015,Leroy.etal.2017}, and this allows us to compare results of our simulations on these scales as well.

Figure~\ref{fig:tH2-local} shows the variation of $\tH2$ in the Milky Way \citep[points;][]{Heiderman.etal.2010,Lada.etal.2010,Vutisalchavakul.etal.2016} and three nearby spiral galaxies \citep[trapezoidal region;][]{Rebolledo.etal.2015} with the molecular gas depletion time on the scale of $40$ pc in our simulations (blue color map) as a function of $\SH2$. For this comparison we only show GMCs in the Milky Way that have sizes of $\gtrsim 10$ pc, to make the scales comparable to the scale probed in our simulations. Different panels show the distribution of the local depletion times in our simulations with different values of the star formation efficiency: $\epsff = 0.01\%$, 1\%, and 100\%. 

As the figure shows, although the observed $\tH2$ vary substantially, their typical values can be reproduced only in runs with $\epsff \sim 1\%$, while runs with too low (high) $\epsff$ significantly overestimate (underestimate) $\tH2$ in star-forming regions. Note that in all runs the distribution of $\tH2$ is bimodal: $\tH2$ is either finite, which corresponds to star-forming gas, or infinitely long, i.e. the gas is non-star-forming. In the figure, $\tH2$ in the latter case is artificially set to 500 Gyr for illustration purposes. Different runs differ by the fraction of the molecular gas in the star-forming state and by the average $\tH2$ of such gas. The fraction of the star-forming gas is the lowest in the run with $\epsff=100\%$, and this gas has depletion times of only $\sim 2\text{--}200\Myr$. These short depletion times of star-forming H$_2$ are averaged with large amounts of the non-star-forming molecular gas in this run, so that the depletion time on $\gtrsim 1$ kpc scales in the $\epsff=100\%$ case is only a factor of two shorter than in the $\epsff=1\%$ run.  This shows that while $\tH2$ on $\gtrsim 1$ kpc scales is relatively insensitive to $\epsff$, its values on the scales of $\lesssim 100$ pc are quite sensitive to the efficiency and can thus be used to constrain it.

\subsubsection{The Scale Dependence of $\tH2$}
\label{sec:observ:KSR:bias}

\begin{figure}
\centering
\includegraphics[width=\columnwidth]{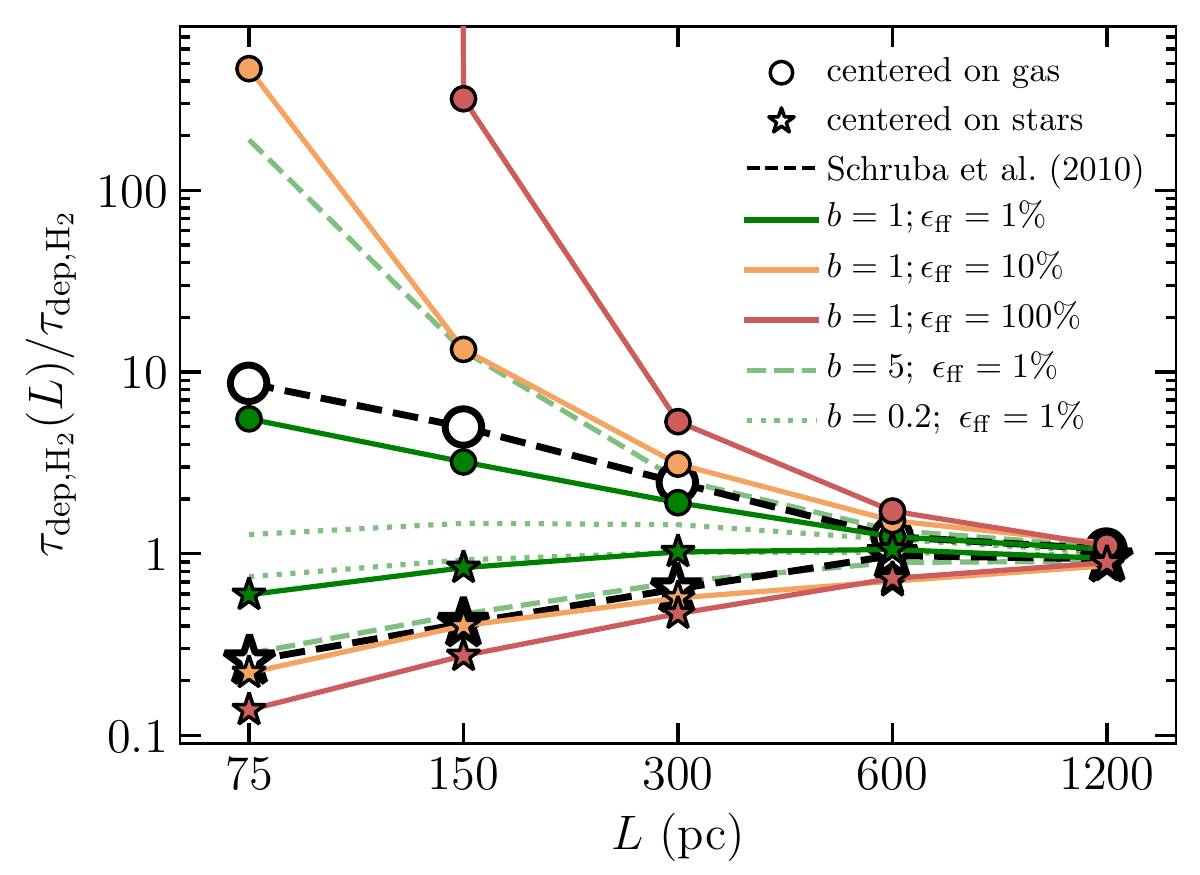}
\caption{\label{fig:scatter} Effect of $\epsff$ and the feedback boost factor, $\b$, on the $\tglob$ bias as a function of the spatial smoothing scale, $L$. The depletion time in a given aperture of size $L$ is defined as $\tL \equiv \Sigma_{{\rm H_2,}L}/\dot\Sigma_{{\star,}L}$, where $\Sigma_{{\rm H_2,}L}$ and $\dot\Sigma_{{\star,}L}$ are the molecular gas and the SFR surface densities smoothed using a Gaussian filter with a width $L$. Star symbols indicate the median depletion time measured in the apertures centered on peaks in $\SSFR$, while circles correspond to the apertures centered on peaks in $\SH2$. To factor out the variation of the global molecular gas depletion time with the feedback strength, we divide $\tL$ by global $\tH2$. Dashed lines show the results obtained for M33 by \citet{Schruba.etal.2010}. To match the temporal averaging of the H$\alpha$ indicator used by \citet{Schruba.etal.2010}, we estimate $\SSFR$ using stars younger than 10 Myr.}
\end{figure}

Results of the previous two sections clearly show that the distribution of $\tH2$ depends on the spatial scale. Indeed, $\tH2$ in a given ISM patch results from averaging over a distribution of gas and stars inside the patch, and thus $\tH2$ depends on the patch size, $L$: $\tL$. The quantity that particularly strongly depends on the spatial scale is scatter: when the size of the patch decreases, patch-to-patch variation of gas and stars contained inside a patch becomes stronger, which leads to a stronger variation of the derived depletion time in each patch and thus larger scatter in $\tH2$. 

Following \citet{Schruba.etal.2010}, one of the ways to express the dependence of scatter on the spatial scale is to consider the scale dependence of the depletion time, $\tL$, measured in patches centered on the peaks of $\SH2$, which thus are biased to long $\tL$, versus those measured in patches centered on the peaks of $\SSFR$,  which are biased to short $\tL$. The difference between these two estimates of $\tL$ is small on large scales, and their values are approximately equal to the global depletion time. At smaller scales, this difference increases, as shown in Figure~\ref{fig:scatter}, which compares $\tL$ observed in M33 by \citet{Schruba.etal.2010} with the results of our simulations. 

In simulations, $\tL$ centered on gas or stars strongly depends on $\epsff$ and the feedback boost factor $\b$ because stronger feedback-induced gas flux results in more expulsive evacuation of the gas from star-forming regions, which leads to a stronger spatial displacement of $\SH2$ and $\SSFR$ peaks. As Figure~\ref{fig:scatter} shows, the fiducial run that satisfied all previous constraints also provides a reasonably good match to the observed $\tL$. Overall, for the fiducial feedback strength, both gas- and star-centered $\tL$ favors $\epsff\lesssim 10\%$. Note, however, that there is a degeneracy between the feedback strength and $\epsff$ value: the simulation with $\epsff=1\%$ and $\b=5$ produces a relation similar to the simulation with $\epsff=10\%$ and $\b=1$.

It is also worth noting that $\SH2$-centered $\tL$ is noticeably more sensitive to  $\epsff$ and $\b$ values. The sensitivity is stronger because at higher $\epsff$ or $\b$ the gas lifetime in the star-forming state is shorter, young stars are more sporadic, and thus it is less probable for a given patch centered on a $\SH2$ peak to contain young stars. As a result, $\tL$ at high $\epsff$ or $\b$ becomes highly biased to very large values. On the contrary, $\SSFR$-centered patches almost always contain molecular gas, because its abundance does not significantly decrease at stronger feedback (see Section~\ref{sec:observ:tH2-fH2}). As a result, for $\SSFR$-centered patches, the bias also increases with $\epsff$ and $\b$ ($\tL$ becomes shorter), but this change is much milder than for $\SH2$-centered patches. 

Such strong dependence of $\SH2$-centered $\tL$ on star formation and feedback parameters can provide tight constraints on these parameters. These constraints can be improved significantly if the scale dependence of $\tH2$ is measured in a larger sample of galaxies. Note, however, that more comprehensive comparison must include the effects of the intrinsic variation of $\epsff$ and the metallicity dependence of the molecular gas fraction on GMC scale, which are not accounted for in our simulations.

\section{Comparison with previous studies}
\label{sec:disc:prev}

In previous sections, we showed that the simple theoretical framework presented in \P1 and Section \ref{sec:model:summary} explains how local star formation and feedback parameters affect the global star formation in our  \Lstar-sized galaxy simulations, both qualitatively and quantitatively. Here we illustrate how our framework can also explain the results of other recent galaxy simulations done with different numerical methods and implementations of star formation and feedback, both in isolated setups and in the cosmological context. Specifically, we will use our model to interpret trends (or lack thereof) of the depletion times with the local star formation efficiency, $\epsff$, the feedback strength, and the adopted star formation thresholds.

For example, our framework predicts that in the simulations that adopt high $\epsff$ values and implement efficient feedback the depletion time is almost completely insensitive to the value of $\epsff$. This is because in this regime $\tglob$ is controlled by the time that gas spends in the non-star-forming state, which does not depend on $\epsff$ explicitly. This explains why $\tglob$ is insensitive to the variation of $\epsff$ in the simulations of \cite{Hopkins.etal.2017}; this behavior is also reproduced in our simulations (see Figures \ref{fig:taudep_fid} and \ref{fig:fb} above). In this regime, our framework also predicts a nearly linear scaling of $\tglob$ with the feedback strength parameter $\xi$, as is indeed observed in simulations \citep[][]{Benincasa.etal.2016,Hopkins.etal.2017,Orr.etal.2017}. 

For smaller values of $\epsff\approx 1\text{--}10\%$, when the two terms in Equation~(\ref{eq:tglob2}) contribute comparably to the total depletion time, the model predicts that $\tglob$ should scale with $\epsff$ weakly (sublinearly). This was indeed observed in a number of simulations carried out in this regime \citep{Saitoh.etal.2008,Dobbs.etal.2011a,Agertz.etal.2013,Agertz.etal.2015,Benincasa.etal.2016}. In this case, sublinear scaling is also expected with the strength of feedback, $\xi$, which is also confirmed by simulations \citep{Hopkins.etal.2011,Agertz.etal.2013,Agertz.etal.2015,Benincasa.etal.2016}. 

For simulations with $\epsff \lesssim 1\%$ or when the feedback implementation is inefficient, $\xi \ll 1$, our model predicts that the depletion time is controlled by the second term in Equation~(\ref{eq:tglob2}) and that it scales inversely with $\epsff$: $\tglob \sim \epsff^{-1}$. Such scaling was observed in the simulations without feedback by \citet{Agertz.etal.2013,Agertz.etal.2015}, while in the simulations using the same galaxy model but with efficient feedback, $\tglob$ was found to be only weakly dependent on $\epsff$.

The weak dependence or complete insensitivity of $\tglob$ to $\epsff$ at intermediate and high $\epsff$ explains why different galaxy simulations with widely different $\epsff \sim 1\%\text{--}100\%$ all produce realistic global depletion times. However, as our results of Section~\ref{sec:observ} show, these simulations make drastically different predictions for the star-forming and molecular gas mass fractions, which can be used to constrain $\epsff$ in this regime (see Section~\ref{sec:observ:global}). A similar idea was reported previously by \citet{Hopkins.etal.2012,Hopkins.etal.2013b}, who showed that the fraction of gas in the dense molecular state with $n>10^4\cc$ strongly depends on the local efficiency $\epsff$ and the feedback implementation. Specifically, simulations with high $\epsff$ and efficient feedback have a small dense gas mass fraction owing to efficient conversion of dense gas into stars and its dispersal by feedback. This effect can explain why in the simulations with $\epsff=100\%$ reported by \citet{Orr.etal.2017} the Kennicutt--Schmidt relation between the surface densities of SFR and dense and cold gas ($n > 10\cc$ and $T < 300\K$) is considerably higher than the observed relation for molecular gas. In these simulations, the SFR is likely realistic because the depletion time of the total gas is expected to be insensitive to $\epsff$. The dense gas fraction, on the contrary, is expected to be small, which leads to the small surface density of such gas and thus high Kennicutt--Schmidt relation as in \citet{Orr.etal.2017}.

Our model also predicts that $\tglob$ depends on the star formation threshold differently in different regimes. For low $\epsff$, $\tglob$ only weakly depends on the threshold value, while at high $\epsff$, $\tglob$ decreases when the threshold encompasses more gas from a given distribution (see top middle and left panels in Figure~\ref{fig:fb}). The former weak trend agrees with the results of \citet{Saitoh.etal.2008}, who found that for $\epsff \sim 1.5\%$ the value of $\tglob$ decreased only by a factor of $\sim 1.5\text{--}2$ when the density threshold was varied from $\nsf = 100$ to $0.1\cc$. Similarly, \citet{Hopkins.etal.2011} and \citet{Benincasa.etal.2016} found almost no dependence of $\tglob$ on $\nsf$. On the contrary, in simulations of \citet{Agertz.etal.2015} with $\epsff=10\%$, $\tglob$ varied relatively strongly with variation of $\nsf$, as expected for high $\epsff$. We note that to observe the effect on $\tglob$ when a combination of thresholds in different physical variables is used, all thresholds must be varied simultaneously. Varying thresholds one by one may not affect $\tglob$ if several thresholds define approximately the same gas as star-forming. This is likely why \citet{Hopkins.etal.2017} found that $\tglob$ is insensitive to variation of star formation thresholds, when thresholds in different variables were changed.

\section{Summary and conclusions}
\label{sec:summary}

Using a simple physical model presented in \citet[][see also Section \ref{sec:model:summary} above]{Semenov.etal.2017} and a suite of \Lstar-sized galaxy simulations, we explored how the global depletion times in galaxies, $\tglob=\Mg/\SFR$, and the gas mass fractions in the star-forming and molecular states depend on the choices of the parameters of local star formation and feedback.

In our model, $\tglob$ is expressed as a sum of contributions from different physical processes, which include dynamical processes in the ISM, the conversion of gas into stars in star-forming regions, and the dispersal of such regions by stellar feedback. Some of these processes explicitly depend on the parameters of the {\it local} star formation and feedback model, such as a star formation efficiency per freefall time, $\epsff$, and a feedback boost factor, $\b$. Others do not have such explicit dependence and may be affected by these parameters only indirectly.  This leads to two distinct regimes, in which terms with and without such explicit dependence dominate.  

We demonstrated these regimes in a suite of \Lstar-sized galaxy simulations, in which we systematically varied $\epsff$, $\b$, and the thresholds used to define the star-forming gas. We also showed that the trends of $\tglob$ and the star-forming gas mass fraction exhibited in the simulations can be reproduced by our model both qualitatively and quantitatively after a minimal calibration of the model parameters. The main results of our simulations and the predictions of our model can be summarized as follows:

\begin{enumerate}

\item When $\epsff$ or $\b$ are large, the contribution of processes without explicit dependence on $\epsff$ dominates and $\tglob$ is insensitive to $\epsff$, which  is usually referred to as ``self-regulation'' in the literature.  However, in this regime, the mass fractions of the star-forming ($\fsf$) and the molecular ($\fH2$) gas {\it do} depend sensitively on $\epsff$, and $\tglob$ scales almost linearly with the feedback strength factor for $\b\gtrsim 1$. 

\item Conversely, when $\epsff$ or $\b$ are sufficiently small, $\tglob$ is dominated by the processes that explicitly depend on the local gas depletion time, $\tdep=\tff/\epsff$ in Equation~(\ref{eq:rhoSFR1}), and thus on $\epsff$, but not on the feedback strength. In this case, the model predicts $\tglob\propto \epsff^{-1}$ and only weak dependence of $\fsf$ and $\fH2$ on $\epsff$, the behavior confirmed by our simulations. 
		
\item The star formation threshold controls the mass fraction of the star-forming gas, the extent of star-forming regions, and their average properties, such as the average freefall time. We find that  when $\epsff$ is small and the threshold is such that only a small fraction of the  ISM gas is star-forming, $\tglob$ and $\fsf$ are sensitive to the threshold value. 

\item When $\epsff$ is large or feedback is efficient (i.e., when the first term in Equation~\ref{eq:tglob2} dominates), $\fsf$ is small and most of the star-forming gas has density or virial parameter close to the star formation threshold. In this case global star formation and the molecular mass fraction, $\fH2$, become sensitive to the value of the threshold.  

\end{enumerate}

The dependence of global star-forming properties of galaxies on the parameters of the local star formation and feedback model can be used to constrain the values of these parameters using observations of global galaxy properties. For example, the global depletion times of the total and molecular gas constrain the feedback strength but cannot constrain the value of $\epsff$ owing to their weak dependence on this parameter. However, the value of $\epsff$ can be constrained using the mass fraction of gas in the star-forming or molecular state. In addition, we showed that $\epsff$ can be constrained using the distribution of local depletion times in star-forming regions and measurements of $\tH2$ for gas patches of different sizes centered on the peaks of the molecular gas surface density. 

Using our simulation suite, we demonstrated that it is possible to find a combination of the local star formation and feedback parameters that satisfies all of these observational constraints. Our fiducial run with $\epsff = 1\%$, the fiducial feedback boost $\b=1$, and the star formation threshold based on the virial parameter, $\avir < \avirsf=10$, is able to match all considered observations reasonably well. The low values of $\epsff\sim 1\%$ are also consistent with previous inferences \citep[e.g.,][and references therein]{Krumholz.etal.2012}. We admit that the obtained constraints on $\epsff$ and other parameters are specific to the scales close to our resolution, i.e. $\sim 40\pc$, and an additional study is required to explore the scale dependence of these constraints on smaller spatial scales. We note, however, that the observed depletion times in GMCs on parsec scales also favor $\epsff \sim 1\%\text{--}10\%$ \citep[e.g.,][]{Heiderman.etal.2010,Gutermuth.etal.2011}, while simulations with a few parsec resolution adopting higher $\epsff$ seem to underpredict the amount of dense star-forming gas (see the end of Section~\ref{sec:disc:prev}).

We also showed that our model explains the results of a number of recent studies that explored the effects of the local star formation and feedback model on the global properties of simulated galaxies. This broad consistency confirms that our model accurately describes the origin of global star-forming properties in galaxy simulations and thus allows us to understand the role played by gas dynamics, star formation, and feedback in shaping these properties. Understanding the role of these processes in simulations also sheds light on their role in real galaxies, which is an essential step toward understanding how real galaxies form and evolve.

\acknowledgments

We thank the anonymous referee for constructive feedback and valuable suggestions. We are also grateful to Robert Feldmann, Romain Teyssier, Philip Hopkins, Clarke Esmerian, and Philip Mansfield, whose comments helped to improve our paper. A.K. and N.G. would like to thank participants and organizers of the Simons symposium series on galactic superwinds for stimulating discussions that played a role in motivating this study. 
A.K. is grateful to the Cosmology Hub at the University College London for hospitality during completion of this paper. 
This work was supported by NASA ATP grant NNH12ZDA001N, NSF grant AST-1412107, and the Kavli Institute for Cosmological Physics at the University of Chicago through grant PHY-1125897 and an endowment from the Kavli Foundation and its founder, Fred Kavli. The simulation and analyses presented in this paper have been carried out using the Midway cluster at the University of Chicago Research Computing Center, which we acknowledge for support. Analyses presented in this paper were greatly aided by the following free software packages: {\tt yt} \citep{yt}, {\tt NumPy} \citep{numpy_ndarray}, {\tt SciPy} \citep{scipy}, {\tt Matplotlib} \citep{matplotlib}, and {\tt GitHub}\footnote{\href{https://github.com}{https://github.com}}. We have also used the Astrophysics Data Service ({ADS}\footnote{\href{http://adsabs.harvard.edu/abstract_service.html}{http://adsabs.harvard.edu/abstract\_service.html}}) and {\tt arXiv}\footnote{\href{https://arxiv.org}{https://arxiv.org}} preprint repository extensively during this project and writing of the paper.

\bibliographystyle{aasjournal}
\bibliography{}

\begin{thebibliography}{}
\expandafter\ifx\csname natexlab\endcsname\relax\def\natexlab#1{#1}\fi

\bibitem[{{Agertz} \& {Kravtsov}(2015)}]{Agertz.Kravtsov.2015}
{Agertz}, O., \& {Kravtsov}, A.~V. 2015, \apj, 804, 18

\bibitem[{{Agertz} \& {Kravtsov}(2016)}]{Agertz.Kravtsov.2016}
---. 2016, \apj, 824, 79

\bibitem[{{Agertz} {et~al.}(2013){Agertz}, {Kravtsov}, {Leitner}, \&
  {Gnedin}}]{Agertz.etal.2013}
{Agertz}, O., {Kravtsov}, A.~V., {Leitner}, S.~N., \& {Gnedin}, N.~Y. 2013,
  \apj, 770, 25

\bibitem[{{Agertz} {et~al.}(2015){Agertz}, {Romeo}, \&
  {Grisdale}}]{Agertz.etal.2015}
{Agertz}, O., {Romeo}, A.~B., \& {Grisdale}, K. 2015, \mnras, 449, 2156

\bibitem[{{Benincasa} {et~al.}(2016){Benincasa}, {Wadsley}, {Couchman}, \&
  {Keller}}]{Benincasa.etal.2016}
{Benincasa}, S.~M., {Wadsley}, J., {Couchman}, H.~M.~P., \& {Keller}, B.~W.
  2016, \mnras, 462, 3053

\bibitem[{{Bertoldi} \& {McKee}(1992)}]{Bertoldi.McKee.1992}
{Bertoldi}, F., \& {McKee}, C.~F. 1992, \apj, 395, 140

\bibitem[{{Bigiel} {et~al.}(2008){Bigiel}, {Leroy}, {Walter}, {Brinks}, {de
  Blok}, {Madore}, \& {Thornley}}]{Bigiel.etal.2008}
{Bigiel}, F., {Leroy}, A., {Walter}, F., {et~al.} 2008, \aj, 136, 2846

\bibitem[{{Bigiel} {et~al.}(2011){Bigiel}, {Leroy}, {Walter}, {Brinks}, {de
  Blok}, {Kramer}, {Rix}, {Schruba}, {Schuster}, {Usero}, \&
  {Wiesemeyer}}]{Bigiel.etal.2011}
{Bigiel}, F., {Leroy}, A.~K., {Walter}, F., {et~al.} 2011, \apjl, 730, L13

\bibitem[{{Bolatto} {et~al.}(2011){Bolatto}, {Leroy}, {Jameson}, {Ostriker},
  {Gordon}, {Lawton}, {Stanimirovi{\'c}}, {Israel}, {Madden}, {Hony},
  {Sandstrom}, {Bot}, {Rubio}, {Winkler}, {Roman-Duval}, {van Loon},
  {Oliveira}, \& {Indebetouw}}]{Bolatto.etal.2011}
{Bolatto}, A.~D., {Leroy}, A.~K., {Jameson}, K., {et~al.} 2011, \apj, 741, 12

\bibitem[{{Bolatto} {et~al.}(2017){Bolatto}, {Wong}, {Utomo}, {Blitz}, {Vogel},
  {S{\'a}nchez}, {Barrera-Ballesteros}, {Cao}, {Colombo}, {Dannerbauer},
  {Garc{\'{\i}}a-Benito}, {Herrera-Camus}, {Husemann}, {Kalinova}, {Leroy},
  {Leung}, {Levy}, {Mast}, {Ostriker}, {Rosolowsky}, {Sandstrom}, {Teuben},
  {van de Ven}, \& {Walter}}]{Bolatto.etal.2017}
{Bolatto}, A.~D., {Wong}, T., {Utomo}, D., {et~al.} 2017, \apj, 846, 159

\bibitem[{{Braun} \& {Schmidt}(2012)}]{Braun.Schmidt.2012}
{Braun}, H., \& {Schmidt}, W. 2012, \mnras, 421, 1838

\bibitem[{{Cen} \& {Ostriker}(1992)}]{Cen.Ostriker.1992}
{Cen}, R., \& {Ostriker}, J.~P. 1992, \apjl, 399, L113

\bibitem[{{Chabrier}(2003)}]{Chabrier.2003}
{Chabrier}, G. 2003, \pasp, 115, 763

\bibitem[{{Colombo} {et~al.}(2018){Colombo}, {Kalinova}, {Utomo}, {Rosolowsky},
  {Bolatto}, {Levy}, {Wong}, {Sanchez}, {Leroy}, {Ostriker}, {Blitz}, {Vogel},
  {Mast}, {Garc{\'{\i}}a-Benito}, {Husemann}, {Dannerbauer}, {Ellmeier}, \&
  {Cao}}]{Colombo.etal.2018}
{Colombo}, D., {Kalinova}, V., {Utomo}, D., {et~al.} 2018, \mnras, 475, 1791

\bibitem[{{Dobbs} {et~al.}(2011){Dobbs}, {Burkert}, \&
  {Pringle}}]{Dobbs.etal.2011a}
{Dobbs}, C.~L., {Burkert}, A., \& {Pringle}, J.~E. 2011, \mnras, 417, 1318

\bibitem[{{Elmegreen}(2015)}]{Elmegreen.2015}
{Elmegreen}, B.~G. 2015, \apjl, 814, L30

\bibitem[{{Elmegreen}(2018)}]{Elmegreen.2018}
---. 2018, \apj, 854, 16

\bibitem[{{Evans} {et~al.}(2014){Evans}, {Heiderman}, \&
  {Vutisalchavakul}}]{Evans.etal.2014}
{Evans}, II, N.~J., {Heiderman}, A., \& {Vutisalchavakul}, N. 2014, \apj, 782,
  114

\bibitem[{{Evans} {et~al.}(2009){Evans}, {Dunham}, {J{\o}rgensen}, {Enoch},
  {Mer{\'{\i}}n}, {van Dishoeck}, {Alcal{\'a}}, {Myers}, {Stapelfeldt},
  {Huard}, {Allen}, {Harvey}, {van Kempen}, {Blake}, {Koerner}, {Mundy},
  {Padgett}, \& {Sargent}}]{Evans.etal.2009}
{Evans}, II, N.~J., {Dunham}, M.~M., {J{\o}rgensen}, J.~K., {et~al.} 2009,
  \apjs, 181, 321

\bibitem[{{Genel} {et~al.}(2013){Genel}, {Vogelsberger}, {Nelson}, {Sijacki},
  {Springel}, \& {Hernquist}}]{Genel.etal.2013}
{Genel}, S., {Vogelsberger}, M., {Nelson}, D., {et~al.} 2013, \mnras, 435, 1426

\bibitem[{{Gentry} {et~al.}(2017){Gentry}, {Krumholz}, {Dekel}, \&
  {Madau}}]{Gentry.etal.2017}
{Gentry}, E.~S., {Krumholz}, M.~R., {Dekel}, A., \& {Madau}, P. 2017, \mnras,
  465, 2471

\bibitem[{{Gentry} {et~al.}(2018){Gentry}, {Krumholz}, {Madau}, \&
  {Lupi}}]{Gentry.etal.2018}
{Gentry}, E.~S., {Krumholz}, M.~R., {Madau}, P., \& {Lupi}, A. 2018, {\mnras},
  submitted (arXiv:1802.06860)

\bibitem[{{Gnedin} \& {Hollon}(2012)}]{Gnedin.Hollon.2012}
{Gnedin}, N.~Y., \& {Hollon}, N. 2012, \apjs, 202, 13

\bibitem[{{Gnedin} \& {Kravtsov}(2011)}]{Gnedin.Kravtsov.2011}
{Gnedin}, N.~Y., \& {Kravtsov}, A.~V. 2011, \apj, 728, 88

\bibitem[{{Gnedin} {et~al.}(2014){Gnedin}, {Tasker}, \&
  {Fujimoto}}]{Gnedin.etal.2014}
{Gnedin}, N.~Y., {Tasker}, E.~J., \& {Fujimoto}, Y. 2014, \apjl, 787, L7

\bibitem[{{Governato} {et~al.}(2010){Governato}, {Brook}, {Mayer}, {Brooks},
  {Rhee}, {Wadsley}, {Jonsson}, {Willman}, {Stinson}, {Quinn}, \&
  {Madau}}]{Governato.etal.2010}
{Governato}, F., {Brook}, C., {Mayer}, L., {et~al.} 2010, \nat, 463, 203

\bibitem[{{Grand} {et~al.}(2017){Grand}, {G{\'o}mez}, {Marinacci}, {Pakmor},
  {Springel}, {Campbell}, {Frenk}, {Jenkins}, \& {White}}]{Grand.etal.2017}
{Grand}, R.~J.~J., {G{\'o}mez}, F.~A., {Marinacci}, F., {et~al.} 2017, \mnras,
  467, 179

\bibitem[{{Grisdale} {et~al.}(2017){Grisdale}, {Agertz}, {Romeo}, {Renaud}, \&
  {Read}}]{Grisdale.etal.2017}
{Grisdale}, K., {Agertz}, O., {Romeo}, A.~B., {Renaud}, F., \& {Read}, J.~I.
  2017, \mnras, 466, 1093

\bibitem[{{Gutermuth} {et~al.}(2011){Gutermuth}, {Pipher}, {Megeath}, {Myers},
  {Allen}, \& {Allen}}]{Gutermuth.etal.2011}
{Gutermuth}, R.~A., {Pipher}, J.~L., {Megeath}, S.~T., {et~al.} 2011, \apj,
  739, 84

\bibitem[{{Heiderman} {et~al.}(2010){Heiderman}, {Evans}, {Allen}, {Huard}, \&
  {Heyer}}]{Heiderman.etal.2010}
{Heiderman}, A., {Evans}, II, N.~J., {Allen}, L.~E., {Huard}, T., \& {Heyer},
  M. 2010, \apj, 723, 1019

\bibitem[{{Hernquist}(1990)}]{Hernquist.1990}
{Hernquist}, L. 1990, \apj, 356, 359

\bibitem[{{Heyer} \& {Dame}(2015)}]{Heyer.Dame.2015}
{Heyer}, M., \& {Dame}, T.~M. 2015, \araa, 53, 583

\bibitem[{{Hopkins} {et~al.}(2014){Hopkins}, {Kere{\v s}}, {O{\~n}orbe},
  {Faucher-Gigu{\`e}re}, {Quataert}, {Murray}, \&
  {Bullock}}]{Hopkins.etal.2014}
{Hopkins}, P.~F., {Kere{\v s}}, D., {O{\~n}orbe}, J., {et~al.} 2014, \mnras,
  445, 581

\bibitem[{{Hopkins} {et~al.}(2013{\natexlab{a}}){Hopkins}, {Narayanan}, \&
  {Murray}}]{Hopkins.etal.2013}
{Hopkins}, P.~F., {Narayanan}, D., \& {Murray}, N. 2013{\natexlab{a}}, \mnras,
  432, 2647

\bibitem[{{Hopkins} {et~al.}(2013{\natexlab{b}}){Hopkins}, {Narayanan},
  {Murray}, \& {Quataert}}]{Hopkins.etal.2013b}
{Hopkins}, P.~F., {Narayanan}, D., {Murray}, N., \& {Quataert}, E.
  2013{\natexlab{b}}, \mnras, 433, 69

\bibitem[{{Hopkins} {et~al.}(2011){Hopkins}, {Quataert}, \&
  {Murray}}]{Hopkins.etal.2011}
{Hopkins}, P.~F., {Quataert}, E., \& {Murray}, N. 2011, \mnras, 417, 950

\bibitem[{{Hopkins} {et~al.}(2012){Hopkins}, {Quataert}, \&
  {Murray}}]{Hopkins.etal.2012}
---. 2012, \mnras, 421, 3488

\bibitem[{{Hopkins} {et~al.}(2017){Hopkins}, {Wetzel}, {Keres},
  {Faucher-Giguere}, {Quataert}, {Boylan-Kolchin}, {Murray}, {Hayward},
  {Garrison-Kimmel}, {Hummels}, {Feldmann}, {Torrey}, {Ma}, {Angles-Alcazar},
  {Su}, {Orr}, {Schmitz}, {Escala}, {Sanderson}, {Grudic}, {Hafen}, {Kim},
  {Fitts}, {Bullock}, {Wheeler}, {Chan}, {Elbert}, \&
  {Narananan}}]{Hopkins.etal.2017}
{Hopkins}, P.~F., {Wetzel}, A., {Keres}, D., {et~al.} 2017, ArXiv e-prints,
  arXiv:1702.06148

\bibitem[{{Hopkins} {et~al.}(2018){Hopkins}, {Wetzel}, {Kere{\v s}},
  {Faucher-Gigu{\`e}re}, {Quataert}, {Boylan-Kolchin}, {Murray}, {Hayward}, \&
  {El-Badry}}]{Hopkins.etal.2017b}
{Hopkins}, P.~F., {Wetzel}, A., {Kere{\v s}}, D., {et~al.} 2018, \mnras, 477,
  1578

\bibitem[{Hunter(2007)}]{matplotlib}
Hunter, J.~D. 2007, Computing In Science \& Engineering, 9, 90

\bibitem[{Jones {et~al.}(2001-2016)Jones, Oliphant, Peterson, {et~al.}}]{scipy}
Jones, E., Oliphant, T., Peterson, P., {et~al.} 2001-2016, http://www.scipy.org

\bibitem[{{Kalberla} \& {Kerp}(2009)}]{Kalberla.Kerp.2009}
{Kalberla}, P.~M.~W., \& {Kerp}, J. 2009, \araa, 47, 27

\bibitem[{{Katz}(1992)}]{Katz.1992}
{Katz}, N. 1992, \apj, 391, 502

\bibitem[{{Kennicutt} \& {Evans}(2012)}]{Kennicutt.Evans.2012}
{Kennicutt}, R.~C., \& {Evans}, N.~J. 2012, \araa, 50, 531

\bibitem[{{Kennicutt}(1989)}]{Kennicutt.1989}
{Kennicutt}, Jr., R.~C. 1989, \apj, 344, 685

\bibitem[{{Kennicutt}(1998)}]{Kennicutt.1998}
---. 1998, \apj, 498, 541

\bibitem[{{Kim} {et~al.}(2016){Kim}, {Agertz}, {Teyssier}, {Butler},
  {Ceverino}, {Choi}, {Feldmann}, {Keller}, {Lupi}, {Quinn}, {Revaz},
  {Wallace}, {Gnedin}, {Leitner}, {Shen}, {Smith}, {Thompson}, {Turk}, {Abel},
  {Arraki}, {Benincasa}, {Chakrabarti}, {DeGraf}, {Dekel}, {Goldbaum},
  {Hopkins}, {Hummels}, {Klypin}, {Li}, {Madau}, {Mandelker}, {Mayer},
  {Nagamine}, {Nickerson}, {O'Shea}, {Primack}, {Roca-F{\`a}brega}, {Semenov},
  {Shimizu}, {Simpson}, {Todoroki}, {Wadsley}, {Wise}, \& {AGORA
  Collaboration}}]{agora2}
{Kim}, J.-h., {Agertz}, O., {Teyssier}, R., {et~al.} 2016, \apj, 833, 202

\bibitem[{{Kravtsov}(1999)}]{Kravtsov.1999}
{Kravtsov}, A.~V. 1999, PhD thesis, NEW MEXICO STATE UNIVERSITY

\bibitem[{{Kravtsov} {et~al.}(2002){Kravtsov}, {Klypin}, \&
  {Hoffman}}]{Kravtsov.etal.2002}
{Kravtsov}, A.~V., {Klypin}, A., \& {Hoffman}, Y. 2002, \apj, 571, 563

\bibitem[{{Kravtsov} {et~al.}(1997){Kravtsov}, {Klypin}, \&
  {Khokhlov}}]{Kravtsov.etal.1997}
{Kravtsov}, A.~V., {Klypin}, A.~A., \& {Khokhlov}, A.~M. 1997, \apjs, 111, 73

\bibitem[{{Kruijssen} \& {Longmore}(2014)}]{Kruijssen.Longmore.2014}
{Kruijssen}, J.~M.~D., \& {Longmore}, S.~N. 2014, \mnras, 439, 3239

\bibitem[{{Krumholz} {et~al.}(2012){Krumholz}, {Dekel}, \&
  {McKee}}]{Krumholz.etal.2012}
{Krumholz}, M.~R., {Dekel}, A., \& {McKee}, C.~F. 2012, \apj, 745, 69

\bibitem[{{Krumholz} {et~al.}(2008){Krumholz}, {McKee}, \& {Tumlinson}}]{KMT1}
{Krumholz}, M.~R., {McKee}, C.~F., \& {Tumlinson}, J. 2008, \apj, 689, 865

\bibitem[{{Krumholz} {et~al.}(2009){Krumholz}, {McKee}, \& {Tumlinson}}]{KMT2}
---. 2009, \apj, 693, 216

\bibitem[{{Lada} {et~al.}(2012){Lada}, {Forbrich}, {Lombardi}, \&
  {Alves}}]{Lada.etal.2012}
{Lada}, C.~J., {Forbrich}, J., {Lombardi}, M., \& {Alves}, J.~F. 2012, \apj,
  745, 190

\bibitem[{{Lada} {et~al.}(2010){Lada}, {Lombardi}, \& {Alves}}]{Lada.etal.2010}
{Lada}, C.~J., {Lombardi}, M., \& {Alves}, J.~F. 2010, \apj, 724, 687

\bibitem[{{Lee} {et~al.}(2016){Lee}, {Miville-Desch{\^e}nes}, \&
  {Murray}}]{Lee.etal.2016}
{Lee}, E.~J., {Miville-Desch{\^e}nes}, M.-A., \& {Murray}, N.~W. 2016, \apj,
  833, 229

\bibitem[{{Leroy} {et~al.}(2008){Leroy}, {Walter}, {Brinks}, {Bigiel}, {de
  Blok}, {Madore}, \& {Thornley}}]{Leroy.etal.2008}
{Leroy}, A.~K., {Walter}, F., {Brinks}, E., {et~al.} 2008, \aj, 136, 2782

\bibitem[{{Leroy} {et~al.}(2013){Leroy}, {Walter}, {Sandstrom}, {Schruba},
  {Munoz-Mateos}, {Bigiel}, {Bolatto}, {Brinks}, {de Blok}, {Meidt}, {Rix},
  {Rosolowsky}, {Schinnerer}, {Schuster}, \& {Usero}}]{Leroy.etal.2013}
{Leroy}, A.~K., {Walter}, F., {Sandstrom}, K., {et~al.} 2013, \aj, 146, 19

\bibitem[{{Leroy} {et~al.}(2016){Leroy}, {Hughes}, {Schruba}, {Rosolowsky},
  {Blanc}, {Bolatto}, {Colombo}, {Escala}, {Kramer}, {Kruijssen}, {Meidt},
  {Pety}, {Querejeta}, {Sandstrom}, {Schinnerer}, {Sliwa}, \&
  {Usero}}]{Leroy.etal.2016}
{Leroy}, A.~K., {Hughes}, A., {Schruba}, A., {et~al.} 2016, \apj, 831, 16

\bibitem[{{Leroy} {et~al.}(2017){Leroy}, {Schinnerer}, {Hughes}, {Kruijssen},
  {Meidt}, {Schruba}, {Sun}, {Bigiel}, {Aniano}, {Blanc}, {Bolatto},
  {Chevance}, {Colombo}, {Gallagher}, {Garcia-Burillo}, {Kramer}, {Querejeta},
  {Pety}, {Thompson}, \& {Usero}}]{Leroy.etal.2017}
{Leroy}, A.~K., {Schinnerer}, E., {Hughes}, A., {et~al.} 2017, \apj, 846, 71

\bibitem[{{Li} {et~al.}(2017{\natexlab{a}}){Li}, {Gnedin}, \&
  {Gnedin}}]{Li.etal.2017b}
{Li}, H., {Gnedin}, O.~Y., \& {Gnedin}, N.~Y. 2017{\natexlab{a}}, ArXiv
  e-prints, arXiv:1712.01219

\bibitem[{{Li} {et~al.}(2017{\natexlab{b}}){Li}, {Gnedin}, {Gnedin}, {Meng},
  {Semenov}, \& {Kravtsov}}]{Li.etal.2017}
{Li}, H., {Gnedin}, O.~Y., {Gnedin}, N.~Y., {et~al.} 2017{\natexlab{b}}, \apj,
  834, 69

\bibitem[{{Licquia} \& {Newman}(2015)}]{Licquia.Newman.2015}
{Licquia}, T.~C., \& {Newman}, J.~A. 2015, \apj, 806, 96

\bibitem[{{Madore}(2010)}]{Madore.2010}
{Madore}, B.~F. 2010, \apjl, 716, L131

\bibitem[{{Martizzi} {et~al.}(2015){Martizzi}, {Faucher-Gigu{\`e}re}, \&
  {Quataert}}]{Martizzi.etal.2015}
{Martizzi}, D., {Faucher-Gigu{\`e}re}, C.-A., \& {Quataert}, E. 2015, \mnras,
  450, 504

\bibitem[{{McKee} \& {Krumholz}(2010)}]{McKee.Krumholz.2010}
{McKee}, C.~F., \& {Krumholz}, M.~R. 2010, \apj, 709, 308

\bibitem[{{Miville-Desch{\^e}nes} {et~al.}(2017){Miville-Desch{\^e}nes},
  {Murray}, \& {Lee}}]{MivilleDeschenes.etal.2016}
{Miville-Desch{\^e}nes}, M.-A., {Murray}, N., \& {Lee}, E.~J. 2017, \apj, 834,
  57

\bibitem[{{Murray}(2011)}]{Murray.2011}
{Murray}, N. 2011, \apj, 729, 133

\bibitem[{{Naab} \& {Ostriker}(2017)}]{Naab.Ostriker.2017}
{Naab}, T., \& {Ostriker}, J.~P. 2017, \araa, 55, 59

\bibitem[{{Navarro} {et~al.}(1996){Navarro}, {Frenk}, \& {White}}]{NFW.1996}
{Navarro}, J.~F., {Frenk}, C.~S., \& {White}, S.~D.~M. 1996, \apj, 462, 563

\bibitem[{{Navarro} {et~al.}(1997){Navarro}, {Frenk}, \& {White}}]{NFW.1997}
---. 1997, \apj, 490, 493

\bibitem[{{Orr} {et~al.}(2017){Orr}, {Hayward}, {Hopkins}, {Chan},
  {Faucher-Gigu{\`e}re}, {Feldmann}, {Kere{\v s}}, {Murray}, \&
  {Quataert}}]{Orr.etal.2017}
{Orr}, M., {Hayward}, C., {Hopkins}, P., {et~al.} 2017, {\mnras}, accepted
  (arXiv:1701.01788)

\bibitem[{{Padoan} {et~al.}(2014){Padoan}, {Federrath}, {Chabrier}, {Evans},
  {Johnstone}, {J{\o}rgensen}, {McKee}, \& {Nordlund}}]{Padoan.etal.2014}
{Padoan}, P., {Federrath}, C., {Chabrier}, G., {et~al.} 2014, Protostars and
  Planets VI, 77

\bibitem[{{Padoan} {et~al.}(2012){Padoan}, {Haugb{\o}lle}, \&
  {Nordlund}}]{Padoan.etal.2012}
{Padoan}, P., {Haugb{\o}lle}, T., \& {Nordlund}, {\AA}. 2012, \apjl, 759, L27

\bibitem[{{Padoan} {et~al.}(2017){Padoan}, {Haugb{\o}lle}, {Nordlund}, \&
  {Frimann}}]{Padoan.etal.2017}
{Padoan}, P., {Haugb{\o}lle}, T., {Nordlund}, {\AA}., \& {Frimann}, S. 2017,
  \apj, 840, 48

\bibitem[{{Rebolledo} {et~al.}(2015){Rebolledo}, {Wong}, {Xue}, {Leroy},
  {Koda}, \& {Donovan Meyer}}]{Rebolledo.etal.2015}
{Rebolledo}, D., {Wong}, T., {Xue}, R., {et~al.} 2015, \apj, 808, 99

\bibitem[{{Rudd} {et~al.}(2008){Rudd}, {Zentner}, \&
  {Kravtsov}}]{Rudd.etal.2008}
{Rudd}, D.~H., {Zentner}, A.~R., \& {Kravtsov}, A.~V. 2008, \apj, 672, 19

\bibitem[{{Safranek-Shrader} {et~al.}(2017){Safranek-Shrader}, {Krumholz},
  {Kim}, {Ostriker}, {Klein}, {Li}, {McKee}, \&
  {Stone}}]{Safranek-Shrader.etal.2017}
{Safranek-Shrader}, C., {Krumholz}, M.~R., {Kim}, C.-G., {et~al.} 2017, \mnras,
  465, 885

\bibitem[{{Saitoh} {et~al.}(2008){Saitoh}, {Daisaka}, {Kokubo}, {Makino},
  {Okamoto}, {Tomisaka}, {Wada}, \& {Yoshida}}]{Saitoh.etal.2008}
{Saitoh}, T.~R., {Daisaka}, H., {Kokubo}, E., {et~al.} 2008, \pasj, 60, 667

\bibitem[{{Schaye} \& {Dalla Vecchia}(2008)}]{Schaye.DallaVecchia.2008}
{Schaye}, J., \& {Dalla Vecchia}, C. 2008, \mnras, 383, 1210

\bibitem[{{Schmidt}(1959)}]{Schmidt.1959}
{Schmidt}, M. 1959, \apj, 129, 243

\bibitem[{{Schmidt} {et~al.}(2014){Schmidt}, {Almgren}, {Braun}, {Engels},
  {Niemeyer}, {Schulz}, {Mekuria}, {Aspden}, \& {Bell}}]{Schmidt.etal.2014}
{Schmidt}, W., {Almgren}, A.~S., {Braun}, H., {et~al.} 2014, \mnras, 440, 3051

\bibitem[{{Schruba} {et~al.}(2010){Schruba}, {Leroy}, {Walter}, {Sandstrom}, \&
  {Rosolowsky}}]{Schruba.etal.2010}
{Schruba}, A., {Leroy}, A.~K., {Walter}, F., {Sandstrom}, K., \& {Rosolowsky},
  E. 2010, \apj, 722, 1699

\bibitem[{{Schruba} {et~al.}(2017){Schruba}, {Leroy}, {Kruijssen}, {Bigiel},
  {Bolatto}, {de Blok}, {Tacconi}, {van Dishoeck}, \&
  {Walter}}]{Schruba.etal.2017}
{Schruba}, A., {Leroy}, A.~K., {Kruijssen}, J.~M.~D., {et~al.} 2017, \apj, 835,
  278

\bibitem[{{Semenov} {et~al.}(2016){Semenov}, {Kravtsov}, \&
  {Gnedin}}]{Semenov.etal.2016}
{Semenov}, V.~A., {Kravtsov}, A.~V., \& {Gnedin}, N.~Y. 2016, \apj, 826, 200

\bibitem[{{Semenov} {et~al.}(2017){Semenov}, {Kravtsov}, \&
  {Gnedin}}]{Semenov.etal.2017}
---. 2017, \apj, 845, 133, (Paper I)

\bibitem[{{Simpson} {et~al.}(2015){Simpson}, {Bryan}, {Hummels}, \&
  {Ostriker}}]{Simpson.etal.2015}
{Simpson}, C.~M., {Bryan}, G.~L., {Hummels}, C., \& {Ostriker}, J.~P. 2015,
  \apj, 809, 69

\bibitem[{{Somerville} \& {Dav{\'e}}(2015)}]{Somerville.Dave.2015}
{Somerville}, R.~S., \& {Dav{\'e}}, R. 2015, \araa, 53, 51

\bibitem[{{Springel} \& {Hernquist}(2003)}]{Springel.Hernquist.2003}
{Springel}, V., \& {Hernquist}, L. 2003, \mnras, 339, 289

\bibitem[{{Stecher} \& {Williams}(1967)}]{Stecher.Williams.1967}
{Stecher}, T.~P., \& {Williams}, D.~A. 1967, \apjl, 149, L29

\bibitem[{{Stinson} {et~al.}(2013){Stinson}, {Brook}, {Macci{\`o}}, {Wadsley},
  {Quinn}, \& {Couchman}}]{Stinson.etal.2013}
{Stinson}, G.~S., {Brook}, C., {Macci{\`o}}, A.~V., {et~al.} 2013, \mnras, 428,
  129

\bibitem[{{Turk} {et~al.}(2011){Turk}, {Smith}, {Oishi}, {Skory}, {Skillman},
  {Abel}, \& {Norman}}]{yt}
{Turk}, M.~J., {Smith}, B.~D., {Oishi}, J.~S., {et~al.} 2011, \apjs, 192, 9

\bibitem[{{Utomo} {et~al.}(2017){Utomo}, {Bolatto}, {Wong}, {Ostriker},
  {Blitz}, {Sanchez}, {Colombo}, {Leroy}, {Cao}, {Dannerbauer},
  {Garcia-Benito}, {Husemann}, {Kalinova}, {Levy}, {Mast}, {Rosolowsky}, \&
  {Vogel}}]{Utomo.etal.2017}
{Utomo}, D., {Bolatto}, A.~D., {Wong}, T., {et~al.} 2017, \apj, 849, 26

\bibitem[{van~der Walt {et~al.}(2011)van~der Walt, Colbert, \&
  Varoquaux}]{numpy_ndarray}
van~der Walt, S., Colbert, S.~C., \& Varoquaux, G. 2011, Computing in Science
  Engineering, 13, 22

\bibitem[{{Vutisalchavakul} {et~al.}(2016){Vutisalchavakul}, {Evans}, \&
  {Heyer}}]{Vutisalchavakul.etal.2016}
{Vutisalchavakul}, N., {Evans}, II, N.~J., \& {Heyer}, M. 2016, \apj, 831, 73

\bibitem[{{Wong} \& {Blitz}(2002)}]{wong_blitz02}
{Wong}, T., \& {Blitz}, L. 2002, \apj, 569, 157

\bibitem[{{Wong} {et~al.}(2013){Wong}, {Xue}, {Bolatto}, {Leroy}, {Blitz},
  {Rosolowsky}, {Bigiel}, {Fisher}, {Ott}, {Rahman}, {Vogel}, \&
  {Walter}}]{Wong.etal.2013}
{Wong}, T., {Xue}, R., {Bolatto}, A.~D., {et~al.} 2013, \apjl, 777, L4

\bibitem[{{Yepes} {et~al.}(1997){Yepes}, {Kates}, {Khokhlov}, \&
  {Klypin}}]{Yepes.etal.1997}
{Yepes}, G., {Kates}, R., {Khokhlov}, A., \& {Klypin}, A. 1997, \mnras, 284,
  235

\end{thebibliography}

\appendix

\section{Summary of model parameters}
\label{app:model}

\begin{deluxetable*}{cclc}
\tablecaption{Definitions of the quantities used in our model
\label{tab:defs}}
\tablewidth{0pt}
\tablehead{
\colhead{Quantity} &
\colhead{Definition} &
\colhead{Meaning} &
\colhead{Model\tablenotemark{a}}
}
\startdata
\sidehead{Modeled properties of the galaxy}
$\tglob$ & $\Mg/\SFR$ & Global depletion time of total gas & Equation~(\ref{eq:tglob2}) \\
$\fsf$ & $\Msf/\Mg$ & Star-forming gas mass fraction & Equation~(\ref{eq:fsf2}) \\
$\tauff$ & $\langle 1/\tff \rangle^{-1}_{\rm sf}$ & Average freefall time in star-forming gas & Equations~(\ref{eq:app:interp}--\ref{eq:app:width})\tablenotemark{b} \\
\sidehead{Model parameters}
$\taust$ & $\Msf/\SFR$ & Global depletion time of star-forming gas & $\tauff/\epsff$ \\
$\tp$ & Equation~(\ref{eq:Fp}) & Dynamical timescale on which non-star-forming gas becomes star-forming & $100\;(\avirsf/10)^{-0.5} \Myr$ \\
$\tmd$ & Equation~(\ref{eq:Fmdyn}) & Timescale on which star-forming gas is dynamically dispersed & Spline\tablenotemark{c} $\tmd(\avirsf)$ \\
$\xi$ & Equation~(\ref{eq:Fmfb}) & Average feedback mass-loading factor on the scale of star-forming regions & $60\;\b^{0.75}$ \\
$\taufflow$ & & $\tauff$ in the dynamics-regulated regime & Spline\tablenotemark{c} $\taufflow(\avirsf)$\\
$\tauffhi$ & & $\tauff$ in the self-regulated regime & Equation~(\ref{eq:app:tauffhi_calib})\\
\sidehead{Simulation parameters controlling local star formation and feedback}
$\epsff$ & Equation~(\ref{eq:rhoSFR}) & Star formation efficiency per freefall time & \\
$\avirsf$ & see Section~\ref{sec:sims:sf-fb} & Star formation threshold in virial parameter, $\avir < \avirsf$ & \\
$\nsf$ & see Section~\ref{sec:sims:sf-fb} & Star formation threshold in gas density, $n > \nsf$ & \\
$\b$ & see Section~\ref{sec:sims:sf-fb} & Boost factor of momentum injected per supernova &
\enddata
\tablenotetext{a}{The last column indicates model predictions for $\tglob$, $\fsf$, and $\tauff$ and calibrated values for model parameters. Listed calibrations are obtained for the $\avir$-based star formation threshold. Calibrations for the density-based threshold are provided at the end of Appendix~\ref{app:model}.}
\tablenotetext{b}{The model predicts the position and the width of $\tauff$ transition between $\taufflow$ and $\tauffhi$.}
\tablenotetext{c}{The values of $\tmd$ and $\taufflow$ are obtained directly from the \ns distribution in our simulation with $\epsff=0.01\%$ and fiducial $\b=1$ and $\avirsf=10$ (see the end of Appendix~\ref{app:model}). }
\end{deluxetable*}

Our model equations,
\begin{align}
\label{eq:app:tglob}
\tglob &= (1 + \xi) \tp + \left( 1 + \frac{\tp}{\tmd} \right) \frac{\tauff}{\epsff},\\
\label{eq:app:fsf}
\fsf &= \frac{1}{\epsff}\frac{\tauff}{\tglob},
\end{align}
are derived from the mass conservation equation between star-forming and non-star-forming states in the ISM, as explained in Section~\ref{sec:model:summary} and \P1. The parameters used in our model and their meanings are summarized in Table~\ref{tab:defs}.

As we showed in Section~\ref{sec:predict:ebdep}, the model equations describe our simulation results even if we assume that all the model parameters, including $\tauff$, are fixed. However, the accuracy of our model can be improved if the variation of $\tauff$ is incorporated. 

\begin{figure}
\centering
\includegraphics[width=\columnwidth]{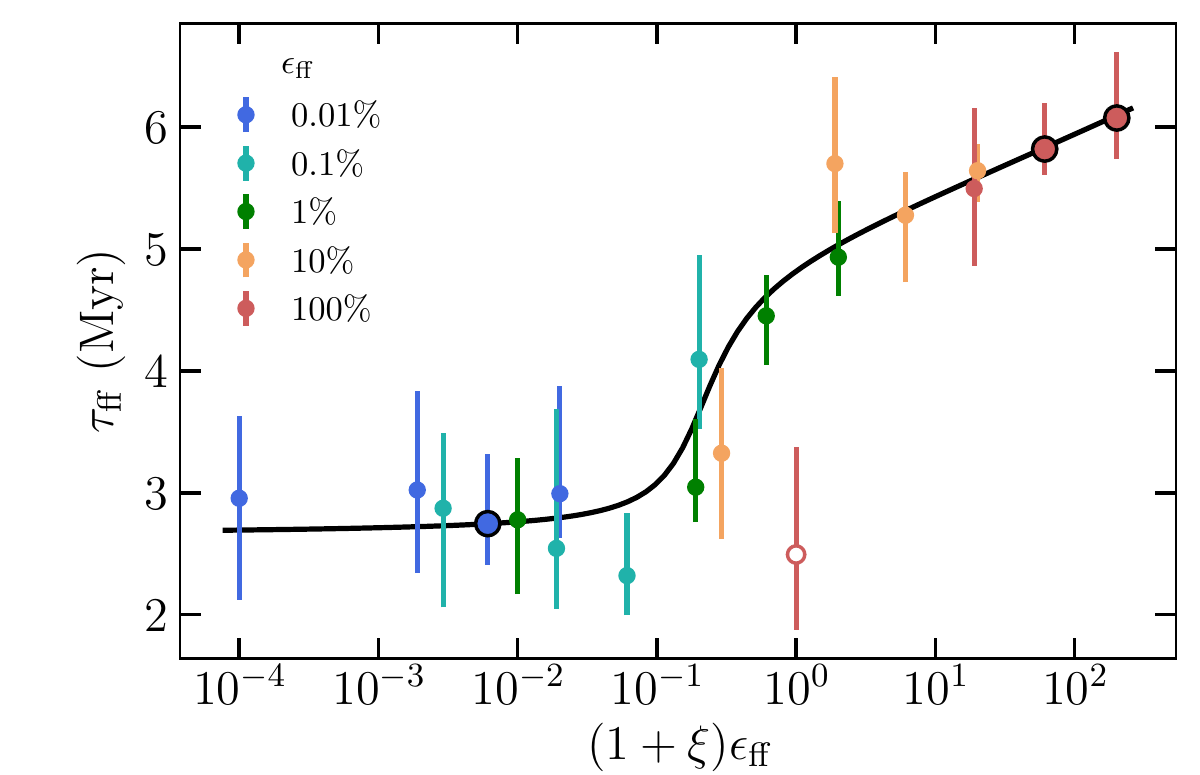}
\caption{\label{fig:tauff} Comparison of our model prediction for the variation of the freefall time in the star-forming gas, $\tauff$, with the results of our simulations. To measure $\tauff$ in the dynamics-regulation (small $(1+\xi)\epsff$) and self-regulation (large $(1+\xi)\epsff$) regimes and the parameters of the transition between these regimes, we use the the same runs that were used to calibrate $(1+\xi)\tp$ and $\tp/\tmd$ in Section~\ref{sec:predict:ebdep} (indicated by circled points). The predictions of our model agree with the results of all simulations, except for the run with $\b=0$ and $\epsff=100\%$ (open circle), which does not remain in equilibrium owing to the rapid global gas consumption.}
\end{figure}

To account for the variation of $\tauff$ with $\xi$ and $\epsff$, we note that star-forming gas is removed at a rate $\SFR+\Fmfb \propto (1+\xi)\epsff$ and therefore $\tauff$ increases from $\tau_{\rm ff}^{\rm dr}$ to $\tau_{\rm ff}^{\rm sr}$ when $(1+\xi)\epsff$ increases and the galaxy switches from the dynamics-regulation (thus the superscript ``dr'') to the self-regulation (``sr'') regime. Note that the dependence of $\tauff$ on the combination $(1+\xi)\epsff$ is itself a prediction of the model. This prediction is confirmed by the simulation results shown in Figure~\ref{fig:tauff}, as $\tauff$ from all simulations with different $\epsff$ and $\xi$ scale as a function of $(1+\xi)\epsff$.

We then can interpolate $\tauff$ between $\taufflow$ and $\tauffhi$ as a function of $\psi \equiv (1+\xi)\epsff$ using a simple fitting formula shown with the solid line in Figure~\ref{fig:tauff}:
\begin{align}
\label{eq:app:interp}
\tauff &= \taufflow + f(\psi)\; (\tauffhi - \taufflow),\\
\label{eq:app:arctan}
f(\psi) &= \frac{1}{\pi} \arctan \left( \frac{\log(\psi)-\log(\psi_{\rm cr})}{w} \right) + \frac{1}{2},
\end{align}
in which the position, $\psi_{\rm cr}$, and the width, $w$, of transition can be predicted by our model. Specifically, from Equation~(\ref{eq:epsffcr}), the transition happens at
\begin{equation}
\label{eq:app:psicr}
\psi_{\rm cr} = \left( 1+\frac{\tp}{\tmd} \right) \frac{ \overline{\tauff} }{\tp},
\end{equation}
where, for simplicity, we assume average $\overline{\tauff} = 4\Myr$, representative of our simulation results. The width of the transition can be estimated assuming that as $(1+\xi)\epsff$ increases from very low values, the transition appears when $\SFR+\Fmfb$ becomes comparable to $\Fmdyn$. This yields $(1+\xi)\epsff \sim \taufflow/\tmd$ and thus the width is
\begin{equation}
\label{eq:app:width}
w = \log(\psi_{\rm cr})-\log({\taufflow}/{\tmd}).
\end{equation}

In the dynamics-regulation regime, i.e. at small $(1+\xi)\epsff$, $\taufflow$ is determined by the high-density tail of the star-forming gas probability density function (PDF) and is independent of the star formation. In the self-regulation regime, i.e. at large $(1+\xi)\epsff$, $\tauffhi$ increases as the high-density tail is dispersed and the star-forming gas stays close to the star formation threshold. These trends of $\tauff$ in the limiting regimes are apparent in the results of our simulation suite shown in Figure~\ref{fig:tauff}.

Equations~(\ref{eq:app:interp}--\ref{eq:app:width}) augment the main equations of our model (\ref{eq:app:tglob} and \ref{eq:app:fsf}) with the variation of $\tauff$ with our model parameters: $\epsff$, $\xi$, $\tp$, $\tmd$, $\tauffhi$, $\taufflow$. To calibrate the dependence of these parameters on our simulation parameters---i.e. local efficiency $\epsff$, feedback boost factor $\b$, and star formation threshold $\avirsf$---we assume
\begin{align}
\label{eq:app:xi}
\xi &= \xi_0 \b^\beta,\\
\label{eq:app:tp}
\tp &= (100\Myr)\;(\avirsf/10)^\gamma,\\
\label{eq:app:tauffhi}
\tauffhi &= (\tauffhi)_0\; (\psi/100)^a (\avirsf/10)^b.
\end{align}
Here we assume that at fiducial $\avirsf=10$, $\tp\sim100\Myr$, as indicated by the results in \P1. The value of $\tp$ does depend on the star formation threshold because the threshold determines when the transition from the non-star-forming to the star-forming state happens in the evolution of each gas parcel. Equation~(\ref{eq:app:tauffhi}) incorporates the dependence of $\tauffhi$ on $\psi \equiv (1+\xi)\epsff$ and star formation threshold discussed above.

Next, as detailed in Sections~\ref{sec:predict:ebdep} and \ref{sec:predict:thresholds}, we use three runs in the self-regulation regime with different feedback boost, $\b$, and threshold, $\avirsf$, to estimate
\begin{align}
(1+\xi)\tp &\sim (6\Gyr)\;\b^{0.75}\;(\avirsf/10)^{-0.5},\\
\label{eq:app:tauffhi_calib}
\tauffhi &\sim (6\Myr)\;(\psi/100)^{0.035}\;(\avirsf/10)^{0.4},
\end{align}
which imply $\xi_0 = 60$, $\beta=0.75$, $\gamma=-0.5$, $(\tauffhi)_0=6\Myr$, $a=0.035$, and $\b=0.4$. Note, in particular, that $\xi \approx 60\; \b^{0.75}$ which implies that our fiducial feedback ($\b=1$) is rather efficient and $\xi \gg 1$ in Equation~(\ref{eq:app:tglob}). 

Finally, the last two parameters, $\taufflow$ and $\tmd$, are measured as functions of $\avirsf$ directly from the \ns distribution in our run with $\epsff = 0.01\%$ (bottom left panel of Figure~\ref{fig:phases_fid}). To this end, we note that because of the dynamics-regulation regime, this distribution would not change if $\avirsf$ were varied. We then measure $\taufflow(\avirsf)$ as $\langle 1/\tff \rangle^{-1}$ in gas with $\avir < \avirsf$ and $\tmd(\avirsf)$ from $\fsf(\avirsf)$ using Equation~(\ref{eq:fsf_lowepsff}): $\tmd = \tp/(1/\fsf-1)$. We spline $\taufflow(\avirsf)$ and $\fsf(\avirsf)$ and show them with blue lines in the bottom two panels of the middle column in Figure~\ref{fig:fb}. For example, at our fiducial threshold of $\avirsf=10$, $\taufflow \sim 2.5\Myr$ and $\fsf \sim 20\%$, which implies $\tmd \sim \tp/4 \sim 25\Myr$.

For the density-based star formation threshold we study only the dependence on $\nsf$ but not on $\b$. In other words, we replace Equations~(\ref{eq:app:xi}--\ref{eq:app:tauffhi}) with 
\begin{align}
\xi &= \xi_0,\\
\tp &= (100\Myr)\;(\nsf/100\cc)^\gamma,\\
\tauffhi &= (\tauffhi)_0\; (\nsf/100\cc)^{-1/2}.
\end{align}
Note that the slope in the last equation is not a parameter because, in contrast to the $\avir$-based threshold, the dependence of $\tauffhi$ on the density threshold follows from definition, since in this regime all star-forming gas has density $\sim \nsf$. For the same reason, $\tauffhi$ does not depend on $\psi$ for a density-based threshold.

The value $(\tauffhi)_0 = 5\Myr$ is measured directly from the simulation with $\epsff = 100\%$ and $\nsf = 100\cc$, and using another run with lower $\nsf$ we get \begin{align}
(1+\xi)\tp &\sim (4.5\Gyr)\;(\nsf/100\cc)^{0.5},
\end{align}
and thus $\xi_0 = 45$ and $\gamma=0.5$.

\section{Model for molecular gas mass fraction}
\label{app:H2}

\begin{figure}
\centering
\includegraphics[width=\columnwidth]{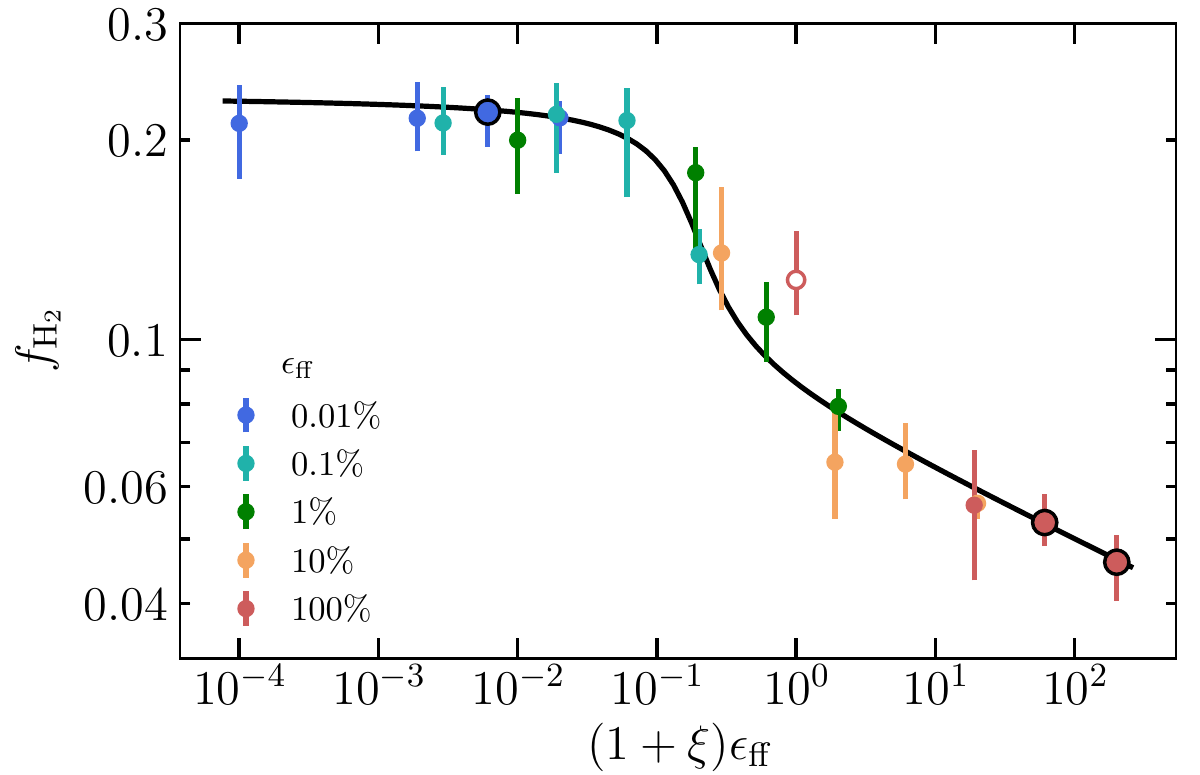}
\caption{\label{fig:epsxi-fH2} Comparison of our model predictions for the variation of the global molecular mass fraction, $\fH2 \equiv \MH2/\Mg$, with the results of the simulations. We average the total molecular mass in the simulations between 300 and 600 Myr, defining it as a sum of molecular masses in individual cells, which are computed using the \citet{KMT2} model (see the beginning of Section~\ref{sec:observ} for details). To obtain model predictions, we interpolate $\fH2$ between its values at low and high $(1+\xi)\epsff$ calibrated using the simulations in corresponding regimes (large circled points). The value of $(1+\xi)\epsff$ at which this transition occurs and the width of the transition are predicted by the model (Equations~\ref{eq:app:psicr} and \ref{eq:app:width}). The open red circle indicates the run with $\b=0$ and $\epsff=100\%$, which does not remain in equilibrium owing to the rapid global gas consumption.}
\end{figure}

Similarly to star-forming gas above a given threshold, molecular gas distribution is also shaped by dynamical and feedback-driven gas flows. Therefore, similarly to Section~\ref{sec:model:summary}, mass conservation can be considered for the molecular state of the ISM to derive the relation between the molecular mass fraction, $\fH2 = \MH2/\Mg$, and the timescales of relevant processes supplying and removing molecular gas. In the equation for total molecular gas mass conservation,
\begin{equation}
\label{eq:MH2cons}
\dot{M}_{\rm H_2} = \Fmolp - \Fmolm - \SFR,
\end{equation}
we parameterize relevant fluxes as 
\begin{align}
\label{eq:Fmolp} \Fmolp &\equiv \frac{(1-\fH2) \Mg}{\tmolp},\\
\label{eq:Fmol} \Fmolm &\equiv \frac{\fH2 \Mg}{\tmolm},\\
\label{eq:SFRapp} \SFR &\equiv \frac{\fsf \Mg}{\taust}.
\end{align}
That is, $\Fmolp$ and $\Fmolm$ are parameterized analogously to $\Fp$ and $\Fm$ in Section~\ref{sec:model:summary} and the equation for $\SFR$ repeats Equation~(\ref{eq:SFR}).

Then, assuming steady state with $\dot{M}_{\rm H_2} \approx 0$, substitution of Equations~(\ref{eq:Fmolp})--(\ref{eq:SFRapp}) into Equation~(\ref{eq:MH2cons}) yields
\begin{equation}
\label{eq:app:fH2}
\fH2 \approx \frac{1 - (\tmolp/\taust) \fsf }{ 1 + (\tmolp/\tmolm)},
\end{equation}
where $\fsf$ can be computed using Equation~(\ref{eq:fsf2}).

At low $\epsff$, $\taust = \tauff/\epsff \to \infty$, and thus $\fH2 \sim [1 + (\tmolp/\tmolm)]^{-1}$, which is analogous to Equation~(\ref{eq:fsf_lowepsff}), with $\tmolp$ and $\tmolm$ independent of star formation and feedback. At high $\epsff$, all terms in Equation~(\ref{eq:app:fH2}) are relevant and $\tmolm$ depends on star formation and feedback parameters in a nontrivial way. This nontrivial dependence is more complex than a simple scaling with local depletion time $\taust$---as was the case for the star-forming gas removal time $\tm\approx\tmfb = \taust/\xi$---because $\tmolm$ also depends on the dynamics of non-star-forming molecular gas and the details of its dissociation. 

Thus, $\tmolm$ cannot be easily related to the parameters of subgrid star formation and feedback, which does not allow to use Equation~(\ref{eq:app:fH2}) for predicting how $\fH2$ depends on the parameters of star formation and feedback. However, this dependence can be calibrated using the same approach that we used to model variation of the freefall time in star-forming gas, $\tauff$ (Appendix~\ref{app:model}).

The approach is similar because the change of both $\tauff$ and $\fH2$ reflects the response of the gas PDF to the changing feedback-induced flux parameterized by $(1+\xi)\epsff$, and thus $\fH2$ variation with $(1+\xi)\epsff$ is qualitatively similar to that of $\tauff$. Indeed, as Figure~\ref{fig:epsxi-fH2} shows, at $(1+\xi)\epsff < 0.1$, the value of $\fH2 \sim 20\%$ remains independent of $\xi$ and $\epsff$ because feedback is too weak to affect the gas PDF. Between $(1+\xi)\epsff \sim 0.1$ and $1$, the value of $\fH2$ decreases by a factor of 2 as feedback clears the high-density tail of the molecular gas distribution, and at $(1+\xi)\epsff>1$ the decrease of $\fH2$ slows down as the non-star-forming molecular gas accumulates above the star formation threshold. As the black curve shows, such variation of $\fH2$ with $(1+\xi)\epsff$ can be approximated by the same fitting formula as the one used for $\tauff$ (Equations~\ref{eq:app:interp}--\ref{eq:app:width}), with the limiting values of $\fH2$ at low and high $(1+\xi)\epsff$ calibrated using the simulations: $f^{\rm dr}_{\rm H_2} = 23\%$ and $f^{\rm sr}_{\rm H_2} = 0.05\;[(1+\xi)\epsff/60]^{-0.1}$.

The discussed effect of star formation and feedback on $\fH2$ also allows us to predict the variation of $\fH2$ with the star formation threshold. Namely, in the dynamics-regulation regime, we expect $f_{\rm H_2} \sim 23\%$ to be independent of the star formation threshold because the ISM gas distribution remains independent of star formation. In the self-regulation regime, $\fH2$ decreases when the threshold is shifted to higher $\avirsf$ or lower $\nsf$, because the region in the \ns plane corresponding to the non-star-forming molecular gas shrinks. 

\end{document}